\documentclass[onecolumn,draftclsnofoot,12pt]{IEEEtran}
\usepackage{amsfonts}
\usepackage{times}
\usepackage{graphicx}
\usepackage{latexsym}
\usepackage{dsfont}
\usepackage{amssymb}
\usepackage{amsmath}
\usepackage{cite}
\usepackage{verbatim}

\newtheorem{algorithm}{Algorithm}

\newcommand{\figref}[1]{{Fig.}~\ref{#1}}

\def\bb0{{\mathbb{0}}}

\def\ba{{\mathbf{a}}}
\def\bb{{\mathbf{b}}}
\def\bc{{\mathbf{c}}}

\def\bff{{\mathbf{f}}}
\def\bg{{\mathbf{g}}}

\def\bm{{\mathbf{m}}}

\def\bv{{\mathbf{v}}}
\def\bw{{\mathbf{w}}}
\def\bx{{\mathbf{x}}}
\def\by{{\mathbf{y}}}
\def\bz{{\mathbf{z}}}
\def\b0{{\mathbf{0}}}

\def\bA{{\mathbf{A}}}
\def\bB{{\mathbf{B}}}

\def\bD{{\mathbf{D}}}

\def\bF{{\mathbf{F}}}

\def\bH{{\mathbf{H}}}
\def\bI{{\mathbf{I}}}

\def\bR{{\mathbf{R}}}

\def\bW{{\mathbf{W}}}
\def\bX{{\mathbf{X}}}
\def\bY{{\mathbf{Y}}}

\def\bbC{{\mathbb{C}}}

\def\bbE{{\mathbb{E}}}

\def\bbR{{\mathbb{R}}}

\def\cA{\mathcal{A}}

\def\cC{\mathcal{C}}
\def\cD{\mathcal{D}}
\def\cE{\mathcal{E}}
\def\cF{\mathcal{F}}

\def\cM{\mathcal{M}}
\def\cN{\mathcal{N}}
\def\cO{\mathcal{O}}
\def\cP{\mathcal{P}}
\def\cQ{\mathcal{Q}}

\def\cS{\mathcal{S}}
\def\cT{\mathcal{T}}

\def\cW{\mathcal{W}}
\def\cX{\mathcal{X}}
\def\cY{\mathcal{Y}}
\def\cZ{\mathcal{Z}}

\def\sfG{\mathsf{G}}

\def\sfd{{\mathsf{d}}}

\def\sf0{{\mathsf{0}}}

\def\kron{\otimes}

\usepackage{epstopdf}
\usepackage{enumerate}
\usepackage{algorithm}
\usepackage{amsmath}
\usepackage[noend]{algpseudocode}
\usepackage{float}
\usepackage{color}
\usepackage{makeidx}
\usepackage{bbm}

\usepackage{cleveref}
\usepackage{url}
\usepackage{steinmetz}
\usepackage{varwidth}

\usepackage{caption}
\usepackage{subcaption}
\newcommand{\sref}[1]{{Section}~\ref{#1}}

\newcommand{\pinv}[1]{\ensuremath{#1^{\dagger}}} 	

\DeclareMathOperator*{\argmax}{arg\,max}

\def\j{\mathrm{j}}

\def \rm {\mathrm}

\algnewcommand{\Initialize}[1]{%
	\State \textbf{Initialize:} \parbox[t]{.8\linewidth}{\raggedright #1}
}

\begin{document}
\title{Millimeter Wave MIMO based Depth Maps \\ for Wireless Virtual and Augmented Reality}
\author{Abdelrahman Taha$^{1}$, Qi Qu$^{2}$, Sam Alex$^{2}$, Ping Wang$^{2}$, William L. Abbott$^{2}$, and \\ Ahmed Alkhateeb$^{1}$\\  $^{1}$ Arizona State University, \{a.taha, alkhateeb\}@asu.edu \\ $^{2}$  Facebook, Inc., \{qqu, sampalex,  pingwang, billabbott\}@fb.com \thanks{Abdelrahman Taha and Ahmed Alkhateeb are  with the School of Electrical, Computer and Energy Engineering, Arizona State University. Qi Qu, Sam Alex, Ping Wang, and Bill Abbott are with Facebook, Inc. This material is based upon work supported by Facebook, Inc.}}
\maketitle

\begin{abstract}
	Augmented and virtual reality systems (AR/VR) are rapidly becoming key components of the wireless landscape. For immersive AR/VR experience, these devices should be able to construct accurate depth perception of the surrounding environment. Current AR/VR devices rely heavily on using RGB-D depth cameras to achieve this goal. The performance of these depth cameras, however, has clear limitations in several scenarios, such as the cases with shiny objects, dark surfaces, and abrupt color transition among other limitations. In this paper, we propose a novel solution for AR/VR depth map construction using mmWave MIMO communication transceivers. This is motivated by the deployment of advanced mmWave communication systems in future AR/VR devices for meeting the high data rate demands and by the interesting propagation characteristics of mmWave signals. Accounting for the constraints on these systems, we develop a comprehensive framework for constructing accurate and high-resolution depth maps using mmWave  systems. In this framework,  we developed new sensing beamforming codebook approaches that are specific for the depth map construction objective. Using these codebooks, and leveraging tools from successive interference cancellation, we develop a joint beam processing approach that can construct high-resolution depth maps using practical mmWave antenna arrays. Extensive simulation results highlight the potential of the proposed solution in building accurate depth maps. Further, these simulations show the promising gains of mmWave based depth perception compared to RGB-based approaches in several important use cases. 
\end{abstract}

\section{Introduction} \label{sec:Intro}
Wireless augmented and virtual reality (AR/VR) applications are recently attracting increasing interest. Realizing wireless AR/VR in practice can open the door for a wide range of interesting applications and use cases. Enabling Immersive AR/VR experience, however, requires high resolution and accurate depth perception. This can potentially allow the wireless AR/VR users to move freely within their indoor or outdoor environment. Current depth perception approaches for AR/VR systems rely mainly on RGB-D (depth) cameras for constructing the depth maps. While RGB-D based depth map construction approaches can generally provide good accuracy, they suffer from critical limitations in scenarios with bright shiny or transparent surfaces, dark objects, and large rooms among others.  
These limitations stem from the fundamental properties of the way visible light propagate and interact with the different surfaces.

In order to overcome these limitations, \textbf{we propose to leverage mmWave systems and signals for improving the depth map estimation accuracy}. This is motivated by the interesting characteristics of mmWave signals and by the note that mmWave systems will be deployed in future AR/VR devices anyway for meeting the wireless communication requirements  \cite{HTC}. In terms of the mmWave signal characteristics, the propagation of these signals is not affects by the interference from the light sources which makes mmWave systems capable of detecting bright and dark objects.  Further, the mmWave diffuse scattering and specular reflection properties could help in detecting transparent objects as well as rough surfaces. These aspects among others motivate exploring the potential of leveraging mmWave transceivers for complementing the RGB-D depth-maps in AR/VR systems, which is the focus of this paper.  

\subsection{Prior Work}

Previous depth map construction approaches focused on leveraging: (i) monocular images using RGB cameras \cite{Hu2019}, (ii) passive/active stereo images using either RGB-D depth cameras \cite{Mal2018,Riegler2019} or  infrared (IR) stereo cameras \cite{Zhang2018,Cheng2019p}, and (iii) gated images using active gated imaging cameras \cite{Gruber2018,Gruber2019}.
In \cite{Hu2019}, a monocular depth estimation approach capable of capturing the object boundaries is proposed. In \cite{Mal2018}, RGB images along with sparse depth samples, acquired from depth cameras or computed via  Simultaneous Localization and Mapping (SLAM) algorithms, are used jointly to reconstruct the depth maps. An alternative approach for depth estimation was proposed in \cite{Riegler2019}, where a monocular structured-light camera --- a calibrated stereo set-up with one camera and one laser projector--- is leveraged for estimating the disparity. As for the active stereo systems, in \cite{Zhang2018}, IR projected pattern from stereo IR cameras is utilized for depth estimation through active stereo matching. The IR images are acquired from the Intel Realsense camera \cite{IntelIRcamera}. Also, the IR pattern characteristics needed for active stereo matching are described in \cite{Cheng2019p}. In addition, high-resolution depth images can be achieved for far objects using active gated imaging systems, as in \cite{Gruber2018,Gruber2019}.

These depth map construction approaches \cite{Hu2019,Mal2018,Riegler2019,Wang2019,Zhang2018,Cheng2019p,Gruber2018,Gruber2019}, however, have several important limitations complications as follows. 
(i) First, these depth map construction approaches normally fail to sense the depth for shiny, dark, transparent, and distant surfaces. While there are some attempts in solving these challenges using IR stereo cameras \cite{Zhang2018} or excessive processing of the RGB-D images \cite{Zhang_2018_CVPR}, there is no complete and general solution yet to this problem. 
(ii) Further, these IR and RGB-D based depth map construction algorithms suffer from a critical limitation, which is the depth ambiguity for far objects/surfaces. The depths for distant surfaces can not be resolved by the algorithms in \cite{Zhang2018,Zhang_2018_CVPR}.
(iii) Another key challenge is the additional bill of materials (BOM) cost incurred from integrating the IR stereo camera systems in the wireless AR/VR device architectures. On the contrary, the existing mmWave systems in the wireless AR/VR device architectures incurs no additional BOM cost when leveraged for depth map estimation purposes jointly with the primary purpose of wireless communications.
(iv) The field of view coverage is also a main challenge. The depth map coverage is limited by the camera field of view. The camera field of view is constrained by the camera lens and by the light sensor. The field of view in mmWave MIMO systems, however, is constrained by the array radiation pattern, as will be explained in \sref{sec:CB_design}. By contrast, the typical field of view in mmWave MIMO systems can be larger than the typical camera field of view.

These challenges motivate the research for other technologies to complement the RGB-D cameras in accurately sensing the VR/AR environment. One promising technology for this goal is employing wireless millimeter wave (mmWave) systems. Since mmWave antenna arrays will be used to satisfy the communication high data rate demands of wireless VR/AR, it is interesting to investigate if they could also be useful for VR/AR-relevant sensing functions, such as depth estimation. 
Initial studies for using mmWave communication arrays for radar and sensing were presented in \cite{8108565,8114253}. These studies, however, focused only on the ranging problem (of one or multiple targets), not on the depth map construction problem.  Other mmWave sensing and tracking work that was not restricted to communications hardware was presented in \cite{Soli,mmTrack}. The research in  \cite{Soli,mmTrack}, though, targeted tracking a single object in a small distance, and cannot be directly applied to depth estimation of surrounding surfaces in VR/AR. Further, the work in \cite{8108565,8114253,Soli,mmTrack}, did not study the trade-offs between estimation accuracy and different system parameters, such as number of antennas and adopted bandwidth, and did not compare between the system performance under transceiver architectures constraints, such as those imposed on the analog phased-array transceiver architectures. By contrast, interesting research challenges are accompanying the mmWave MIMO based scene depth map construction framework ranging from beam codebook design challenges to scene depth estimation challenges. These challenges will be addressed in this work and will be explained in detail in \sref{ssec:mmWave_depth_est}.

\subsection{Contribution}

In this paper, we consider the mmWave MIMO based depth map construction problem for AR/VR systems, adopting mmWave communication hardware and frame structure. The contributions of this paper can be summarized as follows.
\begin{itemize}
	\item \textit{mmWave MIMO depth map construction framework:} We formulate the mmWave MIMO depth map construction problem and propose a general framework for building depth maps under the constraints imposed by mmWave communication hardware and frame structure.  
	
	\item \textit{A design for depth-map suitable sensing beamforming codebook:} We define the characteristics of the desirable mmWave sensing beamforming codebook for efficient depth map construction and develop a codebook construction approach that meets these characteristics.
	
	\item \textit{High-resolution depth map construction approach:} Given the designed beamforming codebook, we develop a novel signal processing approach for jointly processing the signals received by the sensing beams and building high-resolution depth maps. 

\end{itemize}
The proposed solution is extensively evaluated using accurate ray-tracing channels generated from Wireless InSite \cite{Remcom}, and ground truth depth images generated from Blender \cite{Blender}. The simulation results show the promise of mmWave MIMO sensing in becoming a viable depth estimation solution for communication-constrained sensing systems, either as a standalone approach or as an integrated approach with RGB-D depth cameras. These simulation results can be of great usefulness for various applications; they can be generally applied to AR/VR devices, smart home devices, or auto drive devices.

\textbf{Notation:} We use the following notation throughout this paper: $\bA$ is a matrix, $\ba$ is a vector, $a$ is a scalar, and $\cA$ is a set. $\|\ba \|_p$ is the p-norm of $\ba$. $|\bA|$ is the determinant of $\bA$, $\|\bA \|_F$ is its Frobenius norm, whereas $\bA^T$, $\bA^H$, $\bA^*$, $\bA^{-1}$, $\pinv{\bA}$ are its transpose, Hermitian (conjugate transpose), conjugate, inverse, and pseudo-inverse respectively. $\cA_p$ is the $p^\rm{th}$ element of the set $A$. $[\bA]_{r,:}$ and $[\bA]_{:,c}$ are the $r^\rm{th}$ row and $c^\rm{th}$ column of the matrix $\bA$, respectively. $\mathrm{diag}(\ba)$ is a diagonal matrix with the entries of $\ba$ on its diagonal. $\bI$ is the identity matrix. $\mathbf{1}_{N}$ and $\mathbf{0}_{N}$ are the $N$-dimensional all-ones and all-zeros vector respectively. $\bA \kron \bB$ is the Kronecker product of $\bA$ and $\bB$, $\bA \circ \bB$ is their Khatri-Rao product, and $\bA \odot \bB$ is their Hadamard product. $\cN(\bm,\bR)$ is a complex Gaussian random vector with mean $\bm$ and covariance $\bR$. $\bbE\left[\cdot\right]$ is used to denote expectation. $\mathrm{vec}(\mathbf{A})$ is a vector whose elements are the stacked columns of matrix $\mathbf{A}$.

\begin{figure}[t] 
	\centering
	{\includegraphics[width=.9\linewidth]{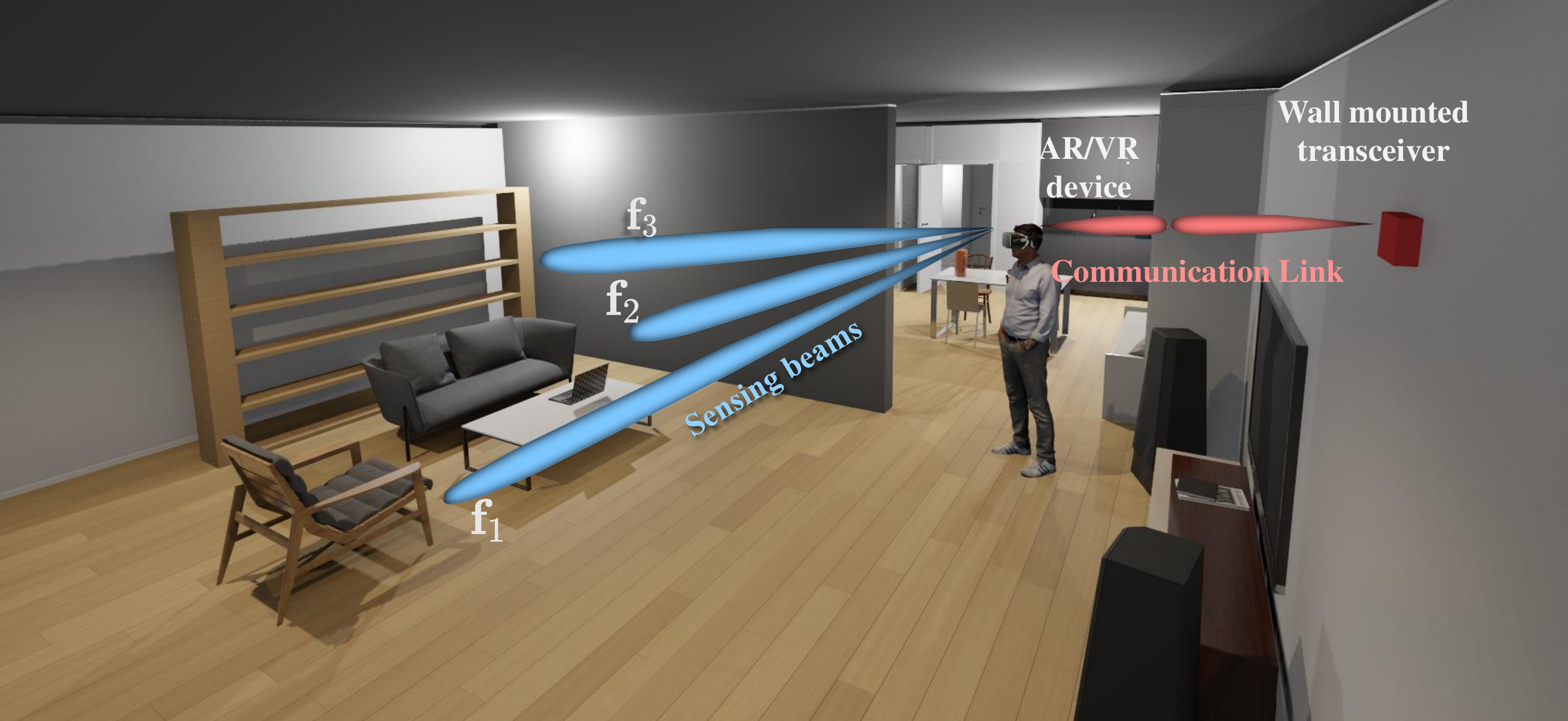}}
	\caption{The considered setup where the mmWave communication system, deployed at the AR/VR device, is jointly leveraged for sensing and depth map construction. This figure is generated using Blender \cite{Blender} with 3D models downloaded from \cite{BlenderKit,BlenderDownloads,TurboSquid,Free3D}. } 
	\label{fig:Sigmodel_2a}
\end{figure}

\section{System and Channel Models} \label{sec:Sys_CH_Models}
In this section, the system model for the adopted communication-constrained sensing framework is first formulated, followed by of the characterization of the adopted channel model. 

\begin{figure}[t] 
	\centering
	{\includegraphics[width=0.9\linewidth]{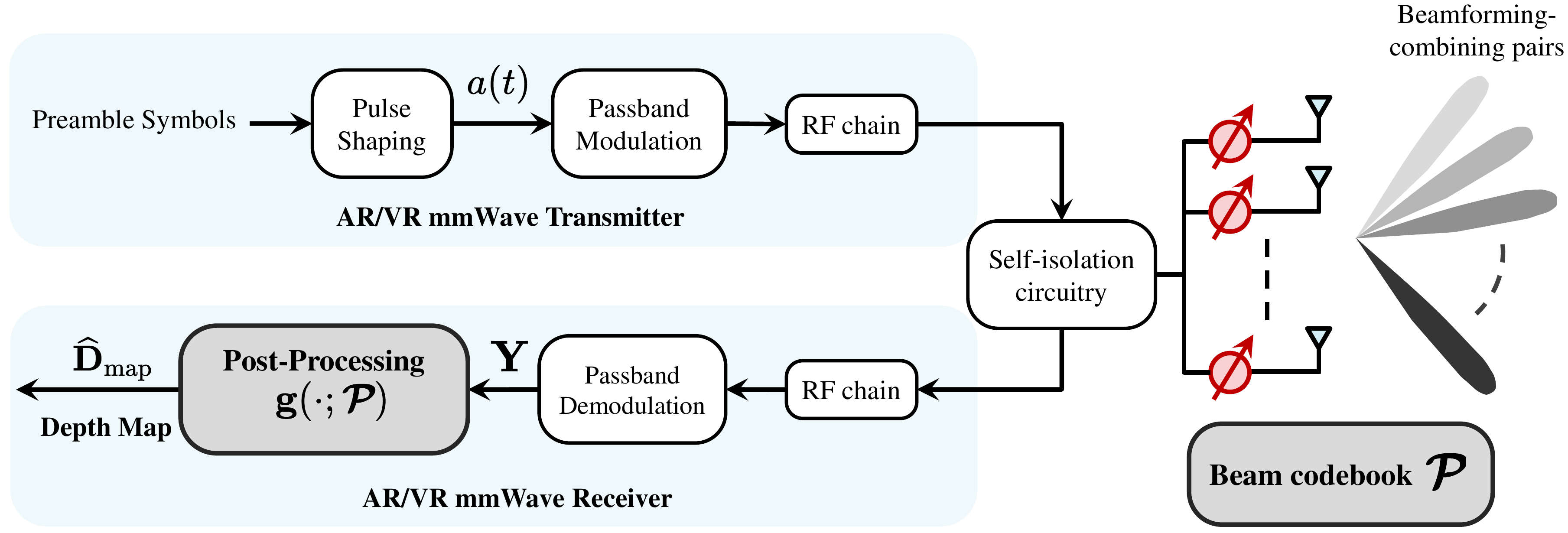}}
	\caption{A block diagram of the communication-constrained sensing model is illustrated.  The sensing framework, $\boldsymbol{\Pi}$, consists of (a) the beam codebook design $\boldsymbol{\cP}$ and (b) the post-processing design $\bg\left(.,\boldsymbol{\cP} \right)$, to estimate the scene depth map $\widehat{\cD}$. The upper path represents the transmitter path while the lower path represents the receiver path.}
	\label{fig:Sigmodel_2b}
\end{figure}

\subsection{System Model} \label{subsec:Sys_model}

In this paper, we propose to reuse the same AR/VR  mmWave communication system/circuits to do the sensing and depth map construction, as shown in \figref{fig:Sigmodel_2a}. Hence, we adopt a sensing model that accounts for the mmWave communication system/circuit constraints. This communication-constrained sensing model consists of a transmitter and a receiver; both are connected through a self-isolation circuitry to a shared $N$ antenna array, as depicted in \figref{fig:Sigmodel_2b}. This type of operation is commonly referred to as MIMO in-band full-duplex operation \cite{Sabharwal2014}.  We assume that the transmitter and receiver chains are well-isolated by an isolation circuitry to avoid any self-interference. This assumption is reasonable with the recent developments of self-interference systems. One example of these systems is the magnetic-free non-reciprocal circulators (i) based on coupled-resonator loops \cite{Estep2014} or (ii) based on CMOS circulators operating in the $28$GHz mmWave band \cite{Dinc2017}. Another example is the receiver with integrated magnetic-free non-reciprocal circulator and baseband self-interference cancellation operating in the Sub-$6$ GHz band \cite{Zhou2016}. A third example is the magnetic-free SOI CMOS circulator operating in the $60$GHz mmWave band \cite{Nagulu2019}. Accounting for this self-interference, however, is an important direction for future extensions.

Further, and for the sake of having low-cost and power consumption mmWave transceivers, we adopt an analog-only architecture for the $N$-antenna array used for transmission and reception, \cite{HeathJr2016, Alkhateeb2014d}, where the beamforming/combining is done in the analog domain using a network of phase shifters. Next, we summarize the transmit and receive signal models.

\textbf{Transmit Signal Model:} 
We consider a wideband single-carrier waveform consisting of multiple time frames. These frames are transmitted over an aggregated time interval of $T$ seconds during which the environment is assumed to be relatively static. This time interval is commonly referred to as a coherent processing interval (CPI) \cite{Richards2010}. Each frame consists of both data and preamble sequences designed for the wireless communication function. The co-existing sensing model also uses these preamble sequences to sense the environment and build the depth maps, as will be explained in detail in the following sections. This can be achieved by either splitting the frames between sensing and communication or by designing the sensing and communication beam training operations to share the same preamble sequences. Next, for ease of exposition, we assume that $M$ frames/preamble sequences are dedicated for sensing. If $s_{m}[n]$ denotes the  $n$th  transmitted symbol at the $m$th frame, with $\bbE \left[ \left|   s_{m}[n]\right|^2 \right]  = 1$, then the complex-baseband representation of the transmit waveform can be written as  \cite{Kumari2019}
\begin{equation} \label{eq:allTX_1}
a(t) =  \sqrt{ \cE_\rm{s} }  \sum_{m =0}^{M -1}  \sum_{n=0}^{N_m-1} s_{m}[n] \ \delta(t -n T_\rm{S} - m T_{\rm{F}}),
\end{equation}
where  $\cE_\rm{s}$ represents the average energy per symbol, $ \ T_\rm{S}$ is the symbol time, and $T_{\rm{F}}$ is the frame duration. $N_{m}$ is the number of symbols in the $m^\rm{th}$ frame, which is divided into a preamble sequence of length $N^{\rm{p}}$ and a set of data symbols of length $N_{m}^{\rm{d}}$. Further, we assume that the same preamble sequence $s^{\rm{p}}[n], n\in \{1, \dotsc, N^{\rm{p}}\}$, is transmitted in the first $N^{\rm{p}}$ symbols of each frame. Note that for the sake of simplifying the transmit and receive signal representation, we incorporated the transmit pulse shaping and receive filtering functions into the channel model. Finally, if a beamforming vector $\bff \in \bbC^{N \times 1}$ is used to transmit the signal at the AR/VR device, the complex-baseband representation of the transmitted signal can expressed as 
\begin{equation}
\bx(t)=\bff a(t). 
\end{equation}
This transmitted signal will interact with the environment (through reflection, scattering, etc.) and will be received back by the AR/VR device. Next, we describe the receive signal model. 

\textbf{Receive Signal Model:} 
Let  $\sfG_{\rm{tar}}$ denote the number of targets/scatterers in the environment.  Then, focusing on the preamble sequence transmission/reception (i.e., the first $N^{\rm{p}}$ symbols of each frame), the receive \textit{sensing} signal of the $m$th frame can be written as 
\begin{equation} \label{eq:FinalRadarRX_pre}
y_{m}[n] = \sum_{g=1}^{\sfG_{\rm{tar}}} \sum_{d=0}^{L_{\sfd}-1} \sqrt{\cE_\rm{s}} \bw^H \bH_{d,g} \bff s^\rm{p}[n-d] + \bw^H \bv_{m}[n], 
\end{equation}
where $\bw\in \bbC^{N \times 1}$ is the combining vector at the AR/VR, and $\bv_{m}[n] \sim \mathcal{N}_\mathbb{C}\left( \b0, \sigma_\rm{n}^2 \bI\right)$ is the receive noise with variance $\sigma_\rm{n}^2$. $\bH_{d,g} \in \bbC^{N \times N}, d\in \{1, \dotsc, L_{\sfd}-1\}$, is the delay-$d$ channel matrix between the transmission from and the reception by the AR/VR antenna array, which is described in the following subsection. 

\subsection{Channel Model} \label{subsec:Ch_model}

Given that the depth sensing problem highly relies on the accurate modeling of the surrounding environment and its geometry, we adopt a geometric channel model in this work. More specifically, we consider the extended Saleh-Valenzuela wideband geometric channel model \cite{Spencer2000,Smulders2009,ElAyach2014,Guerra2017}. Based on that, the $g^\rm{th}$ target contribution in the delay-$d$ channel, $\bH_{d,g}$, can be modeled as 
\begin{equation} \label{eq:radarchannel}
\bH_{d,g} = \sqrt{G_{g}} \sum_{\ell=1}^{L_\rm{ray}} \alpha_{\ell} \rm{e}^{-\j 2\pi f_\rm{c} \tau_{\ell} } p\left(dT_\rm{s} -\tau_{\ell} \right) \ba_{\rm{R}} \left( \phi^{\rm{R}}_{\ell,g}, \theta^{\rm{R}}_{\ell,g} \right)  \ba_{\rm{T}}^H \left( \phi^{\rm{T}}_{\ell,g}, \theta^{\rm{T}}_{\ell,g} \right),
\end{equation}
where $L_\rm{ray}$ is the number of channel clusters; each  cluster is contributing with one ray of complex channel coefficient $\alpha_{\ell}$, time delay $\tau_{\ell}$, and azimuth/elevation angles of departure and arrival,  $\phi^{\rm{T}}_{\ell,g}, \theta^{\rm{T}}_{\ell,g}$ and $\phi^{\rm{R}}_{\ell,g}, \theta^{\rm{R}}_{\ell,g}$, respectively. $\ba_{\rm{T}}^H \left( \phi^{\rm{T}}_{\ell,g}, \theta^{\rm{T}}_{\ell,g} \right)$ and $\ba_{\rm{R}} \left( \phi^{\rm{R}}_{\ell,g}, \theta^{\rm{R}}_{\ell,g} \right)$ represent the transmit and receive array response vectors associated with the angles $\phi^{\rm{T}}_{\ell,g}, \theta^{\rm{T}}_{\ell,g}$ and $\phi^{\rm{R}}_{\ell,g}, \theta^{\rm{R}}_{\ell,g}$. The transmit and receive pulse shaping signals are included within $p(t)$ such that $p(t)= p_\rm{T}(t) \ast p_\rm{R}(t)$. The path gain associated with the $g^\rm{th}$ target is denoted by $G_{g}$ and can be expressed as 
\begin{equation}
 \ G_{g}  = \frac{G_\rm{T} G_\rm{R}  \lambda^2 \sigma^{\rm{RCS}}_{g}} {(4 \pi)^3 {\left( \rho_{g}\right) }^{2\rm{PL}}}, 
\end{equation} 
where $G_\rm{T}$ and $G_\rm{R}$ are the transmitter and receiver gains, $\lambda$ is the operating wavelength, $\rm{PL}$ is the path loss exponent. Finally, $\rho_{g}$ denotes the distance (range) between the AR/VR device and the $g$th target/scatterer and $\sigma^{\rm{RCS}}_{g}$ denotes the radar cross section of this target. 

\section{Problem Definition} \label{sec:Problem_formulation}

\begin{figure}[t] \centerline{\includegraphics[width=.6\columnwidth]{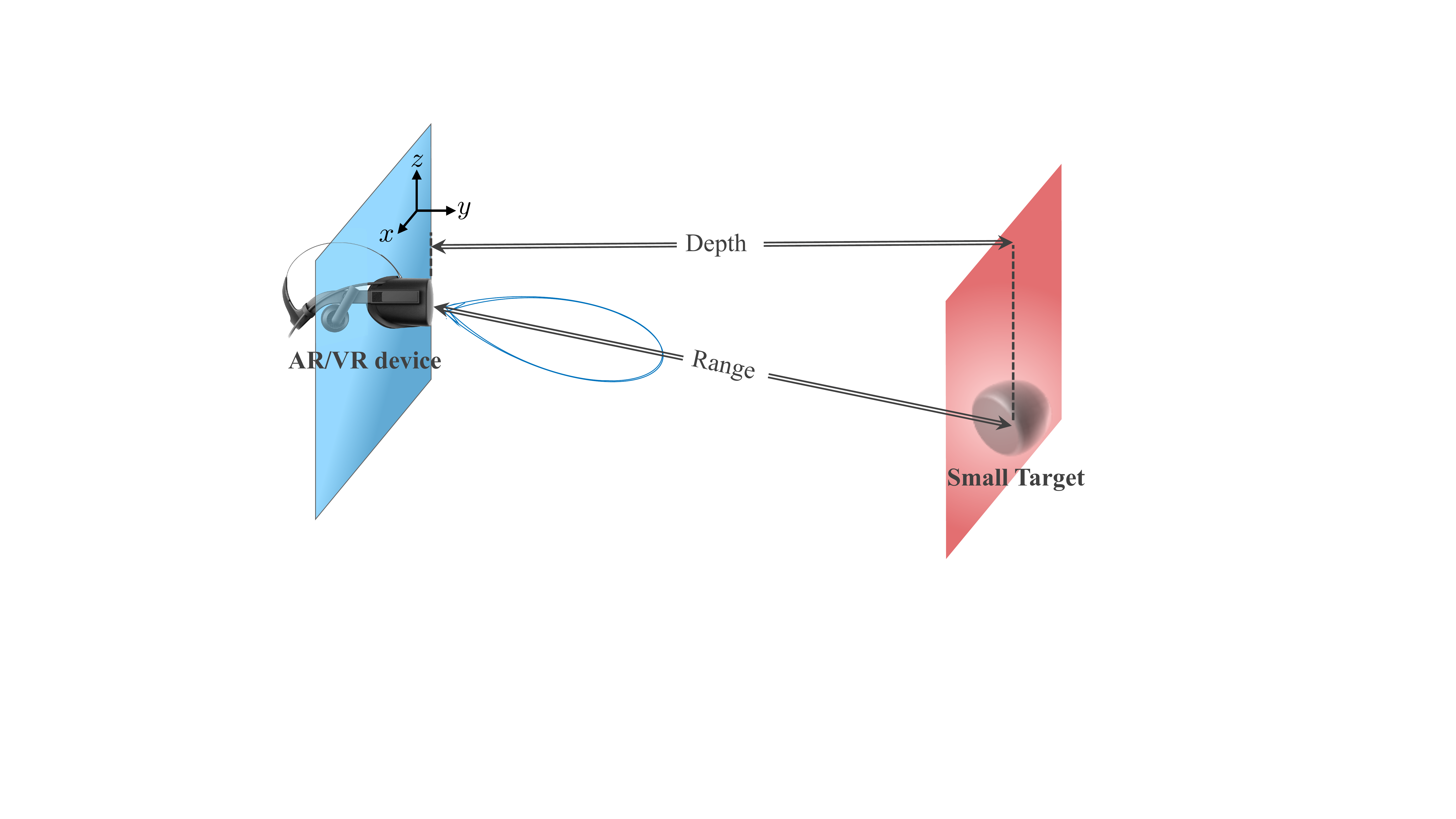}}
	\caption{This figure shows the conventional single target range estimation problem, where one target exists in free space in line-of-sight (LoS) with the AR/VR device. This device steers perfectly one beam towards that target to estimate the range.}
	\label{fig:background}
\end{figure}

Our objective in this paper is to efficiently estimate the depth/range map of the surrounding environment using the communication-constrained mmWave MIMO sensing model in   \sref{sec:Sys_CH_Models}. 
Before delving into the formal problem definition, it is important to distinguish between the range and depth of a certain target. As depicted in \figref{fig:background}, the range of a target with respect to the AR/VR camera (which is aligned with the AR/VR antenna array) is the linear radial distance from the camera center (focal point) to the target. For the depth, it is measured by the z-coordinate of the camera center (focal point) with respect to the x-y plane of the target. Given that the range and depth can be calculated from one another, we focus our formulation on the depth estimation problem. 
Next, we define the depth map of the surrounding environment with respect to an AR/VR device. 

\textbf{Definition (Depth Map):} \textit{We define the depth map $\bD_\rm{map}$ of resolution $M_h \times M_w$ as an image of $M_h$ pixels high and $M_w$ pixels wide, where the value of each pixel represents the smallest depth between the AR/VR device and the targets/objects in this pixel.}

In this paper, we express this depth map as an $M_h \times M_w$ matrix $\bD_\rm{map} \in \mathbb{R}^{M_h \times M_w}$. Further, we use $M_\rm{res}=M_h M_w$ to denote the total number of pixels in the depth map. The range map $\bR_\rm{map}  \in \mathbb{R}^{M_h \times M_w}$ is similarly defined. Now, given the system and channel models in \sref{sec:Sys_CH_Models}, the AR/VR device constructs the estimated mmWave-based depth map through two main steps: (i) sensing the environment using several beamforming and combining sensing vectors and (ii) post-processing the receive sensing signal to construct the estimated depth map. More formally, if a beamforming-combining pair $\left(\bff_m, \bw_m\right)$ is used to transmit and receive the $N^\rm{p}$ symbols of the $m$th preamble sequence, then the receive sensing signal can be expressed as 
\begin{equation}  \label{eq:receive_total}
y_{m}[n] =  \sum_{d=0}^{L_{\sfd}-1} \sqrt{\cE_\rm{s}} \bw_m^H \bH_{d}  \bff_m s^\rm{p}[n-d] + \bw^H \bv_{m}[n], \ \ \ n \in \{0, 1, \dotsc, N^\rm{p}+L_{\sfd}-1\},
\end{equation}
where $\bH_{d} = \sum_{g=1}^{\sfG_{\rm{tar}}} \bH_{d,g}$. By stacking the $N^\rm{p}+L_{\sfd}$ receive symbols (from transmitting the preamble sequence), we get $\by_m=\left[y_m[0], \dotsc, y_m[N^\rm{p}+L_{\sfd}-1]\right]^T$, which represents the receive sensing vector of one preamble sequence using one beamforming-combining pair. Next, if $M$ preamble sequences are used to sense the environment via $M$ beamforming-combining pairs $\left(\bff_m, \bw_m\right), m \in \{1, \dotsc, M\}$, then the aggregated receive sensing signal can be written as 
\begin{equation}  \label{eq:receive_totalF}
\bY=[\by_1, \dotsc, \by_M].
\end{equation}

\noindent For ease of exposition, we define the sensing beamforming codebook $\boldsymbol{\cP}$ as the codebook that includes the $M$ beamforming-combining pairs, i.e.,  $\boldsymbol{\cP} = \left\{  \left(\bff_{m},\bw_{m}\right): m \in \{1, \dotsc, M\} \right\}$. Finally, given the receive sensing matrix $\bY$, a post-processing is applied for estimating the depth map. If $\bg(.)$ denotes the post-processing function, the estimated depth map $\widehat{\bD}_\rm{map} \in \mathbb{R}^{M_h \times M_w}$ can be written as 
\begin{equation}
\widehat{\bD}_\rm{map}=\bg(\bY; \boldsymbol{\mathcal{P}}).
\end{equation}

\noindent Our objective in this paper then is to design the sensing beamforming codebook $\boldsymbol{\mathcal{P}}$ and the post-processing $\bg(\cdot)$ to efficiently  estimate the depth map $\widehat{\bD}_\rm{map}$ to be as close as possible to the actual depth map $\bD_\rm{map}$. To evaluate the performance of the proposed approaches, we will adopt the root-mean squared estimation error (RMSE) and  the mean absolute error (MAE) between the depth maps, which are defined as 
\begin{align} \label{eq:obj}
\Delta_{\rm{RMSE}} & = \sqrt{\frac{1}{K} \left\| {\bD}_\rm{map} -  \bg \left( \bY ;  \boldsymbol{\cP}\right) \right\|_{2}^{2}}, \\
\Delta_{\rm{MAE}} & =  \frac{1}{K} \left\| {\bD}_\rm{map} -  \bg \left( \bY ;  \boldsymbol{\cP}\right) \right\|_{1}^{2}.
\end{align}

In  \sref{ssec:mmWave_depth_est}, we will present the general framework of our proposed depth map estimation approach. This will be followed by a detailed description of the two main components in this framework, namely the  beamforming codebook design $\boldsymbol{\mathcal{P}}$ and the post-processing solution $\bg(.)$, in Sections \ref{sec:CB_design} and  \ref{ssec:scene_range_section}. 

\section{Background} \label{sec:Target_range_est}
Before going into the proposed framework for estimating depth/range maps using mmWave MIMO, we provide a brief background on the basics of the single-target range estimation problem. For a preliminary model, consider one target in the free space with a Line-of-sight (LoS) path to the AR/VR device. Further, consider the case when  one mmWave beam is perfectly steered towards that target, as depicted in \figref{fig:background}. Adopting this preliminary model, the target range estimation accuracy bound will be first examined. Then,  a description of the main algorithms used in the literature to approach this problem is provided.

\subsection{Target Range Estimation Accuracy} \label{subsec:limit_onetarget}
Our main objective is to find the fundamental limit for mmWave MIMO based depth estimation, which can be considered as range estimation at every possible eyesight direction, i.e. at every azimuth angle $\phi \in [0,2\pi[$ and every elevation angle $\theta \in  [0,\pi[$. For the range estimation accuracy, one useful metric is the Cramer-Rao lower bound (CRLB) on the range estimation. For white Gaussian noise, the CRLB provides a lower bound on the mean-squared-error of any unbiased estimator, hence it is used as a benchmark for the performance analysis of parameter estimation\cite{Kay1993}. Considering the case of range estimation for a single target, the CRLB of this single target range is formulated as \cite{Kay1993, Richards2010,Kumari2018}
\begin{equation} \label{eq:crlb_range}
\sigma^2_{\hat{\rho}} \geq  \frac{{\varsigma}^2}{8 P_\rm{int} \ \eta^2 B^2 \ \rm{SNR}_{\rm{rad}} },
\end{equation}
where $\varsigma$ is the speed of light, $B$ is the transmission bandwidth, $P_\rm{int}$ is the integration gain and is equal to the number of symbols used for preamble estimation, and $\eta$ depends on the power spectral density shape of $a(t)$ over the preamble duration. Under the assumption of a flat spectral density for $a(t)$, $\eta^2=\left( 2\pi \right)^2 /12$.  The radar signal-to-noise ratio for this target can then be expressed as $\rm{SNR}_{\rm{rad}} = \cE_\rm{s} G_{\rm{rad}} /\sigma_\rm{n}^2$, where $G_{\rm{rad}}$ denotes the path gain associated with the target.

\subsection{Target Range Estimation Algorithms} \label{subsec:Range_est_Alg}
Estimating the round trip delay $\hat{\tau}$ is equivalent to finding the range estimate $\hat{\rho}$, since they are directly related through $\hat{\tau} = 2\hat{\rho}/\varsigma$. Given the extensive research on delay estimators in the literature \cite{Richards2010,Bidigare2012}, we will restrict the scope of this paper on the magnitude based delay estimators in \cite{Herschfelt2018} for simplicity. In a general sense, given a known transmit preamble sequence $x_0[n]$ and the received baseband sequence $z[n]$, the receiver can  estimate the round-trip delay by maximizing an objective function, the cross-correlation function between the two time-sequences, over a range of possible delays. Based on this notion, two delay estimators are formulated as follows \cite{Herschfelt2018}.

\subsubsection{Basic Correlator} \label{sec:BC}
The basic correlator is a coarse delay estimator that performs the maximization at the same sampling frequency, $f_\rm{S}$, tuned by the AR/VR communication system. Assume that the length of the received baseband sequence, $z[n]$, is $L_\rm{z}$ samples, where the last $N_\rm{z}$ samples are non-zeros. The range estimate can then be formulated as
\begin{equation}
\hat{\rho}^\rm{BC} = \frac{\varsigma T_\rm{S}}{2}  \argmax_{q:q \in \cQ } \ \left| \sum_{n=L_\rm{z}-N_\rm{z}}^{L_\rm{z}-1} x_0[n]  \times z^*[n-q] \right|^{2}. 
\end{equation}
where $T_\rm{S}=\frac{1}{f_\rm{S}}=\frac{1}{B}$ denotes the sampling time, $\cQ$ represents the set of possible discrete sample delays, and the optimal $q$ solution is denoted by $q^\rm{BC}$. Unfortunately, the accuracy of this range estimate is limited by the sampling frequency $f_\rm{S}$. One attempt of improving the estimation accuracy is by performing the maximization at a higher sampling frequency. This attempt, however, increases the computational complexity dramatically, which motivates the role of the upcoming delay estimator, the massive correlator \cite{Herschfelt2018}. 

\subsubsection{Massive Correlator} \label{subsubsec:MC}
The primary function of the massive correlator is to perform the maximization of the objective function at a higher sampling frequency without the computational burden of computing the shift in real time. For this reason, \cite{Herschfelt2018,Herschfelt2019D} introduced the solution of pre-designing a specific correlator bank that contains shifted versions of the reference sequence, $x_0[n]$. The receiver will then multiply the received sequence by the correlator bank to compute the objective function.

We describe the steps of the massive correlator algorithm as follows \cite{Herschfelt2018,Herschfelt2019D}. (a) Upsample $x_0[n]$ with a sampling frequency higher than $f_\rm{S}$, denoted as $f_\rm{est}$. (b) Define the correlator bank matrix, $\bX_0$, where each row of this matrix is a shifted version of the upsampled $x_0[n]$. Let the number of rows in $\bX_0$ be equal to $\left( 2\delta+1\right)$, where $\delta$ is the largest lag/advance discrete fractional delay in the receive sequence, such that $\delta= \frac{f_\rm{est}}{2f_\rm{S}}$. (c) Downsample independently each row of the correlator matrix to the lower sampling frequency $f_\rm{S}$; let the resulting matrix be named as $\bB_0$. The reason for this step is to test delays at the higher sampling frequency, $f_\rm{est}$, but only apply multiplications at the lower sampling frequency, $f_\rm{S}$. (d)  Shift back the receive sequence, $z[n]$, with the coarse discrete delay estimate, $q^\rm{BC}$, such that $\bar{z}[n]=z[n+q^\rm{BC}]$, and then concatenate the sequence into one row vector, $\bar{\bz}$. (e) Calculate the fractional range estimate, $\hat{\rho}'$, such that $\hat{\rho}'= \frac{\varsigma}{2 f_\rm{est}} \left(-(\delta+1) + \argmax_{q'} \left[ \bg \right]_{q'}  \right)$, where $\bg=\bar{\bz}\times\bB_0^H$. (f) Calculate the fine range estimate, $\hat{\rho}^\rm{MC}$, such that $\hat{\rho}^\rm{MC}=\hat{\rho}^\rm{BC}+\hat{\rho}'$. 

With an end goal of constructing depth maps, these range estimation algorithms will then be leveraged by the mmWave MIMO based depth estimation as explained in the upcoming sections.  In the next section, we formulate a general framework for scene depth estimation.

\section{General Framework for Scene Depth Estimation} \label{ssec:mmWave_depth_est}

\begin{figure}[t] \centerline{\includegraphics[scale=0.45]{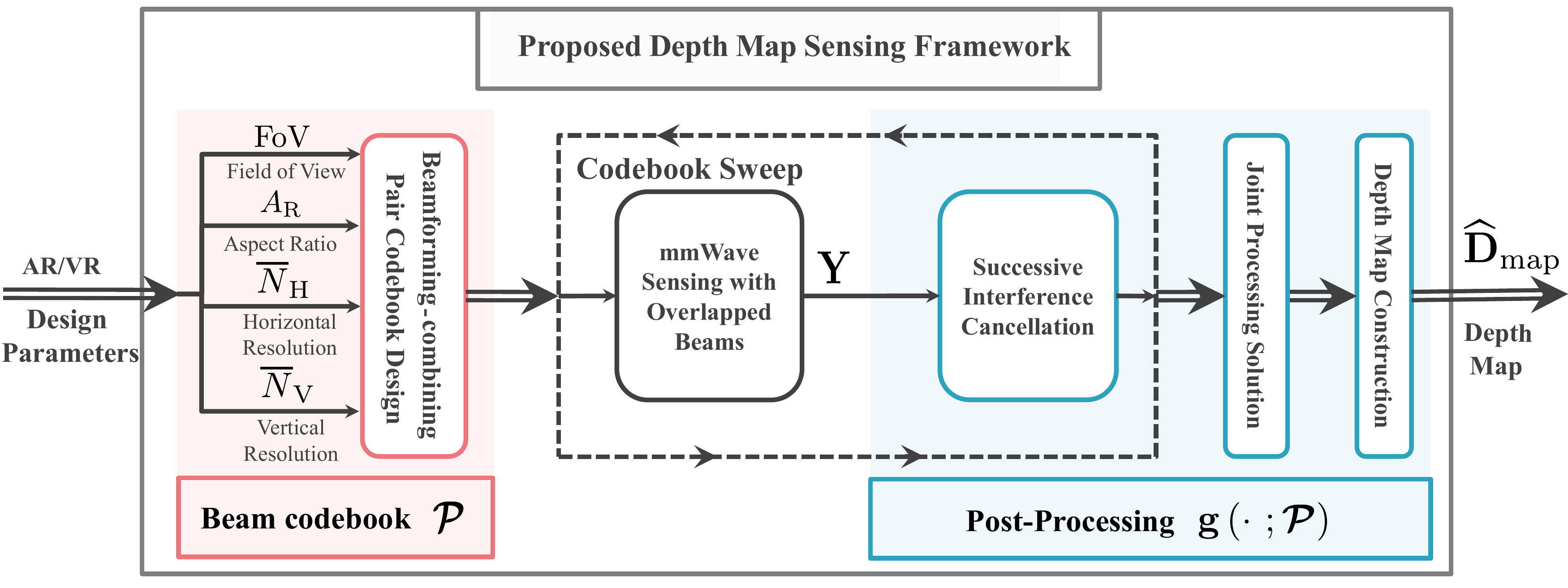}}
	\caption{The figure summarizes the proposed sensing framework for mmWave MIMO based depth estimation, which involves sensing the scene using the designed beamforming codebook $\boldsymbol{\mathcal{P}}$ and applying the proposed post-processing operations $\bg(.; \boldsymbol{\mathcal{P}})$  to the receive signal to construct the estimated depth map $\widehat{\bD}_\rm{map}$. }
	\label{fig:mmWave_depth_alg}
\end{figure}

In this section, we highlight the key elements of the proposed depth map estimation approach, namely the sensing beamforming codebook $\boldsymbol{\mathcal{P}}$ and post-processing $\bg(.)$, and discuss the challenges associated with designing these elements. As depicted in \figref{fig:mmWave_depth_alg}, we first design the sensing beamforming codebook  $\boldsymbol{\mathcal{P}}$  offline based on the desired  AR/VR properties such as the field of view, the scene aspect ratio, and the number of horizontal and vertical beams covering the scene view. To build the depth map of a certain scene, the beam pairs of the designed codebook are used to sense the environment and acquire the receive sensing matrix $\bY$ in \eqref{eq:receive_total}.  This receive signals are then jointly processed using the post-proposed approach to build the depth map. In the remaining of this section, we explain the challenges associated with designing the codebook and the post-processing operations. Then, we will present how our proposed solutions overcome these challenges in Sections  \ref{sec:CB_design} and \ref{ssec:scene_range_section}.

\subsection{Codebook Design Challenges} \label{ssec:BF_challenges}
\begin{figure}[t] 
	\centering
	\begin{subfigure} [t]{.57\textwidth}
		\includegraphics [width=1\columnwidth]{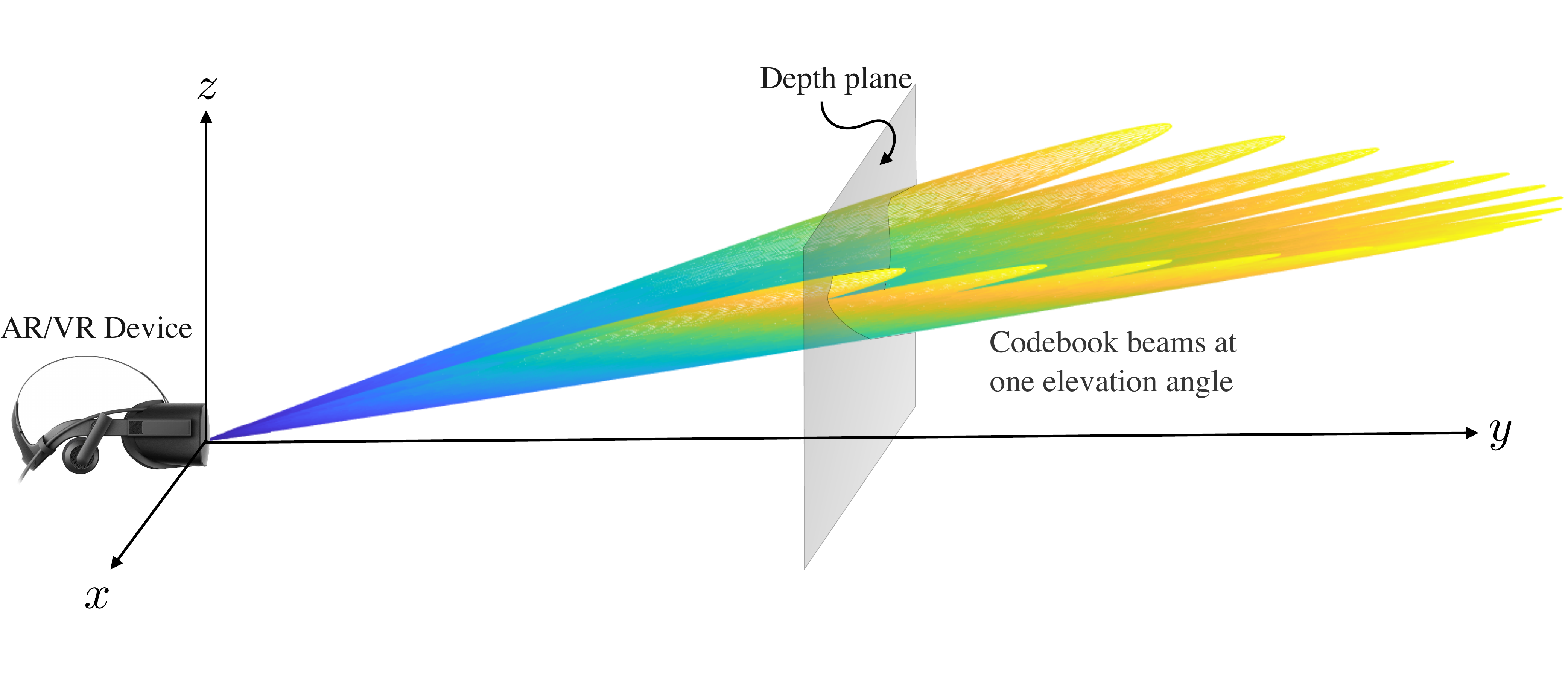}
		\caption{Parabolic shape of the classical codebook grid}
	\end{subfigure}
	\begin{subfigure}[t]{.41\textwidth}
		\includegraphics[width=1\columnwidth]{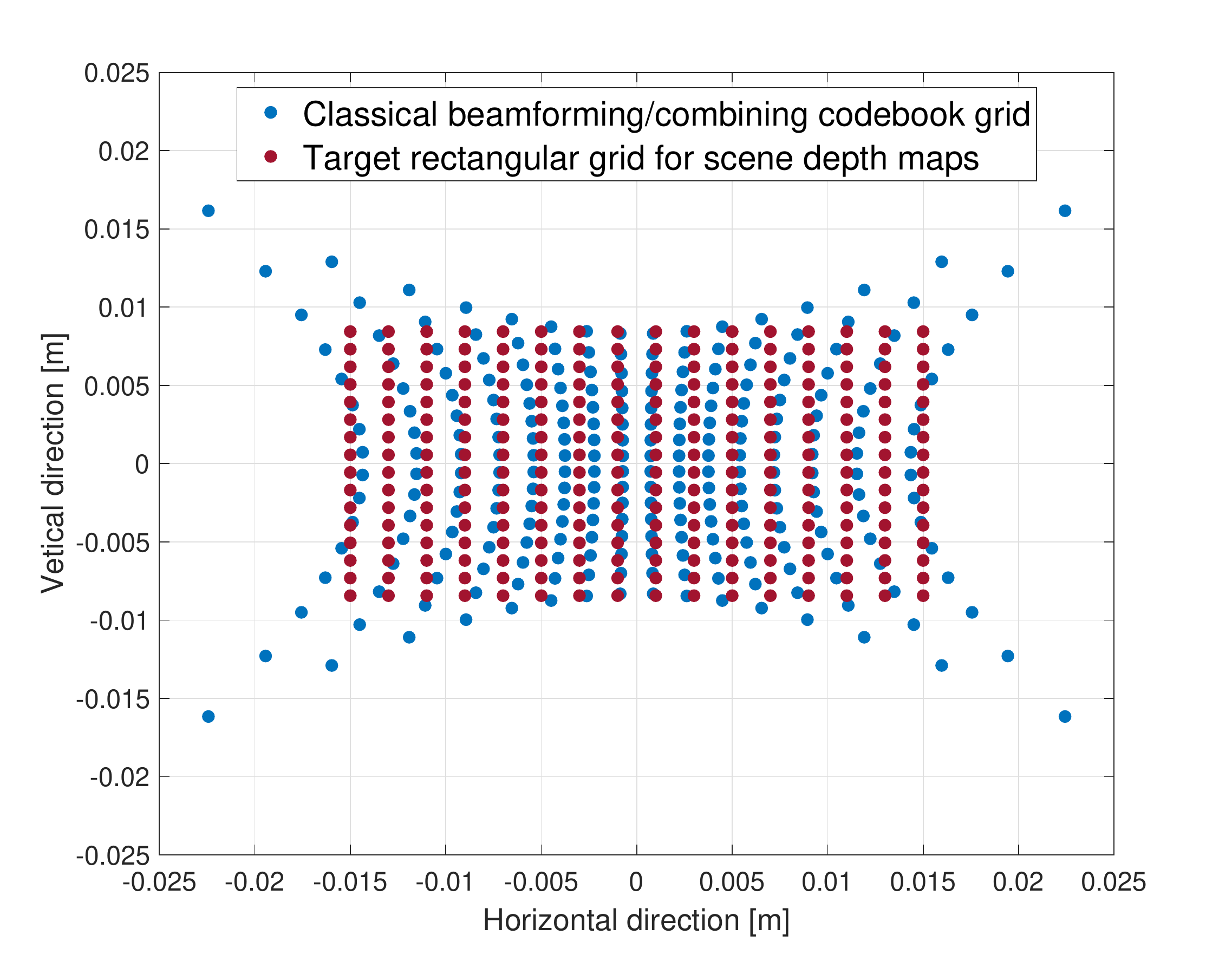}
		\caption{Classical codebook grid mismatch}
	\end{subfigure}
	\caption{(a) The intersections between the classical codebook beam directions and the $x$-$z$ depth plane form the parabolic shape of the classical codebook grid. (b) The mismatch between the classical codebook grid of a $16\times16$ UPA and the desirable rectangular grid for a depth map is illustrated at a $y=13.32$mm depth plane, for a scene of $100^\circ$ field of view, $16/9$ aspect ratio.} 
	\label{fig:BF_codebook_mismatch}
\end{figure}

To effectively sense the surrounding environment and build efficient depth maps, the beams of the sensing codebooks should be designed to scan the full scene. Since the mmWave MIMO based depth maps will likely complement the RGB-D based maps, our objective is to build a beamforming codebook that scans the full rectangular grid of the typical depth sensors of the AR/VR cameras. However, the classical beam steering codebooks such as the DFT codebooks \cite{Alkhateeb2015a}, that independently sample the azimuth and elevation directions, do not normally fit a rectangular grid.  They instead form a parabolic grid, i.e.,  for a fixed elevation angle, the grid line of these codebook beams are parabolic curves as shown in \figref{fig:BF_codebook_mismatch}(a). This mismatch between the mmWave MIMO-based and camera-based depth grids could lead to clear distortion in the joint mmWave/RGB-D depth map construction and make it hard to complement the RGB-D depth map using mmWave MIMO sensing.

One possible solution is to estimate the depths on the parabolic grid using the classical beamforming codebook and then interpolate/extrapolate to calculate the rectangular depth map. The main disadvantage of this solution, however, is that the interpolation can potentially lead to considerable loss in the depth map accuracy as the changes of the depth are not normally smooth in nature.  Hence, in order to avoid the interpolation loss, the more persuasive solution is to  develop a depth map compatible beamforming codebook that fits exactly the desirable rectangular sensor grid. With this motivation, we propose a beamforming/combining design approach in \sref{sec:CB_design} to overcome the codebook mismatch challenge.

\subsection{Scene Depth Estimation Challenges} \label{ssec:scene_depth_challenges}
The  sensing beamforming codebook is used to sense the surrounding environment. Now,  given the receive sensing matrix $\bY$, the objective of the post-processing is to construct an accurate depth map of the facing scene. This process, however, has several challenges. In order to explain these challenges, let's first consider the case when the environment has only a single target. In this case, the sensing/scanning beam that is directed towards the region that includes this target will result in some backscattering signal. This signal can be used for calculating the round-trip time of flight and consequently the range of this target, leveraging the MIMO radar concepts  \cite{Richards2010, Melvin2013} and the algorithms detailed in \sref{subsec:Range_est_Alg}. In terms of the range/depth map, the pixel that includes the region of this target will simple have the value of the estimated range/depth.  In practice, however, the environment has several targets/surfaces and the mmWave arrays have strict constraints on their hardware: power budget, computational complexity, etc. These limitations lead to critical challenges for our objective of building accurate depth and range maps of the environment. More specifically, if we adopt the approach that scans the surrounding scene using a beamforming codebook and processed the receive sensing signal of each beam independently to estimate the depth of the region defined by this beam, then this approach will have the following key drawbacks.

\begin{figure}[t] \centerline{\includegraphics[width=.5\columnwidth]{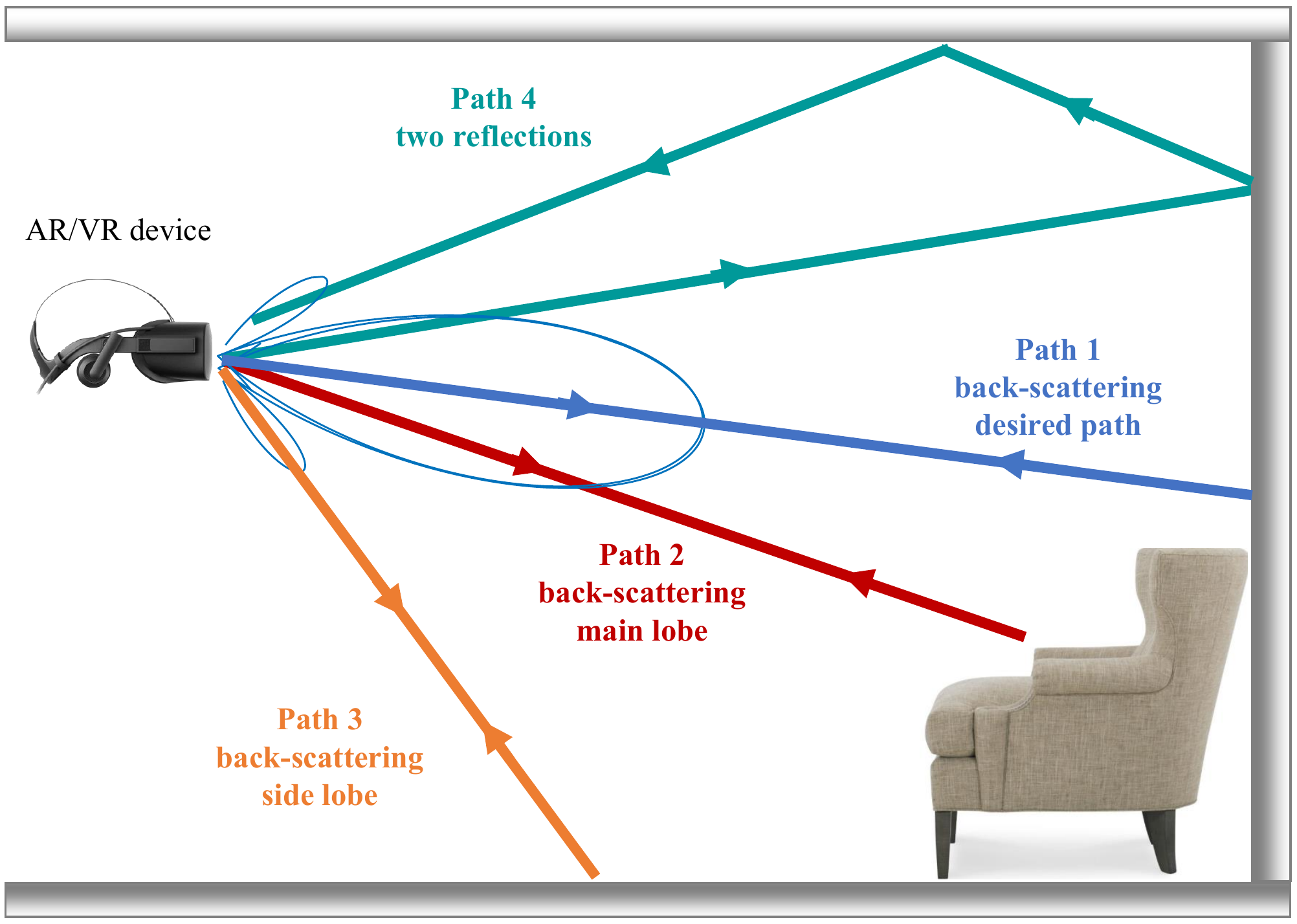}}
	\caption{The multipath estimation challenge for scene range estimation is illustrated. The design challenge is how the sensing framework can detect and estimate the range through the desired channel path (path $1$ in blue) and avoid making faulty estimation because of the other undesired paths (paths $2$-$4$) in the environment.}
	\label{fig:Soln_key_idea}
\end{figure}

\begin{itemize}
	\item \textbf{Low-resolution depth-maps:}
	The  low resolution drawback is mainly due to (i) the limitation on the number of AR/VR antennas, which is controlled by many factors in the AR/VR device such as the device dimensions, computational complexity, circuit routing, power consumption, etc., and (ii) the number of beams in the sensing codebook $\boldsymbol{\mathcal{P}}$, which is limited by the time allocated for the depth estimation process. 
	
	\item \textbf{Inter-target interference:}
	The constraints on the number of antennas at the AR/VR device limit the system spatial resolution. This makes it hard to differentiate between the ranges/depths of the different targets/surfaces that are close to each other. In other words, when measuring the depth of the object in a particular region/ direction, multiple objects/surfaces may reflect the incident signal at the same time. The interference between these reflected/scattered signals may highly affect the accuracy of the range/ depth estimation.  Hence, if a certain pixel has multiple objects/surfaces, it will be difficult to estimate the shortest depth of the objects in this pixel (to follow the depth map definition in \sref{sec:Problem_formulation}).  
	
	\item \textbf{Inter-path interference:}
	When sensing the range/depth of a certain target, the optimal situation (in terms of depth estimation accuracy) is when the target backscatters a single ray to the receiving array. In practice, however, the signal incident on a certain target may experience more than one phenomenon, such as scattering, reflection, diffraction, etc., which results in multiple rays. More than one of these rays could traverse the environment in different ways/directions, especially in indoor environments, before reaching the receiver. This means that they may reach the receiving array from multiple angles and with different time of flights. This causes an inter-path interference which makes it hard to accurately estimate the range/depth of the target of interest. For example, if the receiver estimates the range/depth based on a wrong path, this may noticeably degrade the accuracy of the depth map estimation. This challenge is depicted in \figref{fig:Soln_key_idea}.  As illustrated, the challenge is how to design the sensing framework (the codebook and post-processing) to detect the desired channel path (the path in blue) while filtering out all the undesired channel paths. Examples of undesired paths are the paths $2$-$4$. Path $2$ is transmitted and received within the main lobe. Path $3$ is transmitted and received within the side lobe. Path $4$ experiences multiple reflections instead of back-scattering, before reaching back the receiver.	 
\end{itemize}

In the next two sections, we efficiently design the two elements of our proposed depth map sensing framework, namely the sensing codebook and the post-processing, to address these challenges. 

\section{Depth Map Based Design for Sensing Codebooks} \label{sec:CB_design}

As discussed in \sref{ssec:BF_challenges}, our objective is to design a sensing beamforming codebook that fits the rectangular grid of the depth camera.  In this section,  we first present our codebook design that achieves this objective. Then, we incorporate a new side-lobe reduction approach to ameliorate the inter-path interference problem.

\subsection{Proposed Codebook Design} \label{ssec:CBDesign}
Since the objective from the beamforming-combining pair codebook design is for the codebook grid to match the desired rectangular grid of a range/depth scene, we start with the relevant camera geometry equations. 
The scene definition starts by defining the key quantities of the field of view, $\rm{FoV}$, and the scene aspect ratio, $A_\rm{R}$. Let the field of view be centered around the boresight antenna array direction. 
It is worth noting that the separation distance of the camera plane away from the antenna array reference point, aka the focal length, is irrelevant in our codebook design. This is based on the notice that the beamforming/combining codebook design normally depends on angles rather than distances.

In a general sense, for any chosen value of focal length, the sensor grid points' coordinates are first calculated to determine the codebook angles accordingly. More specifically, assume that the focal length is set to a certain value, $F_\rm{L}$. The camera plane width, aka the sensor grid width in the horizontal dimension, $S_\rm{H}$, and camera plane height, aka the sensor grid height in the vertical dimension, $S_\rm{V}$, can be calculated as 
\begin{equation}
S_\rm{H}=2F_\rm{L}\tan\left(\rm{FoV}/2\right)\text{, and} \  S_\rm{V}=S_\rm{H}/A_\rm{R}.
\end{equation}
For designing a beamforming-combining pair codebook, let $N={N}_\rm{V} \times {N}_\rm{H}$, where ${N}_\rm{V}$ and ${N}_\rm{H}$ denote the number of UPA antennas on the elevation (vertical) and azimuth (horizontal) dimensions, respectively. Consider an oversampled beamforming codebook of $M=\overline{N}_\rm{V}\overline{N}_\rm{H}$ beams, where  $\overline{N}_\rm{V}={N}_\rm{V} F_\rm{V}^\rm{OS}$ and $\overline{N}_\rm{H}={N}_\rm{H} F_\rm{H}^\rm{OS}$ with ${F}_\rm{V}^\rm{OS}$ and ${F}_\rm{H}^\rm{OS}$ denoting the oversampling factors in the  elevation and azimuth dimensions, respectively. The grid spacing in the vertical and horizontal directions are expressed as $Q_\rm{V}=S_\rm{V}/ \overline{N}_\rm{V}$ and $Q_\rm{H}=S_\rm{H}/\overline{N}_\rm{H}$. Notice that the codebook resolution of $\overline{N}_\rm{V}\overline{N}_\rm{H}$ beams will be mapped at the end to the desired up-scaled depth image resolution of $M_\rm{res}=M_h \times M_w$ pixels. 

Let the $x$- and $z$-axes be in the direction of the sensor grid width and height, respectively, and let the $y$-axis be the direction of the depth. The $\left(x,y,z\right)$ rectangular coordinates of the sensor grid points on the camera plane can then be defined as
\begin{equation} \label{eq:rect_coordinates}
\begin{split}
&\left(x,y,z\right) \in \cC, \ \  \cC = \cX \times \cY \times \cZ, \\
& \cX = \left\lbrace x: x \in \left\lbrace \tfrac{-S_\rm{H}}{2} + \tfrac{Q_\rm{H}}{2},\tfrac{-S_\rm{H}}{2} + \tfrac{3Q_\rm{H}}{2},\dotsc,\tfrac{S_\rm{H}}{2} - \tfrac{Q_\rm{H}}{2}   \right\rbrace   \right\rbrace  ,  \\
& \cY = \left\lbrace  y: y = F_\rm{L} \right\rbrace  , \\
& \cZ = \left\lbrace   z: z \in \left\lbrace \tfrac{-S_\rm{V}}{2} + \tfrac{Q_\rm{V}}{2},\tfrac{-S_\rm{V}}{2} + \tfrac{3Q_\rm{V}}{2},\dotsc,\tfrac{S_\rm{V}}{2} - \tfrac{Q_\rm{V}}{2}   \right\rbrace   \right\rbrace,
\end{split}
\end{equation}
where we note that $\left|\mathcal{C}\right|= \overline{N}_\rm{V}  \overline{N}_\rm{H}= M$. After defining the $\left(x,y,z\right)$ coordinates of every grid point on the camera plane, their $M$ corresponding $\left( \theta_\rm{z},\theta_\rm{x}\right)$ angles with respect to the $z$- and $x$-axes can now be calculated using the mapping from rectangular to spherical coordinates, such that
\begin{equation}
\cO = \left\lbrace   \left(\theta_\rm{z},\theta_\rm{x}\right): \theta_\rm{z} = \left[  \tfrac{\pi}{2} - \arctan\left(\tfrac{z}{\sqrt{x^2 + y^2}} \right) \right],
\theta_\rm{x} = \left[ \tfrac{\pi}{2} - \arctan\left(\tfrac{x}{\sqrt{(-z)^2 + y^2}} \right) \right],  \left(x,y,z\right) \in  \cC \right\rbrace.
\end{equation}

\begin{figure}[t] \centerline{\includegraphics[scale=0.7]{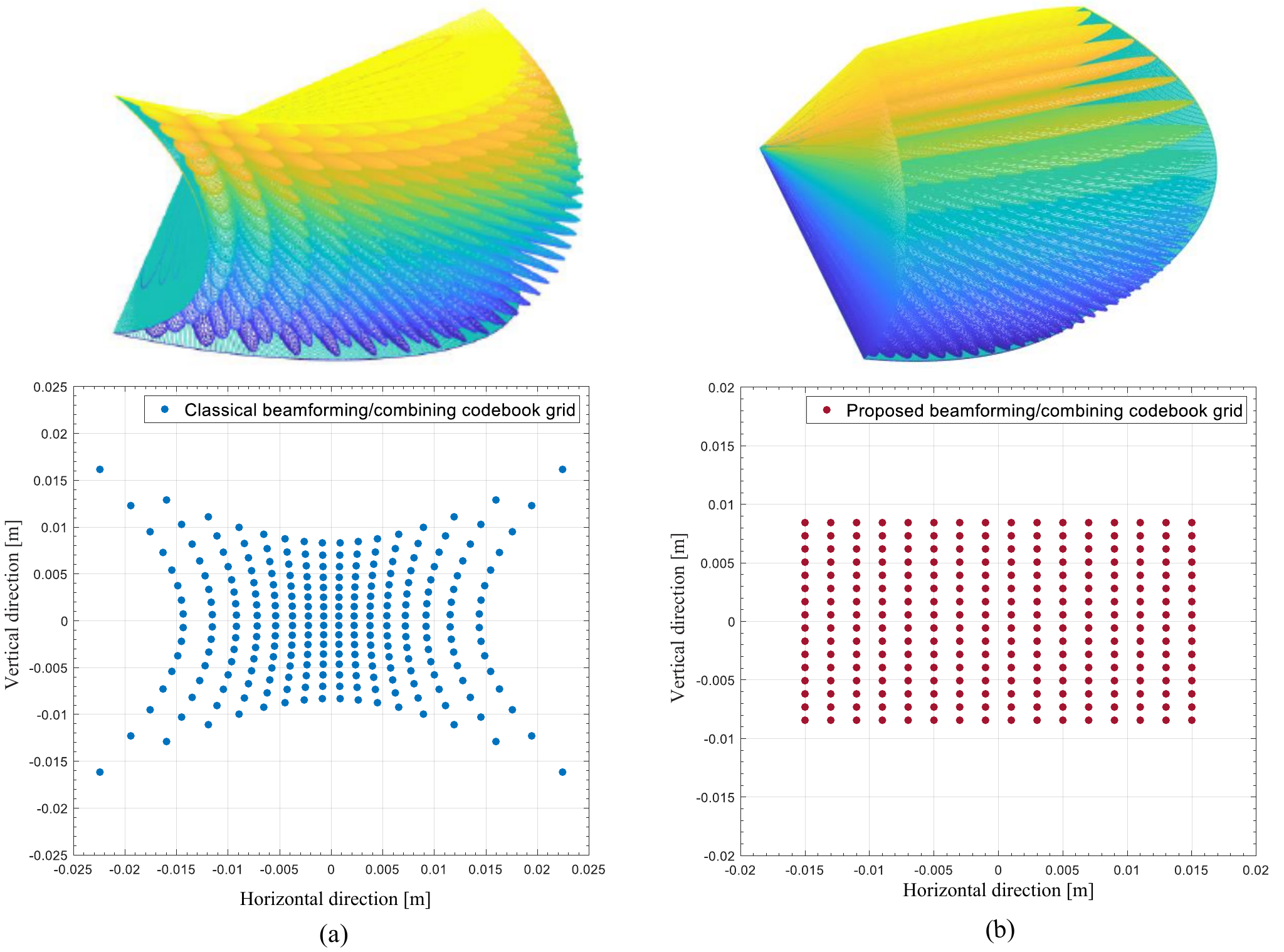}}
	\caption{The comparison between (a) the classical (on the left side) and the proposed (on the right side) beam codebook design is demonstrated for a scene of $100^\circ$ field of view and $16/9$ aspect ratio, using $16 \times 16$ UPAs. The proposed codebook eliminates any grid mismatch distortion. The top figures are the 3D codebook radiation patterns while the bottom figures are the 2D codebook grids at a plane within $13.32$mm depth.} 
	\label{fig:BF_codebook}
\end{figure}

Finally, after calculating the $\left( \theta_\rm{z},\theta_\rm{x}\right)$ angles for each and every grid point, the beamforming codebook, $\boldsymbol{\cF}$, for an ${N}_\rm{H} \times {N}_\rm{V}$ transmit UPA, is then expressed as
\begin{align}
\boldsymbol{\cF} &= \left\lbrace \bff \in \bbC^{N \times 1} : \bff = \widetilde{\bb}_\rm{V} \left(\theta_\rm{z} \right) \circ  \widetilde{\bb}_\rm{H}\left(\theta_\rm{x} \right), \left( \theta_\rm{z},\theta_\rm{x}\right) \in \cO \right\rbrace, \nonumber \\
\widetilde{\bb}_\rm{V} \left(\theta_\rm{z} \right) &= \left[ 1, \rm{e}^{-\j \kappa d_s \cos \left(\theta_\rm{z} \right) }, \rm{e}^{- \j 2\kappa d_s \cos \left(\theta_\rm{z} \right) },  \dotsc, \rm{e}^{-\j \left( {N}_\rm{V} -1\right) \kappa d_s \cos \left(\theta_\rm{z} \right) } \right] ^T, \\
\widetilde{\bb}_\rm{H} \left(\theta_\rm{x} \right) &= \left[ 1, \rm{e}^{-\j \kappa d_s \cos \left(\theta_\rm{x} \right) }, \rm{e}^{- \j 2\kappa d_s \cos \left(\theta_\rm{x} \right) },  \dotsc, \rm{e}^{-\j \left( {N}_\rm{H} -1\right) \kappa d_s \cos \left(\theta_\rm{x} \right) } \right] ^T, \nonumber
\end{align}
where $\kappa=\tfrac{2\pi}{\lambda}$ is the wave number, $\lambda$ is the operating wavelength, and $d_s$ is the antenna element spacing between adjacent UPA elements in meters. $\widetilde{\bb}_\rm{H} \in \bbC^{{N}_\rm{H} \times 1}$ and  $\widetilde{\bb}_\rm{V} \in \bbC^{{N}_\rm{V} \times 1}$ are the horizontal and vertical basic vectors used for constructing the beamforming codebook. We will call these vectors, $\widetilde{\bb}_\rm{H}$ and  $\widetilde{\bb}_\rm{V}$,  the \textit{constituent} horizontal and vertical beamforming vectors, respectively. 
In our depth estimation problem, the receive combining codebook, $\boldsymbol{\cW}$, can be similarly defined for the ${N}_\rm{H} \times {N}_\rm{V}$  receive UPA. For such case, the cardinalities of the sets are equal, $\left| \boldsymbol{\cW} \right| = \left| \boldsymbol{\cF} \right|=\left| \cC \right|=\left| \cO \right|=M$. Further, let $\bF \in \mathbb{C}^{N \times M}$ and $\bW \in \mathbb{C}^{N \times M}$ be the matrices that consist of the codebooks beams of $\boldsymbol{\mathcal{F}}$ and $\boldsymbol{\mathcal{W}}$. Then, the proposed sensing beamforming-combining pair codebook $\boldsymbol{\cP}$ can be expressed as
\begin{equation} \label{eq:codebook_pair}
\boldsymbol{\cP} = \left\lbrace  \left(\bff_{m},\bw_{m}\right) \in \bbC^{N \times 1} \times \bbC^{N \times 1} : \bff_{m} = \left[\bF\right]_{:,m}, \bw_{m} =\left[\bW\right]_{:,m}, m \in \{1,\dotsc, M \}   \right\rbrace.
\end{equation}
A comparison between the classical and the proposed beam codebook design is demonstrated in \figref{fig:BF_codebook} for a scene of $100^\circ$ field of view and $16/9$ aspect ratio, using $16 \times 16$ UPAs. The top figures are the 3D codebook radiation patterns while the bottom figures are the 2D codebook grids at a plane within $13.32$mm depth. As shown, the proposed beam codebook eliminates any grid mismatch distortion.

\subsection{Sidelobe Reduction Approach} \label{ssec:Sidelobe}
As discussed in \sref{ssec:scene_depth_challenges}, to rectify the inter-path interference problem,  the sensing framework needs to filter out the undesired channel paths. As illustrated in \figref{fig:Soln_key_idea}, one type of undesired channel paths is the type of paths transmitted from/received by the sidelobes of a codebook beam. For this reason, we propose an efficient sidelobe reduction (SLR) approach. In \cite{Dessouky2006,Albagory2007}, an SLR approach was proposed for low sidelobe beamforming in uniform circular arrays. Inspired by their work, we propose a new efficient sidelobe reduction approach to uniform planar arrays (UPAs) to reduce beamforming/combining sidelobe levels.

The key idea of this approach is when applying different weights on the beamforming/combining vector elements, the sensing framework can control the beam radiation pattern in a way to increase the power difference between the mainlobe and the sidelobes. Specifically, let ${\bc}_\rm{H}\in \bbR^{{N}_\rm{H} \times 1}$ and ${\bc}_\rm{V}\in \bbR^{{N}_\rm{V} \times 1}$ represent the horizontal and vertical weight vectors for sidelobe reduction. Let ${\bb}_\rm{H}$ and ${\bb}_\rm{V}$ denote the horizontal and vertical constituent beamforming vectors after sidelobe reduction. The updated beamforming codebook, $\boldsymbol{\cF}$, for an ${N}_\rm{H} \times {N}_\rm{V}$ transmit UPA, can then be rewritten as
\begin{equation}
\begin{split}
\boldsymbol{\cF} =& \ \left\lbrace \bff \in \bbC^{N \times 1} : \bff = {\bb}_\rm{V} \left(\theta_\rm{z} \right) \circ  {\bb}_\rm{H}\left(\theta_\rm{x} \right), \left( \theta_\rm{z},\theta_\rm{x}\right) \in \cO \right\rbrace, \\
& {\bb}_\rm{V} \left(\theta_\rm{z} \right) = \widetilde{\bb}_\rm{V} \left(\theta_\rm{z} \right) \odot {\bc}_\rm{V}, \ {\bb}_\rm{H} \left(\theta_\rm{x} \right) = \widetilde{\bb}_\rm{H} \left(\theta_\rm{x} \right) \odot  {\bc}_\rm{H}, \\
\left[ {\bc}_\rm{V}\right]_{r_\rm{V}} =& \ \rm{e}^{\frac {-(r_\rm{V} - \mu_{\rm{V}})^{2}}{2 \sigma_\rm{V}^{2}}}, \ \mu_{\rm{V}} =  \frac{{N}_\rm{V}}{2}, \ \sigma_\rm{V} =  \frac{{N}_\rm{V}}{\delta_{\rm{V}}}, r_\rm{V} \in\{ 1,2,\dotsc,{N}_\rm{V}\},  \\
\left[ {\bc}_\rm{H}\right]_{r_\rm{H}} =& \ \rm{e}^{\frac {-(r_\rm{H}-\mu_{\rm{H}})^{2}}{2 \sigma_\rm{H}^{2}}}, \ \mu_{\rm{H}} =  \frac{{N}_\rm{H}}{2}, \ \sigma_\rm{H} =  \frac{{N}_\rm{H}}{\delta_{\rm{H}}}, r_\rm{H} \in \{1,2,\dotsc,{N}_\rm{H}\},
\end{split}
\end{equation}
where ${\delta_{\rm{H}}},{\delta_{\rm{V}}}$ are the sidelobe reduction control variables; the higher the values, the greater reduction in the sidelobe power levels compared to the mainlobe power level. 

\begin{figure}[t] \centerline{\includegraphics[scale=0.3]{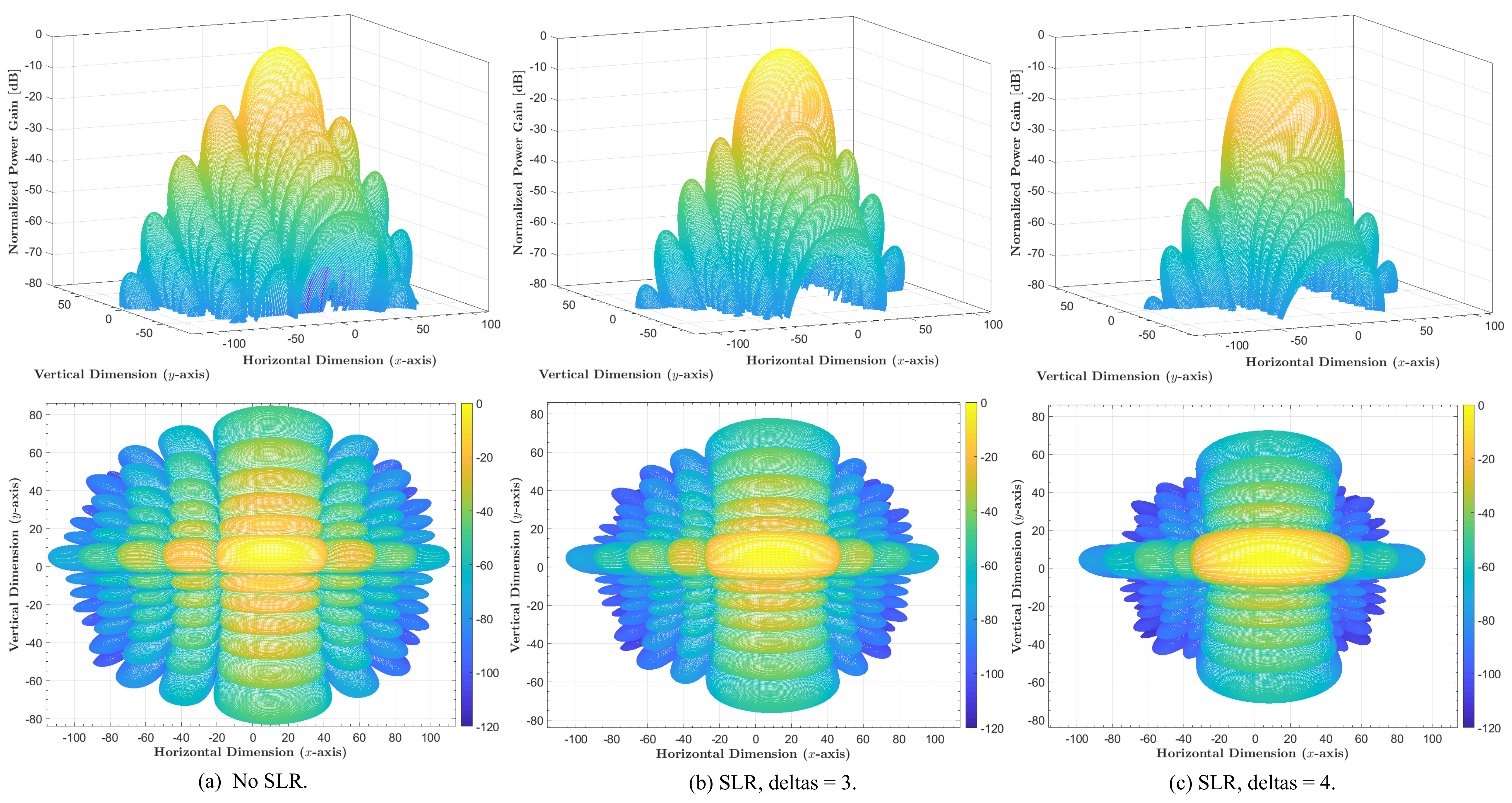}}
	\caption{Normalized power radiation pattern comparison between (a) the case without the sidelobe reduction (SLR) approach, (b) the case with the SLR approach where ${\delta_{\rm{H}}}={\delta_{\rm{V}}}=3$, and (c) where ${\delta_{\rm{H}}}={\delta_{\rm{V}}}=4$. As shown, increasing the values of the control variables (the deltas) increases the gap between the mainlobe level and the sidelobes levels. The top figures are the 3D views of the patterns while the bottom figures are the top views.}
	\label{fig:SLR}
\end{figure}

\begin{figure}[t] 
	\centering
	\begin{subfigure} [t]{.38\textwidth}
		\captionsetup{justification=centering}
		\centering
		\includegraphics [width=0.99\columnwidth]{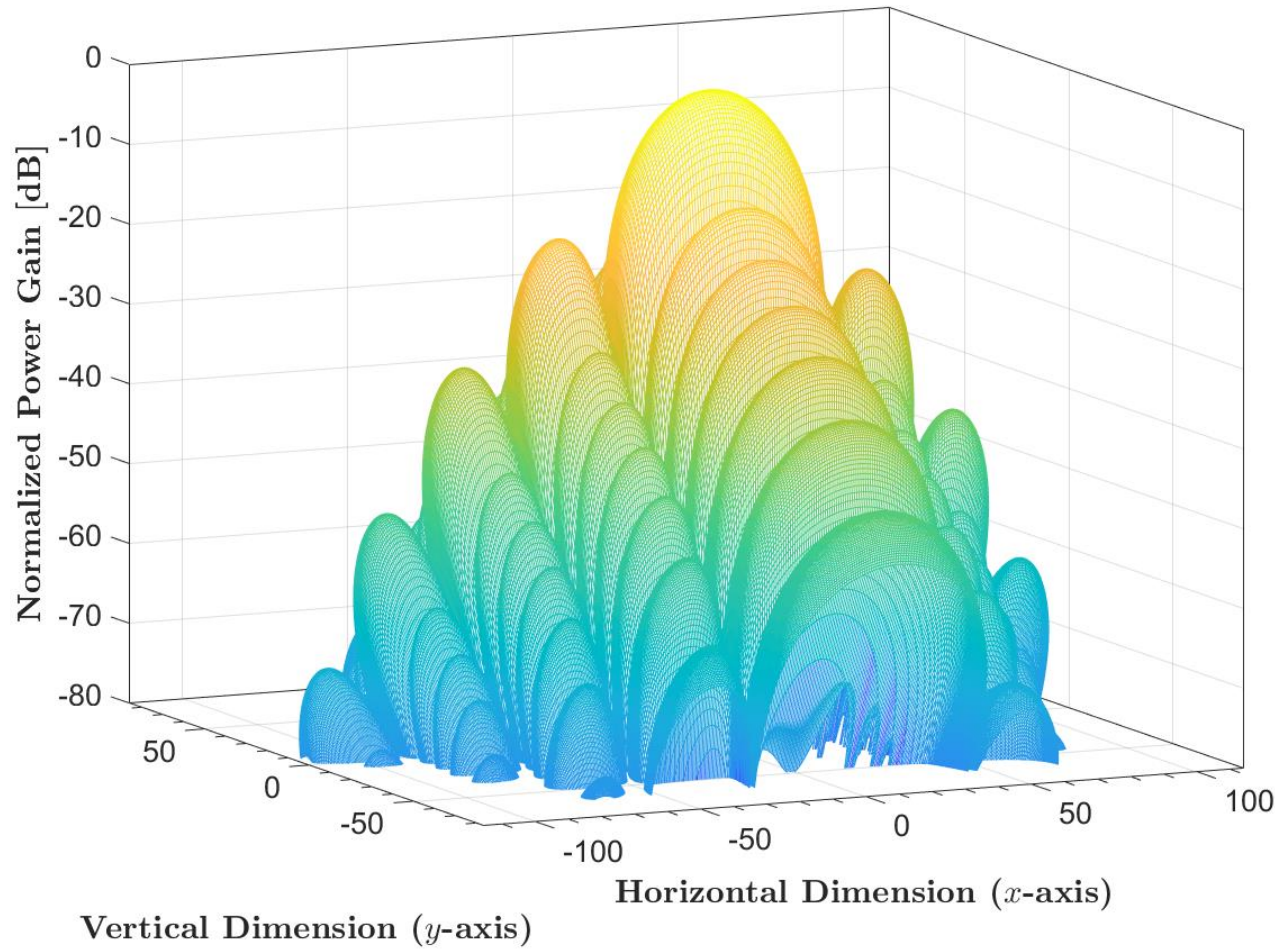}
		\caption{No SLR and continuous phase shifts}
	\end{subfigure}
	\begin{subfigure}[t]{.38\textwidth}
		\captionsetup{justification=centering}
		\centering
		\includegraphics[width=0.99\columnwidth]{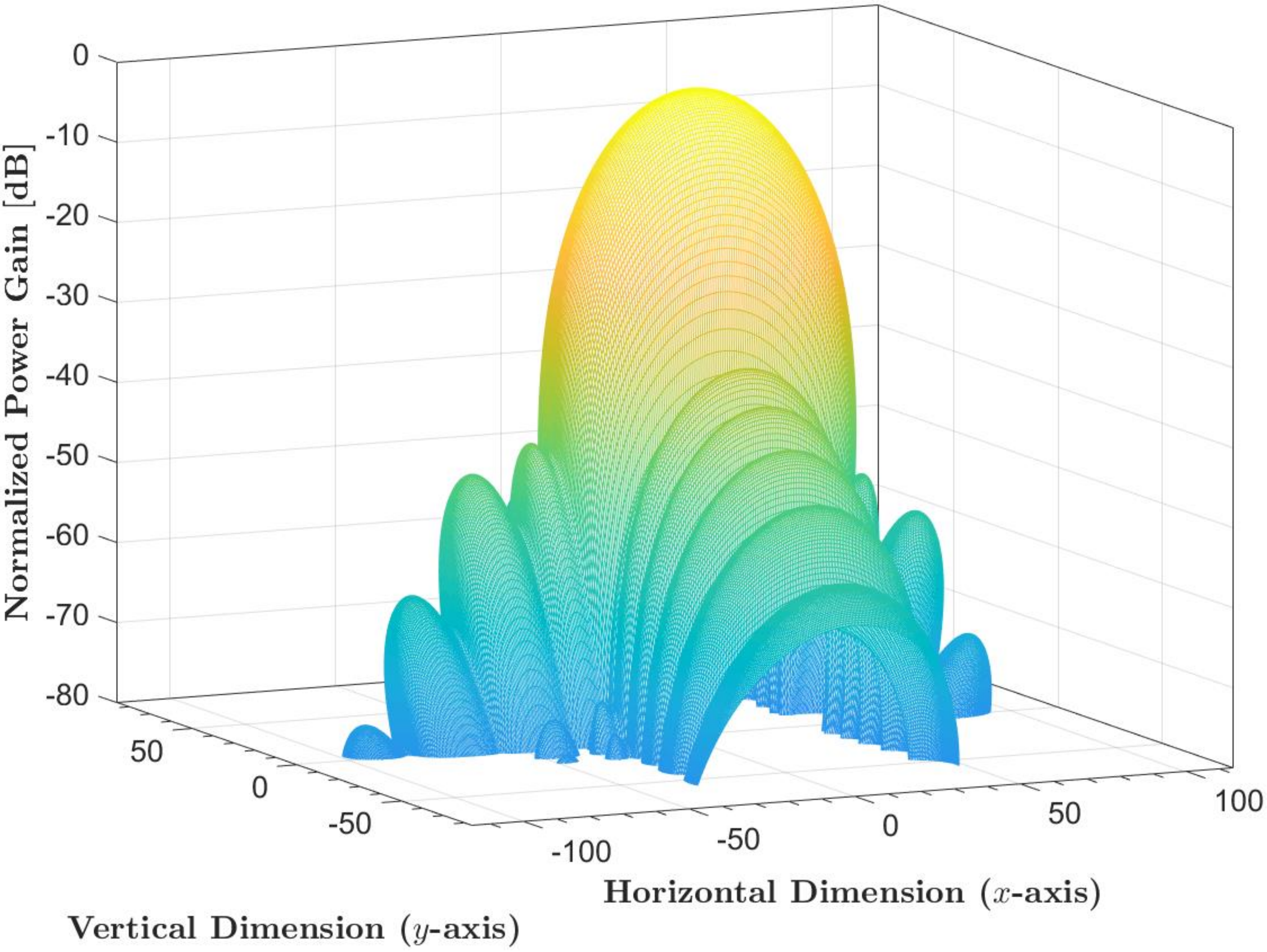}
		\caption{SLR deltas $=4$ and continuous phase shifts}
	\end{subfigure}
	\begin{subfigure} [t]{.38\textwidth}
		\captionsetup{justification=centering}
		\centering
		\includegraphics [width=0.99\columnwidth]{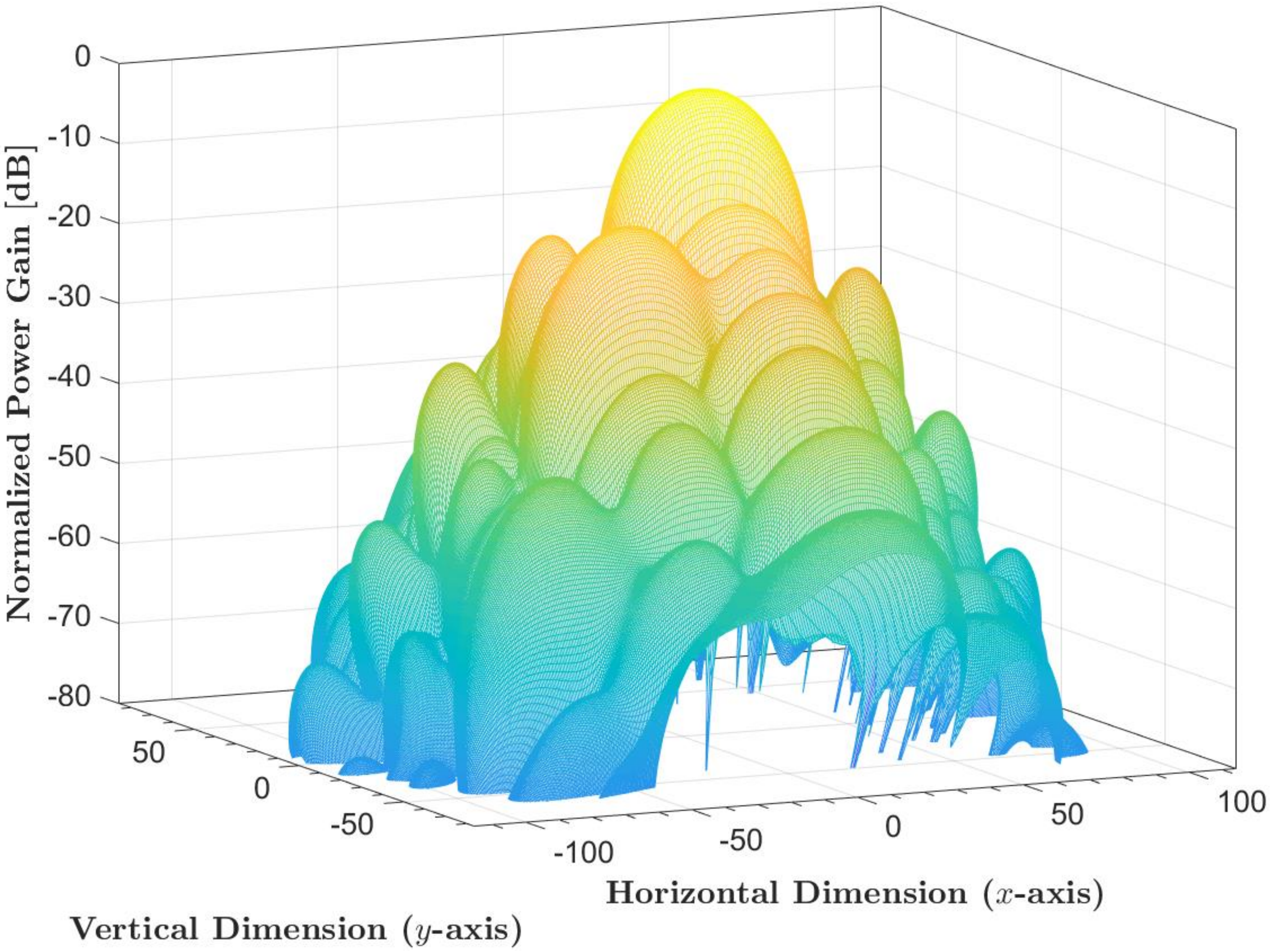}
		\caption{No SLR and 2-bit phase shifts}
	\end{subfigure}
	\begin{subfigure} [t]{.38\textwidth}
		\captionsetup{justification=centering}
		\centering
		\includegraphics [width=0.99\columnwidth]{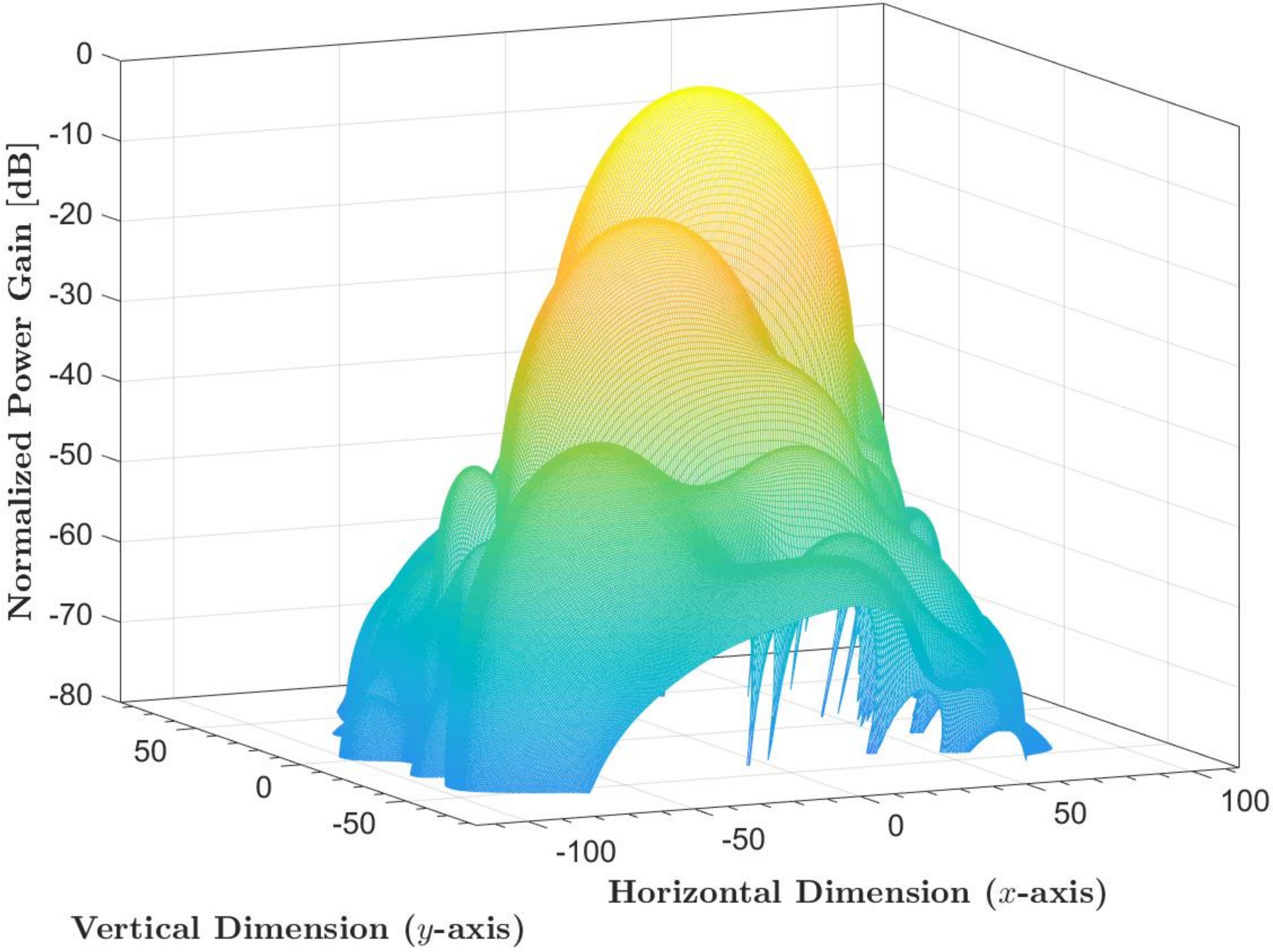}
		\caption{SLR deltas $=4$ and 2-bit phase shifts}
	\end{subfigure}
	\caption{Normalized power radiation pattern comparison between the case with no phase quantization and the case with $2$-bit phase quantization, for two scenarios: without the sidelobe reduction (SLR) approach and with the SLR approach where ${\delta_{\rm{H}}}={\delta_{\rm{V}}}=4$. The phase quantization affects the beam pattern shape of the sidelobes.} 
	\label{fig:SLR_Quantized_PS}
\end{figure}

\figref{fig:SLR} illustrates the radiation pattern in dB for one beamforming vector out of the updated beamforming codebook, $\boldsymbol{\cF}$, for different values of the sidelobe reduction control variables, ${\delta_{\rm{H}}},{\delta_{\rm{V}}}$. As depicted, increasing the values of the control variables increases the power gap between the mainlobe level and the sidelobes levels.
To take into consideration the phase quantization of the RF phase shifters in the AR/VR transceiver architecture previously shown in \figref{fig:Sigmodel_2b}, we examine the effect of $2$-bit phase quantization on the power radiation pattern. The $2$-bit discrete phase shift set is $\left\lbrace0,\frac{\pi}{2},{\pi},\frac{3\pi}{2} \right\rbrace$. \figref{fig:SLR_Quantized_PS} compares the power radiation pattern between the case of continuous phase shifts and the case of $2$-bit quantized phase shifts. As depicted, the phase quantization affects the beam pattern shape of the sidelobes.

One main advantage of this approach is its computational efficiency; as formulated, only two element-wise multiplication between the weight vectors and the constituent beamforming vectors, $\widetilde{\bb}_\rm{H},\widetilde{\bb}_\rm{V}$, are needed to update the beam radiation pattern. By contrast, reducing the sidelobe levels dramatically increases the beamwidth of the mainlobe, as depicted in \figref{fig:SLR}. 
The increased mainlobe beamwidth, however, can be mitigated by the other solutions proposed for rectifying the inter-path interference  problem, e.g. the successive interference cancellation (SIC) algorithm and the joint processing (JP) solution, as will be described in the following section.

\section{Proposed Scene Range/Depth Estimation} \label{ssec:scene_range_section}

In this section, given a pre-designed beamforming/combining codebook, $\boldsymbol{\cP}$, we propose an efficient approach for the scene range/depth estimation in AR/VR devices. As depicted in \figref{fig:mmWave_depth_alg}, once the beamforming/combining codebook has been designed, the AR/VR transmits the sensing signal while sweeping over all the beamforming-combining vector pairs. Specifically, for a beamforming-combining vector pair $\left( \bff_{m},\bw_{m}\right)$, where $m\in \{1,\dotsc,M\}$, the receive \textit{sensing} signal, $\by_m \in \bbC^{N^\rm{p}+L_{\sfd}}$, can be modeled as in \eqref{eq:receive_total} and  \eqref{eq:receive_totalF}.  After reception, the acquired sensing signals are  processed to estimate the range and depth maps, as will be thoroughly explained in this section. Our proposed post-processing solution has three main elements: (i) The use of oversampled/overlapped beams, (ii) the successive interference cancellation based management of  inter-target and inter-path interference, and (iii) the joint processing of the signals received using the codebook beams to realize high-resolution and accurate depth maps.  Next, we explain these three elements in Sections  \ref{ssec:overlappped_beams}-\ref{ssec:JP_soln} before presenting the scene range/depth map construction approach in \sref{ssec:scene_depth}.

\subsection{Overlapped Beams} \label{ssec:overlappped_beams}

With the objective of increasing the resolution of the mmWave MIMO based depth maps, we propose to adopt oversampled sensing codebooks to scan the surrounding environment. In particular, for the sensing codebook, we adopt the developed codebook in \sref{ssec:CBDesign} with oversampling factors of $ F_\rm{H}^\rm{OS}$ and $F_\rm{V}^\rm{OS}$ in the azimuth and elevation directions. While the oversampled codebook has the potential of enhancing the depth map resolution, it is important to note that advanced post-processing (for the receive signals using these oversampled beams) needs to be incorporated to achieve this goal. The reason mainly goes back to the wide beamwidth (and low spatial resolution) of the codebook beams, which is fundamentally limited by the number of AR/VR antennas. This wide beamwidth leads to a number of challenges: (i) The spatial regions scanned by the oversampled beams have high overlap. This makes it hard to differentiate between the depths of the different objects in the depth map pixels, which challenges the objective of realizing high-resolution depth maps. (ii) Since the codebook beams still have wide beamwidth, the inter-target interference problem discussed in \sref{ssec:scene_depth_challenges} still exists. 

To address these challenges, we propose a novel post-processing approach based on successive interference cancellation and joint-beam processing. This approach in summarized in two main steps as follows. In the first step, a successive interference cancellation (SIC) based algorithm is used to detect the most dominant channel paths contributing to the range/depth estimation of the region covered by each codebook beam. These paths form a set of candidate ranges/depths for the scene range/depth estimation. In the second step, a developed joint-beam processing solution selects one range/depth out of the set of candidate ranges/depths formed by the SIC algorithm. These two sequential algorithms are discussed in detail in the following two subsections.
\begin{algorithm}[t]
	\caption{Successive Interference Cancellation}
	\label{SIC_alg}
	\begin{algorithmic}[1]
		
		\Require{Receive sensing signal $y_{m}[n]$, transmit preamble signal $s^\rm{p}[n]$, threshold level $A_\rm{TH}$, beamforming-combining pair codebook $\boldsymbol{\cP}$.}
		\Ensure{Candidate delay set for each beam, $\cT_{m}, \forall m \in \{1,\dotsc,M\},$.} 
		
		\For{$m=1$ \textbf{to}  $M$} \Comment{For each beamforming and combining vector pair $\left(  \bff_{m}, \bw_{m} \right)$.}	
		
		\Initialize{ Updated signal $\tilde{y}_{m}[n] \gets y_{m}[n]$, solution set $\cT_{m} \gets \emptyset, \forall m$, and $\tilde{A} \gets A_\rm{TH}$.}	
		
		\While{ $\tilde{A} \geq A_\rm{TH}$ } \Comment{Compute the delay of each dominant channel path.}	
		
		\State 
		Calculate the delay of the path with the maximum cross-correlation,
		\begin{align*}
		\tilde{q} \gets &  \argmax_{q:q \in \cQ } \left| \sum_{n=L_\rm{y}-N_\rm{y}}^{L_\rm{y}-1} s^\rm{p}[n]  \times \left( \tilde{y}_{m}[n-q]\right)^*   \right|^{2}. \\
		\tilde{A} \gets  &  \left[  \sum_{n=L_\rm{y}-N_\rm{y}}^{L_\rm{y}-1} s^\rm{p}[n]  \times \left( \tilde{y}_{m}[n-\tilde{q}]\right)^*  \right].
		\end{align*}

		\State \begin{varwidth}[t]{\columnwidth} 
			Add the candidate delay to the solution set, $ \cT_{m} \gets  \cT_{m} \cup \left\lbrace \tilde{q} \right\rbrace.$
		\end{varwidth}
		
		\State 
		Calculate the energy of the transmit signal up to this delay value,
		\begin{equation*}
		E_\rm{Q} \gets \sum_{n=L_\rm{y}-N_\rm{y}}^{L_\rm{y}-\tilde{q}-1} \left|  s^\rm{p}[n] \right|^2. 
		\end{equation*}

		\State 
		Perform interference cancellation at this delay value,
		\begin{equation*}
		\tilde{y}_{m}[n] \gets \tilde{y}_{m}[n] - \frac{\tilde{A}}{E_\rm{Q}}  s^\rm{p}[n-\tilde{q}].
		\end{equation*}

		\State  
		Calculate the next maximum in the cross-correlation,
		\begin{equation*}
		\tilde{A} \gets  \max_{q:q \in \cQ } \left| \sum_{n=L_\rm{y}-N_\rm{y}}^{L_\rm{y}-1} s^\rm{p}[n]  \times \left( \tilde{y}_{m}[n-q]\right)^*   \right|^{2}.
		\end{equation*}

		\EndWhile
		\EndFor
		
	\end{algorithmic}
\end{algorithm}

\subsection{Successive Interference Cancellation} \label{ssec:sic}

\begin{figure}[t] \centerline{\includegraphics[scale=0.53]{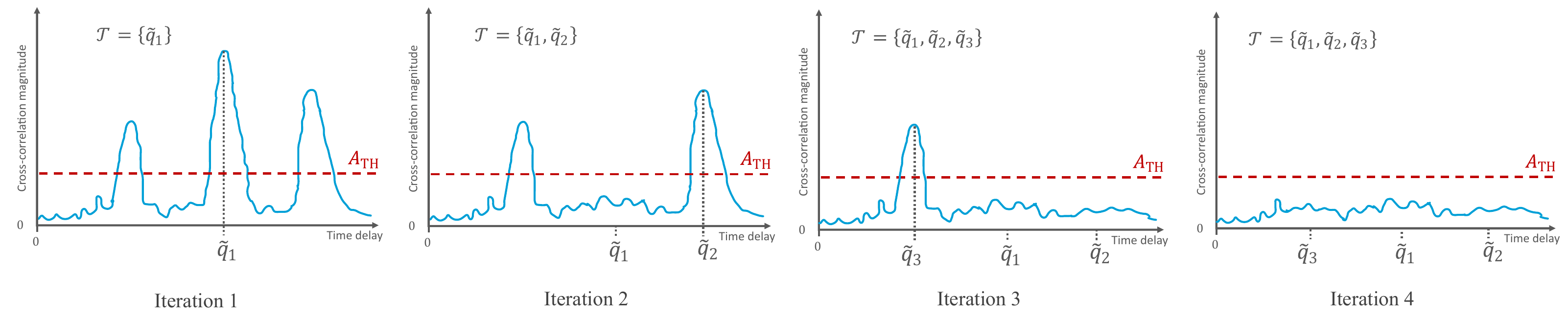}}
	\caption{The operation of the successive interference cancellation (SIC) algorithm is illustrated. The delay position of the maximum cross-correlation is first detected. The SIC algorithm then encodes a signal shifted at this delay position and subtracted it from the receive signal. After that, the algorithm repeats itself until all the local maxima above the threshold value are detected.}
	\label{fig:SIC_soln}
\end{figure}

The main goal of the successive interference cancellation (SIC) algorithm is to detect all the dominant paths that might contribute to the range estimation of the region of interest. This is motivated by its good performance in multi-target detections problems \cite{Grossi2019}. The SIC algorithm is applied in the discrete-time domain and is summarized in Algorithm \ref{SIC_alg}. The algorithm is described as follows. Let the length of the receive sensing sequence $y_{m}[n]$ be $L_\rm{y}=N^\rm{p}+L_{\sfd}$ symbols.  First, as shown in \figref{fig:SIC_soln}, for every codebook beam, the delay position of the maximum cross-correlation magnitude value is detected. $\cQ$ is the set of possible delays. Second, the SIC algorithm encodes the transmit preamble signal to be shifted to this delay position and subtracted it from the received signal. Afterwards, the algorithm repeats itself to detect the second local maximum above the threshold value. Finally, The SIC algorithm stops iterating when all the local maxima above the threshold value are detected. The output of this algorithm is a set of candidate delays for every codebook beam. These sets pass as input to the next algorithm, the joint processing solution, as will be explained in the next part.  In \figref{fig:SIC_soln}, note that the cross-correlation magnitude plot appears to be drawn as a continuous plot, only for illustration purposes. The actual cross-correlation magnitude, however, is expressed in discrete time delays.

\subsection{Joint Processing Solution} \label{ssec:JP_soln}

\begin{figure}[t] \centerline{\includegraphics[scale=0.45]{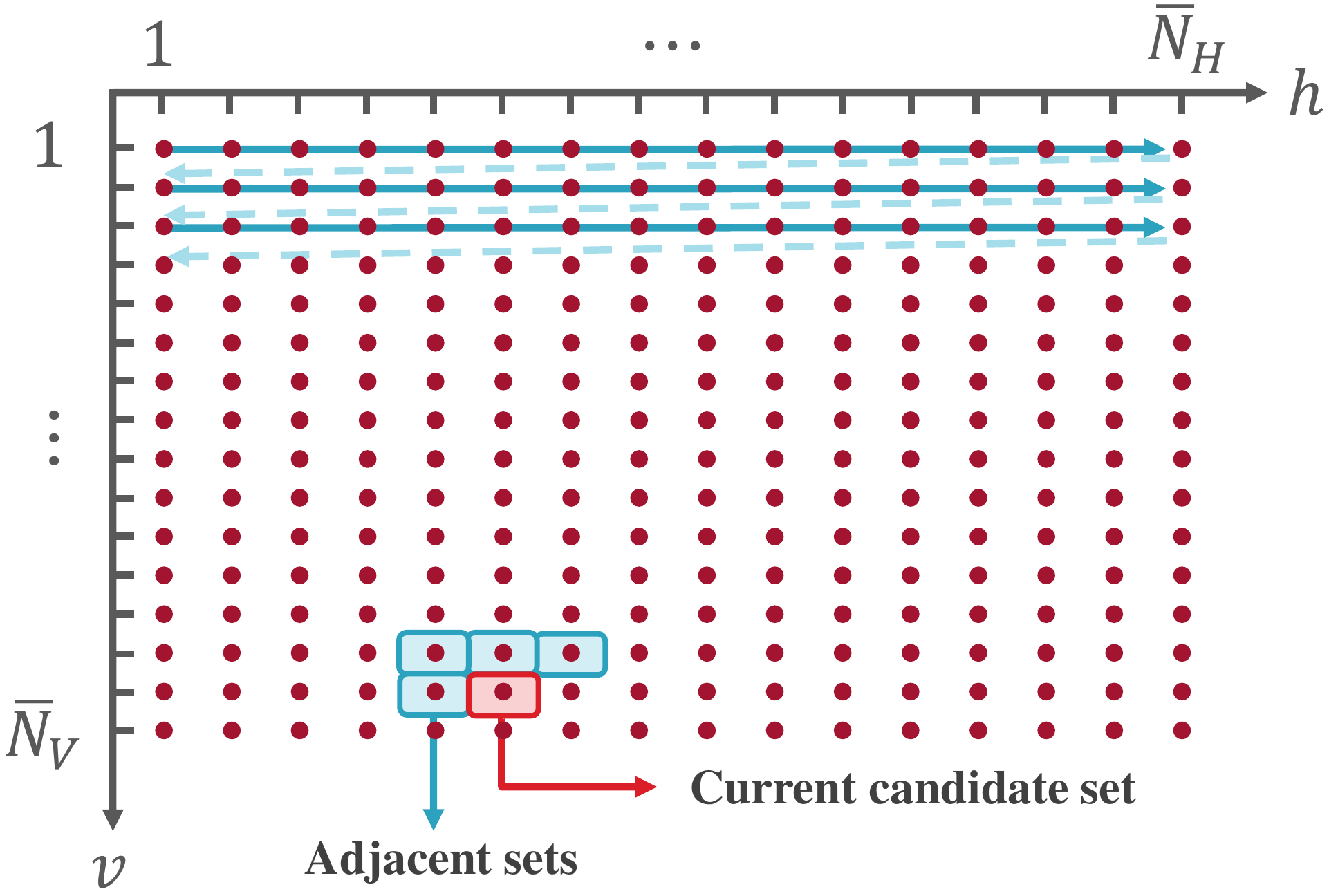}}
	\caption{This figure illustrates the basic operation of the joint processing (JP) solution for overlapped beams. The JP solution sweeps from left to right, then from top to bottom. The JP solution decides on which path to choose from the current candidate set by a simple comparison with the sets of the surrounding grid points.}
	\label{fig:JP_soln}
\end{figure}

The purpose of the joint processing (JP) solution between the overlapped beams is to estimate the transitions in depth/range maps more accurately. The proposed JP solution is summarized in Algorithm \ref{scene_range_alg}. The algorithm is described in detail as follows. First, the JP solution works on the candidate delay sets, the output from the SIC algorithm, $\left\lbrace \cT_{m}\right\rbrace_{m=1}^{M}$, to choose one range estimate out of the candidate delay set. This processing, however, is employed relative to the 2D codebook grid, as illustrated in \figref{fig:JP_soln}. 
Following this notion, the linear indices in $\cT_{m}$ is now converted into matrix subscripts $\cT_{h,v}$ through the transformation $m=(v-1)\overline{N}_\rm{H}+h$, such that $\cT_{m}=\cT_{h,v}$, where $v$ is the elevation beam index (vertical grid index) and $h$ is the azimuth beam index (horizontal grid index). The objective is to calculate the scene range estimates across all beam directions, $\left( \hat{\rho}_{h,v}\right)^\rm{SRE}, \forall h,v$. 

\begin{algorithm}[t]
	\caption{Joint Processing Solution}
	\label{scene_range_alg}
	\begin{algorithmic}[1]
		
		\Require{Candidate delay set for each beam, $\cT_{h,v},\forall h\in\{1,\dotsc,\overline{N}_\rm{H}\},\forall v\in\{1,\dotsc,\overline{N}_\rm{V}\}.$}
		\Ensure{Scene range estimate $\left( \hat{\rho}_{h,v}\right)^\rm{SRE}, \forall h,v.$} 
		
		\Initialize{Common adjacent set $\cN_{h,v} \gets \emptyset, \forall h,v$, difference set $\cM_{h,v} \gets \emptyset, \forall h,v.$}

		\For{ $v=1$ \textbf{to}  $\overline{N}_\rm{V}$ }	
		\For{ $h=1$ \textbf{to}  $\overline{N}_\rm{H}$ }	
		
		\State \begin{varwidth}[t]{\columnwidth} 
			Construct the common adjacent set, $ \cN_{h,v} \gets \! \left( \cT_{h-1,v} \cup \cT_{h,v-1} \cup  \cT_{h-1,v-1} \cup  \cT_{h+1,v-1} \right) \! .$		
		\end{varwidth}
		
		\State \begin{varwidth}[t]{\columnwidth} 
			Construct the difference set, $ \cM_{h,v} \gets  \cT_{h,v} \setminus \cN_{h,v}.$
		\end{varwidth}
		
		\If{$\cM_{h,v} \neq \emptyset$} 
		
		\State \begin{varwidth}[t]{\columnwidth} 
			Choose the least delay from the difference set, $\left( \hat{\rho}_{h,v}\right)^\rm{SRE} \gets \frac{\varsigma T_\rm{S}}{2}  \min \cM_{h,v}.$		
		\end{varwidth}
		
		\Else		
		
		\State \begin{varwidth}[t]{\columnwidth} 
			Choose the least delay from the candidate set, $\left( \hat{\rho}_{h,v}\right)^\rm{SRE} \gets \frac{\varsigma T_\rm{S}}{2}  \min \cT_{h,v}.$
		\end{varwidth}
		
		\EndIf
		\EndFor
		\EndFor		
	\end{algorithmic}
\end{algorithm}

As shown in \figref{fig:JP_soln}, the JP solution sweeps from left to right, then from top to bottom. For each grid point, the JP solution uses (i) the set of the current grid point, named as the "current set", and (ii) the sets of the previous adjacent grid points to construct a "common adjacent set". This common adjacent set is the union of the sets of all previous adjacent grid points. Then, to investigate if a new object/surface transition appears, this current set is compared with the common adjacent set to detect if there is any set difference. This is based on the notion that the difference set can probably be the new edges that will appear in the range map while sweeping. If the set difference is not empty, then the solution chooses the path with the least time-of-flight from the set difference. Otherwise, if the set difference is empty, then the solution chooses the path with the least time-of-flight from the current set. 

\subsection{Range/Depth Map Construction} \label{ssec:scene_depth}

\begin{algorithm}[t]
	\caption{mmWave MIMO Sensing Based Range/Depth Estimation Framework}
	\label{alg1}
	\begin{algorithmic}[1]
		
		\Require Field of view $\! \rm{FoV}$, aspect ratio $\! A_\rm{R}$, number of horizontal and vertical beams $\! \overline{N}_\rm{H},\overline{N}_\rm{V}$.
		\Ensure Range map $\widehat{\bR}_\rm{map}$, depth map $\widehat{\bD}_\rm{map}$.
		
		\State \begin{varwidth}[t]{\columnwidth} 
			Design the beamforming-combining pair codebook, $\boldsymbol{\cP}$, following \sref{sec:CB_design}. 		
		\end{varwidth}

		\For{$m=1$ \textbf{to}  $M$} \!\!\!\!  \Comment{For each beamforming and combining vector pair $ \left(  \bff_{m}, \bw_{m}\right)$.}

		\State \begin{varwidth}[t]{\columnwidth} 
			Acquire receive \textit{sensing} signal, $y_{m}[n], \forall n\in \{0, 1, \dotsc, N^\rm{p}+L_{\sfd}-1\}$, as in \eqref{eq:receive_total}. 
		\end{varwidth} 
		
		\EndFor
		
		\State \begin{varwidth}[t]{\columnwidth} 
			Calculate the candidate delay set for each beam, $\! \cT_{m}, \forall m\in \{1,\dotsc,M\}$, as in Algorithm \ref{SIC_alg}.
		\end{varwidth}
		
		\State \begin{varwidth}[t]{\columnwidth} 
			Calculate the scene range estimate for each beam, $\left( \hat{\rho}_{m}\right)^\rm{SRE}, \forall m$, as in Algorithm \ref{scene_range_alg}.
		\end{varwidth}
		
		\State \begin{varwidth}[t]{\columnwidth} 
			Calculate fine range estimates, 
			$\left( \hat{\rho}_{m}\right)^\rm{MC} \gets \left( \hat{\rho}_{m}\right)^\rm{SRE}+\left( \hat{\rho}_{m}\right)', \forall m$, following \sref{subsubsec:MC}.
		\end{varwidth} 
		
		\State \begin{varwidth}[t]{\columnwidth} 
			Construct the range map, $ \widehat{\bR}_\rm{map}$, from \eqref{eq:range_map}.
		\end{varwidth}
		
		\State \begin{varwidth}[t]{\columnwidth} 
			Construct the depth map, $\widehat{\bD}_\rm{map}$, from \eqref{eq:depth_map}.	
		\end{varwidth}
		
	\end{algorithmic}
\end{algorithm}

In this section, we formulate the depth map construction approach, the final step in \figref{fig:mmWave_depth_alg}. In summation of the broader view, the mmWave MIMO sensing based range/depth map estimation framework is outlined in Algorithm \ref{alg1}. The algorithm steps are summarized as follows.  \textit{Step 1} refers to the design of the beamforming-combining pair codebook $\boldsymbol{\cP}$ was covered in \sref{sec:CB_design}. \textit{Step 4} refers to the successive interference cancellation described in \sref{ssec:sic}. \textit{Step 5} refers to the joint processing solution detailed in \sref{ssec:JP_soln}. After that, in \textit{Step 6}, the fine range estimate can be calculated, such that $\left( \hat{\rho}_{m}\right)^\rm{MC}=\left( \hat{\rho}_{m}\right)^\rm{SRE}+\left( \hat{\rho}_{m}\right)'$, where $\left( \hat{\rho}_{m}\right)'$ is computed from the algorithm described in \sref{subsubsec:MC}.
Next, after calculating the range estimates, the upcoming steps (\textit{Steps 7,8}) are focused on constructing the range and depth maps. Note that the range to an object is actually the radial distance in spherical coordinates. Fortunately, the $\left(x,y,z\right)$ rectangular coordinates of the sensor grid points on the camera plan were already calculated for the design of the beamforming-combining pair codebook using \eqref{eq:rect_coordinates}. These rectangular coordinates in \eqref{eq:rect_coordinates} can then be converted to spherical coordinates, such that
\begin{equation} \label{eq:Sph_coordinates}
\cS =  \Biggl\lbrace \left(\theta_\rm{z},\Phi\right): \theta_\rm{z} = \left[ \tfrac{\pi}{2} - \arctan\left(\tfrac{z}{\sqrt{x^2 + y^2}} \right)\right],\Phi = \arctan\left(\tfrac{y}{x} \right), \left(x,y,z\right) \in \cC \Biggl\rbrace.
\end{equation}

In order to construct the matrices for the range and depth maps, let $\boldsymbol{\Theta},\boldsymbol{\Phi} \in \bbR^{\overline{N}_\rm{V} \times \overline{N}_\rm{H}} $ be the matrices that represent the angles of the spherical coordinates $\left(\theta_\rm{z},\Phi\right) \in \cS$, respectively. Following \textit{Step 6} in Algorithm \ref{alg1}, the range map estimate $\widehat{\bR}_\rm{map} \in \bbR^{\overline{N}_\rm{V} \times \overline{N}_\rm{H}}$ can be expressed as  
\begin{equation} \label{eq:range_map} 
\left[ \widehat{\bR}_\rm{map}\right]_{v,h}  = \left( \hat{\rho}_{m}\right)^\rm{MC}, \ m=(v-1)\overline{N}_\rm{H}+h , \ \forall m \in \{1,\dotsc,M\},
\end{equation}
where $h\in \{1,\dotsc,\overline{N}_\rm{H}\}, v\in \{1,\dotsc,\overline{N}_\rm{V}\}$. Given the angles in spherical coordinates and the range map estimate, the depth map estimate $\widehat{\bD}_\rm{map} \in \bbR^{\overline{N}_\rm{V} \times \overline{N}_\rm{H}}$ can then be expressed as
\begin{equation} \label{eq:depth_map}
\left[ \widehat{\bD}_\rm{map}\right]_{v,h} = \left| \hat{\rho} \sin\left(\theta_\rm{z}\right) \sin\left(\Phi\right) \right|, \ 
\hat{\rho} = \left[ \widehat{\bR}_\rm{map}\right]_{v,h}, \ \theta_\rm{z} = \left[ \boldsymbol{\Theta}\right]_{v,h},\ \Phi = \left[ \boldsymbol{\Phi}\right]_{v,h}, \ \forall v,h.
\end{equation}

Finally, since the range and depth map resolutions are set to $\overline{N}_\rm{H} \times \overline{N}_\rm{V}$, two-dimensional image interpolation can be employed to scale the maps to the desired resolutions, $M_h \times M_w$. Examples of interpolation methods are the nearest neighbor interpolation and the bicubic interpolation. Although the bicubic interpolation can probably be the interpolation method of choice for achieving more estimation accuracy, the nearest neighbor interpolation is more computationally efficient. In the simulation results of \sref{sec:Sim_Results}, we evaluate the  two interpolation approaches for our mmWave MIMO based depth map construction problem.

\begin{figure}[t] \centerline{\includegraphics[scale=0.5]{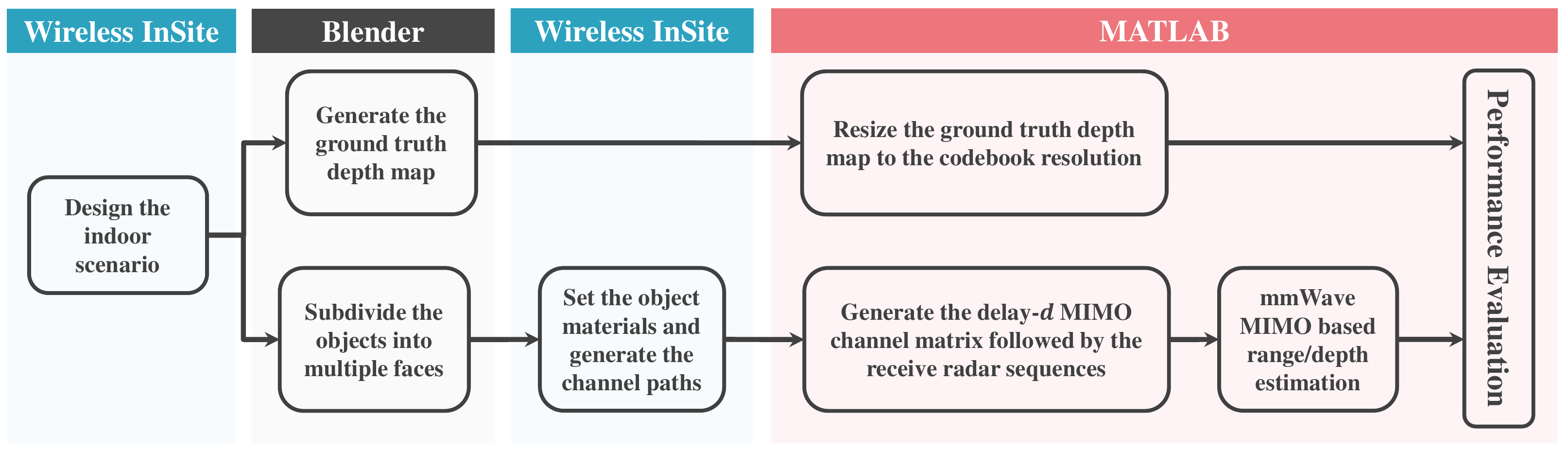}}
	\caption{This figure demonstrates the adopted simulation framework for scene depth estimation. The framework consists of: designing the indoor setup, generating the ground truth range/depth maps, and constructing the estimated maps for performance evaluation. For more complex setups, designing the indoor scenarios jointly in Wireless InSite and Blender can be more effective.} 	
	\label{fig:sim_framework}
\end{figure}

\section{Simulation Results}  \label{sec:Sim_Results}
In this section, we evaluate the performance of the proposed mmWave based depth estimation approach. First, we  describe the adopted simulation framework in  \sref{sec:sim_framework} before extensively studying the estimation accuracy of the proposed approach under various scenarios and system parameters. The simulation results presented can be of great usefulness for various applications; they can be generally applied to AR/VR devices, smart home devices, or auto drive devices.

\subsection{Simulation Framework}  \label{sec:sim_framework}
Since the depth estimation heavily depends on the environment under test, it is crucial to evaluate the performance of the proposed solution based on realistic channels. This motivates using channels generated by accurate ray-tracing  to capture the sensing dependence on the environment geometry, scatterers' materials, AR/VR position, etc. This is why we designed the simulations models using  Remcom Wireless InSite \cite{Remcom}, which is an accurate 3D ray-tracing simulator. Further, to efficiently incorporate diffuse scattering models, we need to have highly detailed floor plans with a sufficient number of faces. To achieve this objective, we resorted to the high-fidelity game engine, Blender \cite{Blender},  to build accurate floor plans. These plans/models are then exported to Wireless InSite to obtain the ray-tracing outputs, and finally to MATLAB to construct the channel models in \eqref{eq:radarchannel} and implement the proposed depth estimation approach. The proposed evaluation framework is illustrated in \figref{fig:sim_framework}. For benchmarking, we also use the Blender floor plans to obtain the ground truth depth maps, which are essential to evaluate the accuracy of our solutions. The ground truth maps are generated by placing a Blender camera at the same position of the UPA reference antenna element, and adjusting the Blender camera parameters to capture the same field of view.

\textbf{Signal model:}
We adopt the signal model described in \sref{sec:Sys_CH_Models} with a focus on the sensing system performance. The AR/VR device is assumed to be fixed in position. Unless otherwise mentioned, the UPA size is $16 \times 16$ antennas (${N}_\rm{H}={N}_\rm{V}=16$) at the mmWave $60$GHz operating band with transmission bandwidth of $2$GHz. The antenna elements have a gain of $0$dBi with half-wavelength antenna spacing. The transmit power is set to $30$dBm. The preamble sequence is the same as the one in the single carrier PHY packet preamble of the IEEE 802.11ad standard (3328 symbols). $M$ preamble sequences are used to sense the environment via $M$ beamforming-combining pairs. For the sake of calculating a rough estimate of the time allocated for environment sensing through transmission and reception, assume that all the $M$ preamble sequences are transmitted sequentially with guard intervals in between. The highest $M$ value reported in the upcoming simulation results is $4096$ beams. Assuming a sampling rate of $2$Gsps, the estimate of the longest sensing time is then $\approx7$ms.

\begin{table}[t]
	\caption{The adopted diffuse scattering parameters for different materials}
	\begin{center}
		\begin{tabular}{ | c | c | c | c | c | c | c | }
			\hline
			\textbf{Diffuse Scattering Parameter} & \textbf{Concrete}  & \textbf{Ceilingboard} & \textbf{Wood} & \textbf{Floorboard}  & \textbf{ Drywall} & \textbf{Glass}  \\ \hline \hline
			Scattered to incident electric field ratio & $40\%$ & $30\%$ & $15\%$ & $15\%$ & $10\%$ & $0\%$  \\ \hline
			Forward to backward scattering power ratio  & $75\%$ & $75\%$ & $75\%$ & $75\%$ & $75\%$ & $75\%$ \\ \hline
			Cross-polarization ratio  & $40\%$ & $40\%$ & $40\%$ & $40\%$ & $40\%$ & $40\%$ \\ \hline
			Narrowness of the scattering lobes & $40\%$ & $40\%$ & $40\%$ & $40\%$ & $40\%$ & $40\%$ \\ \hline
		\end{tabular}
	\end{center}
	\label{table:materials}
\end{table}

\textbf{Channel generation:} 
The channel matrix, $\bH_{d}$, is generated in two steps. The first step is generating the channel rays using the ray-tracing software, Wireless InSite. The Wireless InSite propagation model is set to 'X3D' with $0.1^\circ$ ray-spacing and enabled mode of diffuse scattering. Up to three reflections, one diffraction, and one transmission properties are allowed for each ray in the Wireless InSite simulation. The diffuse scattering model used is ``directive with backscatter"; this model is fixed across all materials in all the testing scenarios. The chosen diffuse scattering model creates two scattering lobes; a forward lobe of diffuse scattered power centered on the direction of specular reflection and a backward lobe centered on the opposite direction of incidence. The diffuse scattering parameters of the different materials are summarized in Table \ref{table:materials}. The values reported in Table \ref{table:materials} follow the ITU default parameter values at $60$GHz. The second step in the sensing channel generation is calculating the delay-$d$ channel matrix out of the channel paths using the DeepMIMO dataset generation code \cite{DeepMIMO2019}. Using these channels and following \eqref{eq:radarchannel}-\eqref{eq:FinalRadarRX_pre}, the noisy receive sensing sequences are generated. The noise power is calculated based on a $2$GHz bandwidth and a receiver noise figure of $7$dB. 

\begin{figure}[t] \centerline{\includegraphics[scale=0.55]{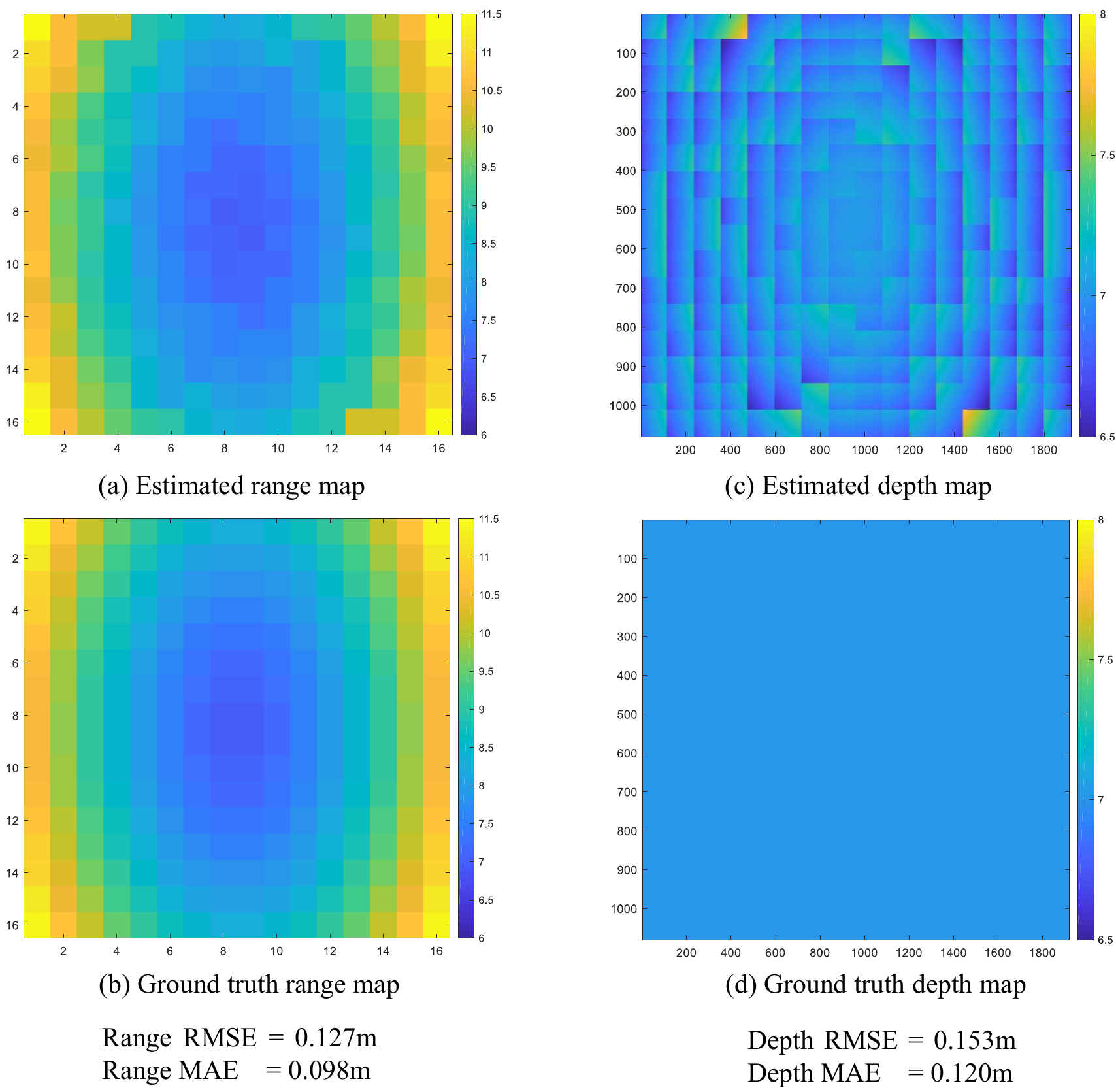}}
	\caption{The maps for the one wall scenario are depicted for a separation distance of $7$ meters from the AR/VR device with $16\times 16$ UPAs. The depicted maps are
	 the estimated maps (at the top), ground truth maps (at the bottom), range maps (on the left side), and $1080\rm{p}$ depth maps (on the right side). Comparing (a) with (b), the range map estimation error: MAE $= 0.098$m. Comparing (c) with (d), the depth map estimation error: MAE $= 0.12$m.}
	\label{fig:Onewall_maps}
\end{figure}

\textbf{mmWave based depth estimation parameters:} The beamforming-combining pair codebook is designed based on a $100^{\circ}$ field of view centered around the antenna array boresight, a $16/9$ scene aspect ratio, and horizontal and vertical oversampling factors of unity. The ground truth depth maps are generated from Blender using a Blender camera with a $100^{\circ}$ field of view, a focal length of $13.43$mm corresponding to a sensor width of $32$mm. The ground truth depth map image quality is set to $1080\rm{p}$ resolution; i.e., $1920 \times 1080$ pixels. Concerning the massive correlator, $f_\rm{est}$ is set to $100$ multiple of the sampling frequency $f_\rm{S}$; i.e., $\delta= \frac{f_\rm{est}}{2f_\rm{S}}=50$. Unless mentioned otherwise, the massive correlator is adopted for range estimation. Throughout this paper, two performance metrics are used: (i) root-mean-square-error (RMSE) between the estimated map and the ground truth map to indicate the standard deviation of the estimation error, and (ii) mean-absolute-error (MAE) to denote the expected value of the estimation error. The two metrics are defined in \eqref{eq:obj}. Next, we evaluate the performance of our proposed mmWave MIMO depth estimation approach in four main scenarios: (i) A one wall scenario in \sref{subsec:one_wall}, (ii) a two walls scenario in \sref{subsec:two_walls}; (iii)) a room with two pillars scenario in \sref{subsec:indoor_room}, and (iv) a conference room scenario in \sref{subsec:conference_room}. 

\begin{figure}[t] \centerline{\includegraphics[scale=0.75]{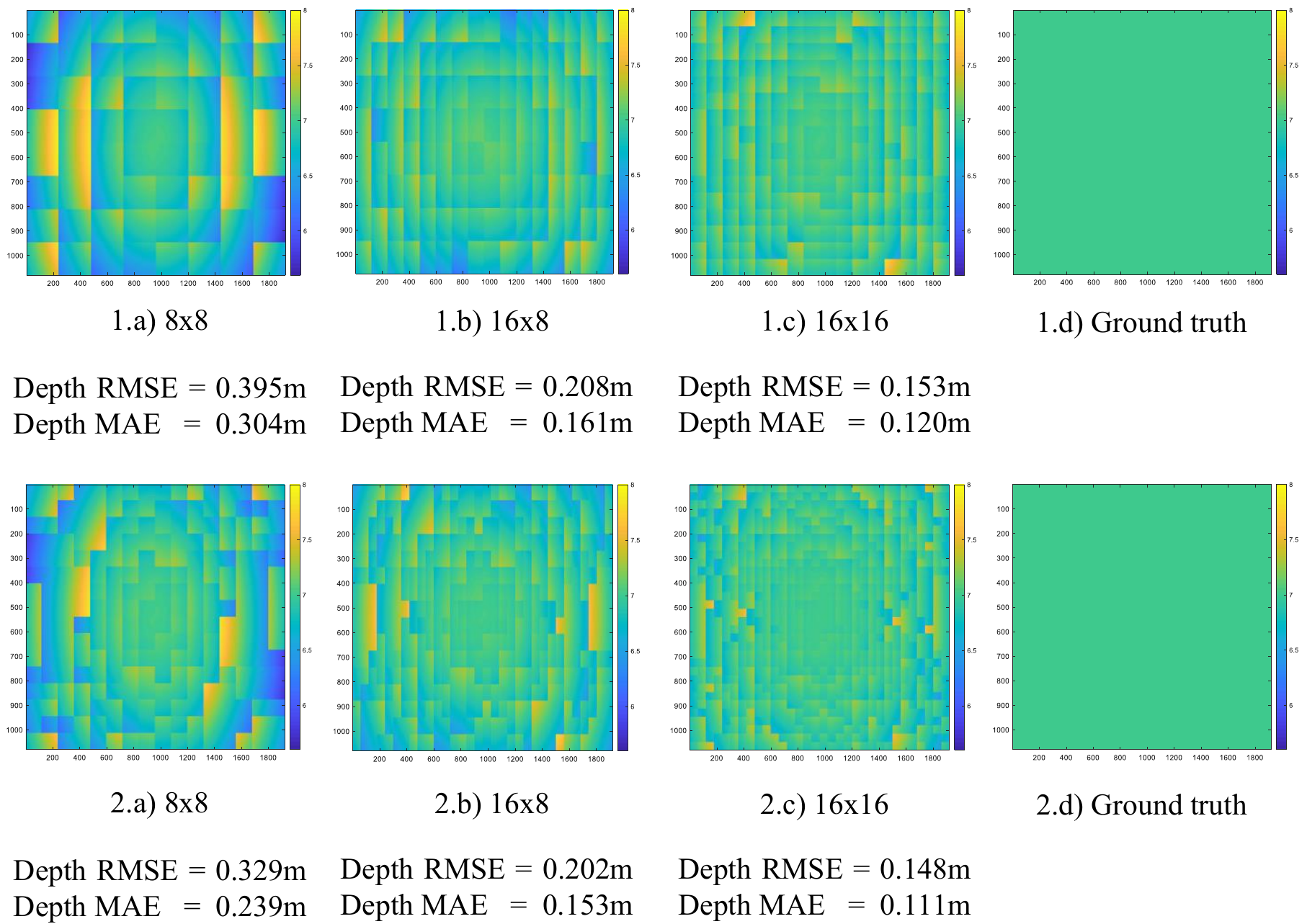}}
	\caption{The depth maps for the one wall scenario are depicted for different antenna configurations and codebook resolutions, for a separation distance of $7$ meters. Figures (a), (b), and (c) illustrate the estimated $1080\rm{p}$ maps for $8\times 8$, $16\times 8$, and $16 \times 16$ UPAs. Figures (d) illustrate the ground truth maps. The top maps are with no codebook oversampling while the bottom maps are with codebook oversampling factors of two.}
	\label{fig:Onewall_maps_var_antennas}
\end{figure}

\begin{figure}[t] 
	\centering
	\begin{subfigure} [t]{.24\textwidth}
		\captionsetup{justification=centering}
		\includegraphics [width=0.99\columnwidth]{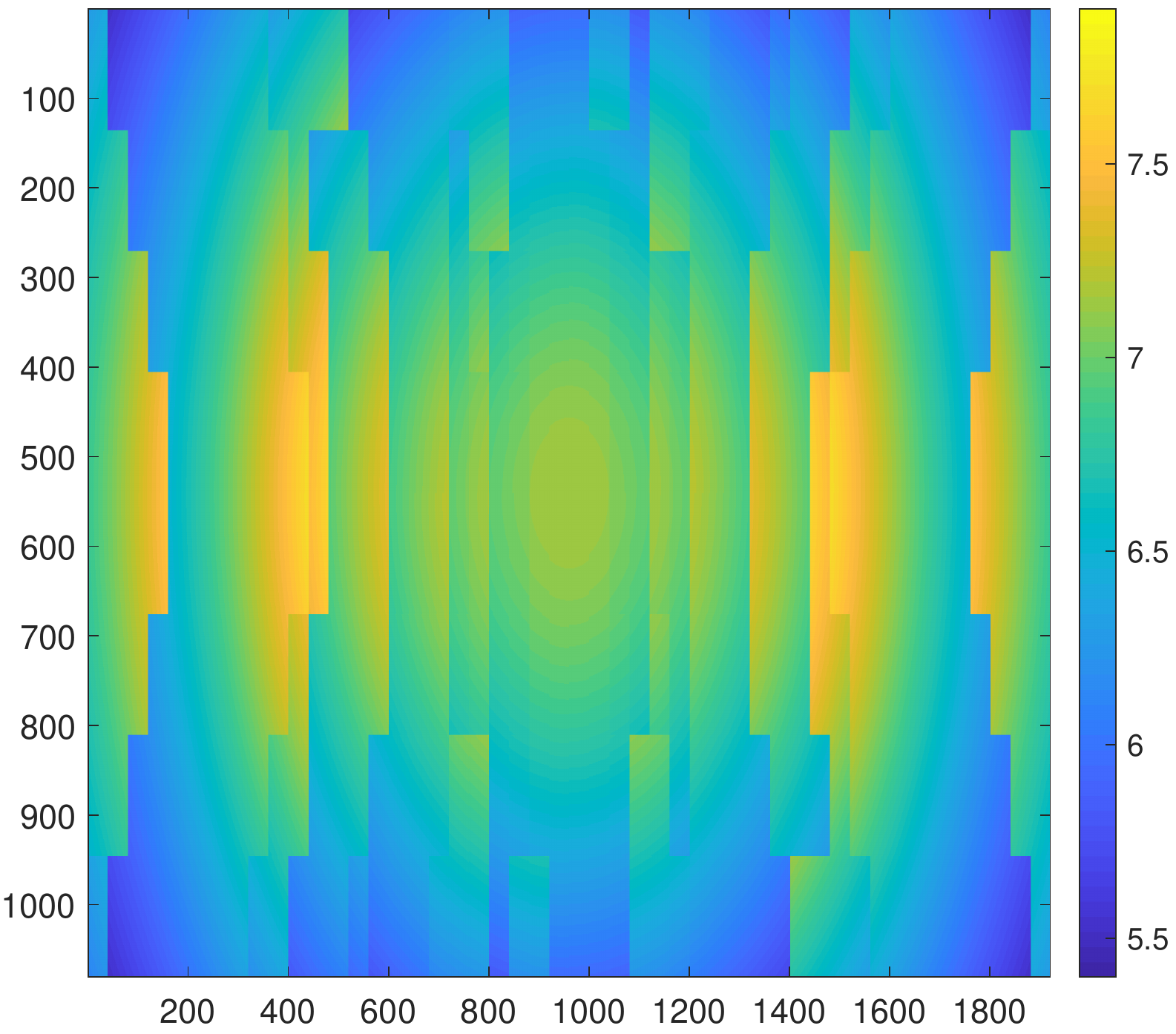}
		\caption{12x2 \linebreak Depth RMSE $=0.505$m \linebreak Depth MAE $=0.392$m}
	\end{subfigure}
	\begin{subfigure}[t]{.24\textwidth}
		\captionsetup{justification=centering}
		\includegraphics[width=0.99\columnwidth]{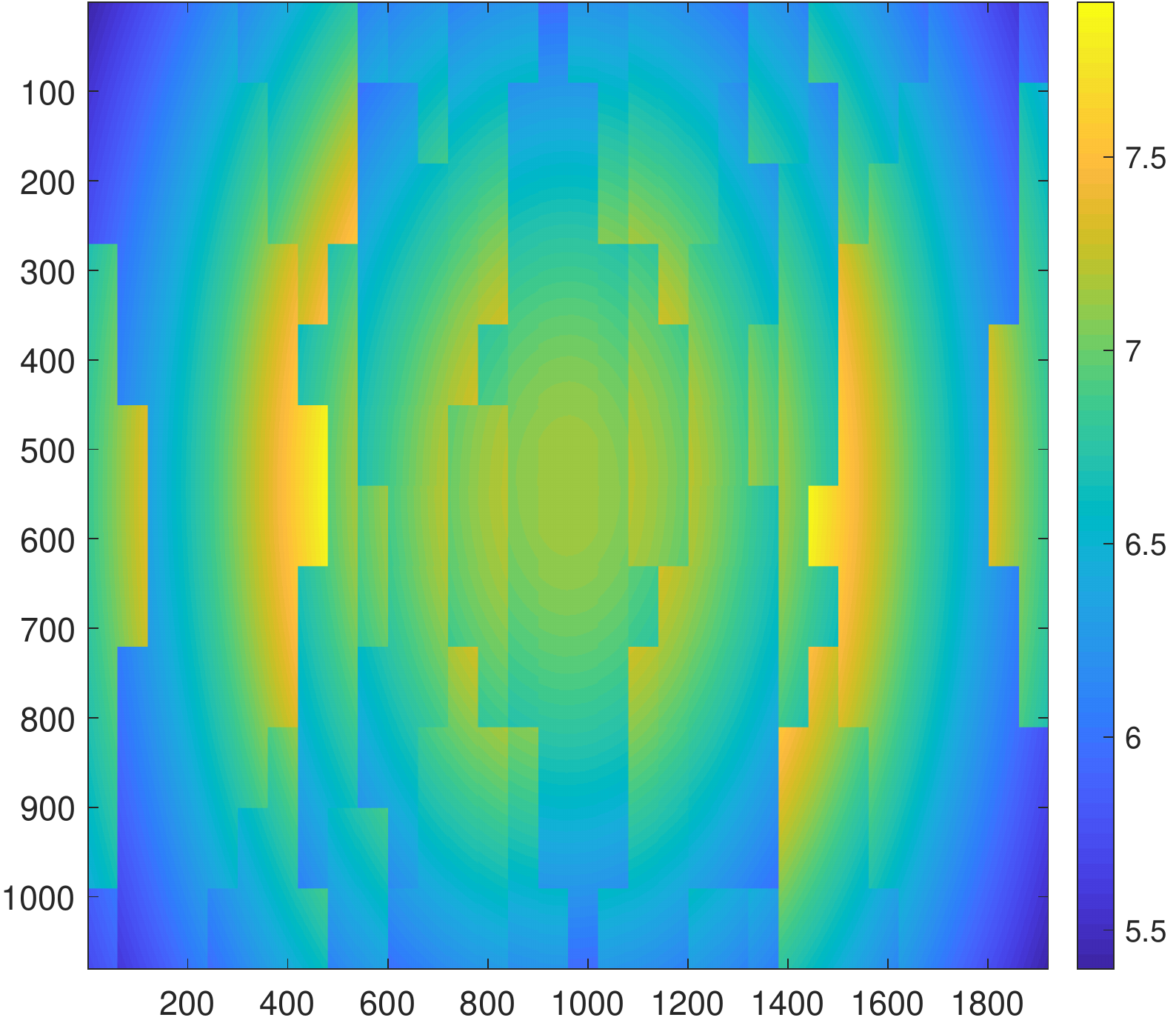}
		\caption{8x3 \linebreak Depth RMSE $=0.512$m \linebreak Depth MAE $=0.397$m}
	\end{subfigure}
	\begin{subfigure} [t]{.24\textwidth}
		\captionsetup{justification=centering}
		\includegraphics [width=0.99\columnwidth]{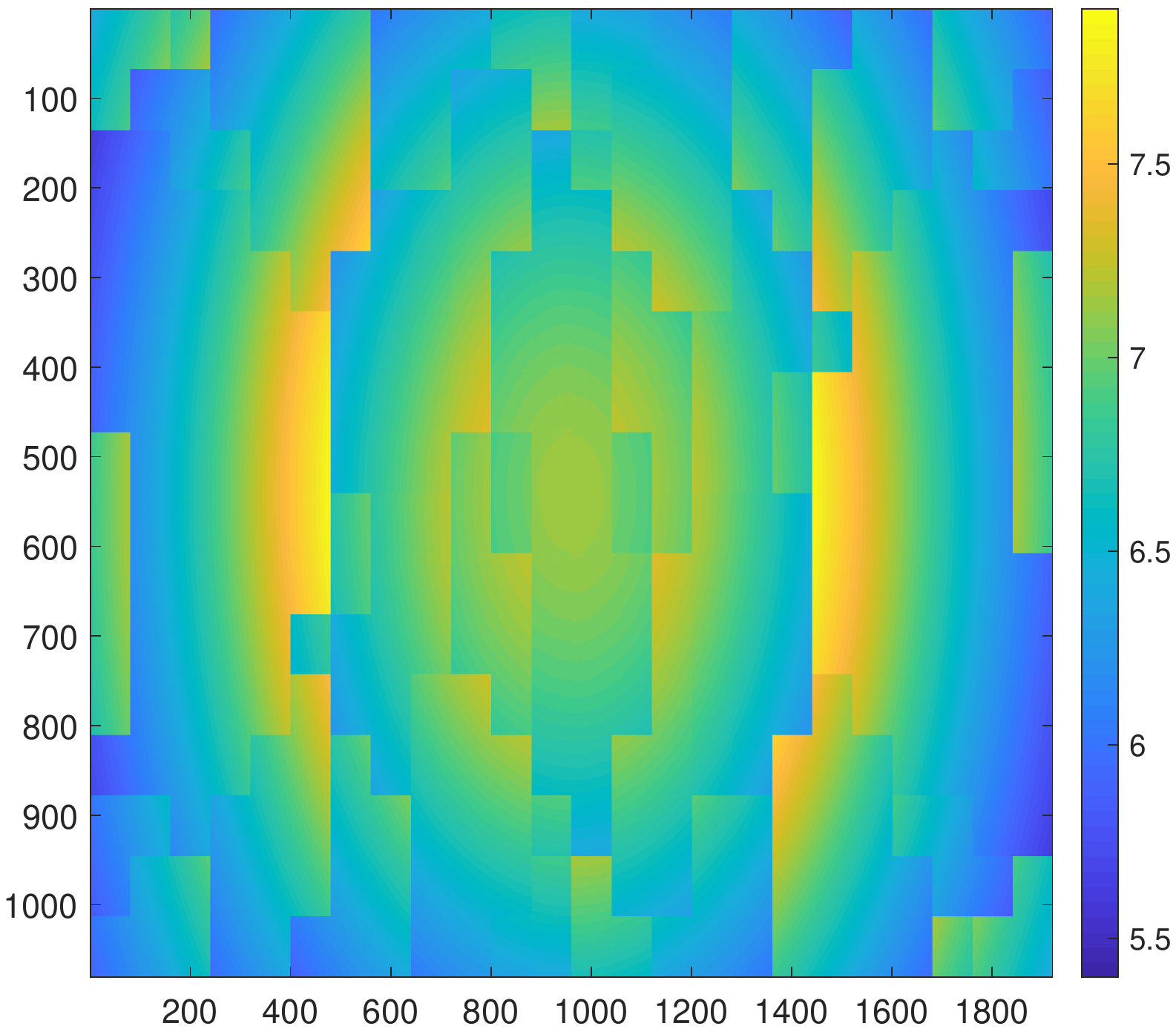}
		\caption{6x4 \linebreak Depth RMSE $=0.473$m \linebreak Depth MAE $=0.372$m}
	\end{subfigure}
	\begin{subfigure} [t]{.24\textwidth}
		\captionsetup{justification=centering}
		\includegraphics [width=0.99\columnwidth]{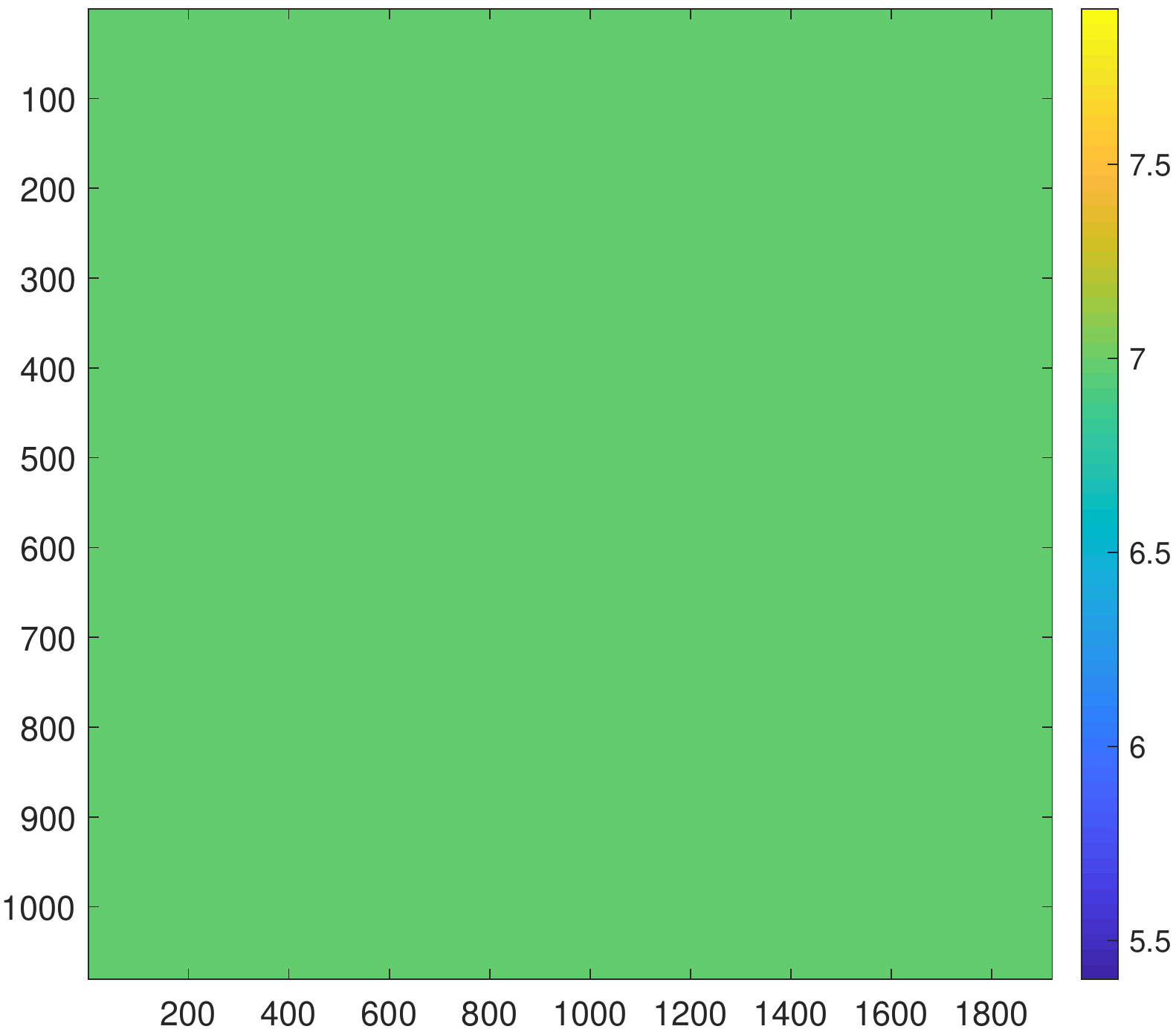}
		\caption{Ground truth}
	\end{subfigure}
	\caption{The $1080$p depth maps for the one wall scenario are depicted at different antenna configurations, for a separation distance of $7$ meters. The same number of antenna elements is used (24 elements) and codebook oversampling factors of four are employed. Figures (a), (b), and (c) illustrate the estimated maps for $12\times 2$, $8\times 3$, and $6 \times 4$ UPAs. Figures (d) illustrate the ground truth depth map.} 
	\label{fig:Onewall_maps_practicalUPAsize}
\end{figure}

\begin{figure}[t] 
	\centering
	\begin{subfigure} [t]{.29\textwidth}
		\captionsetup{justification=centering}
		\includegraphics [width=0.99\columnwidth]{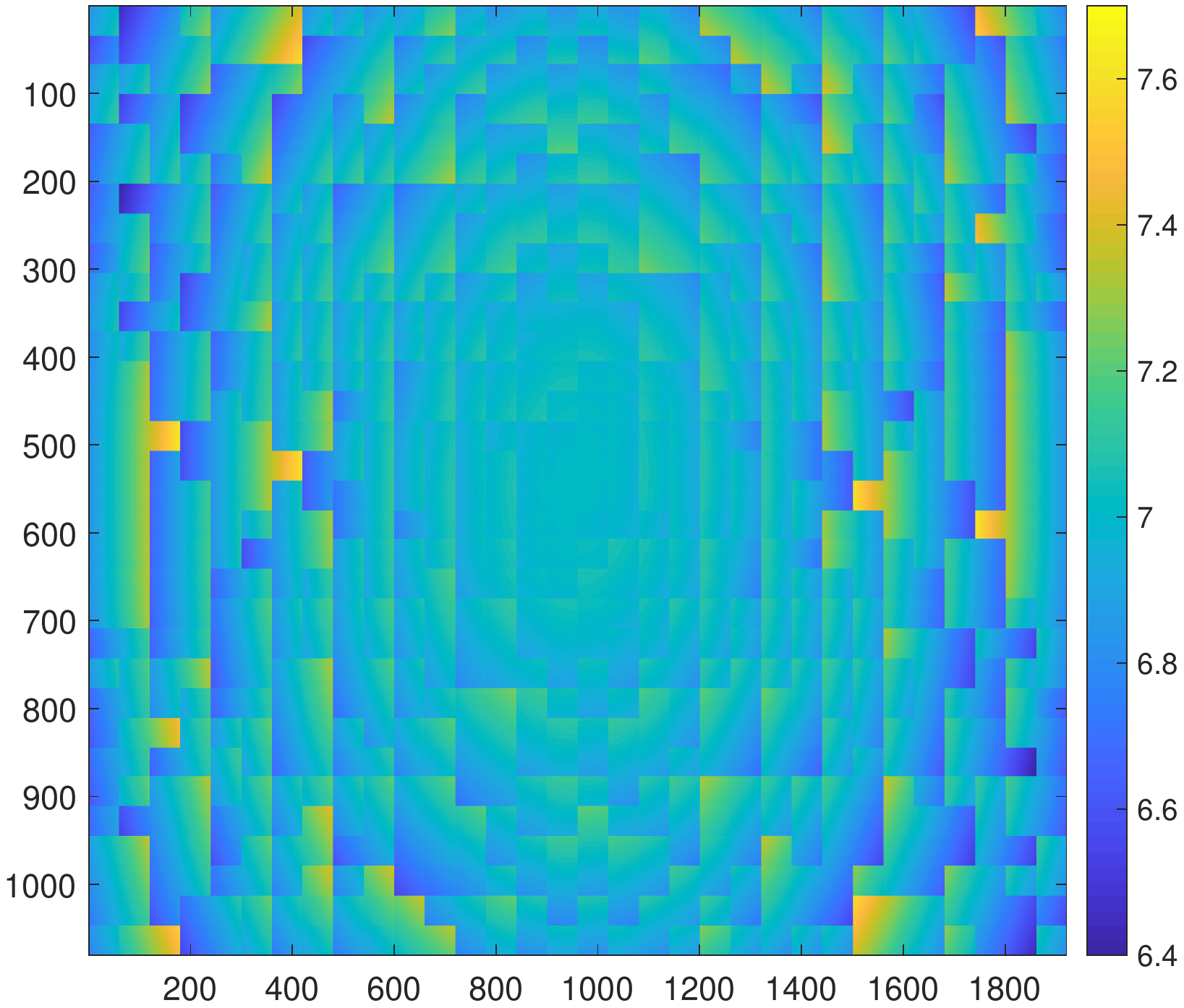}
		\caption{Continuous phase shifts \linebreak Depth RMSE $=0.148$m \linebreak Depth MAE $=0.1106$m}
	\end{subfigure}
	\begin{subfigure}[t]{.29\textwidth}
		\captionsetup{justification=centering}
		\includegraphics[width=0.99\columnwidth]{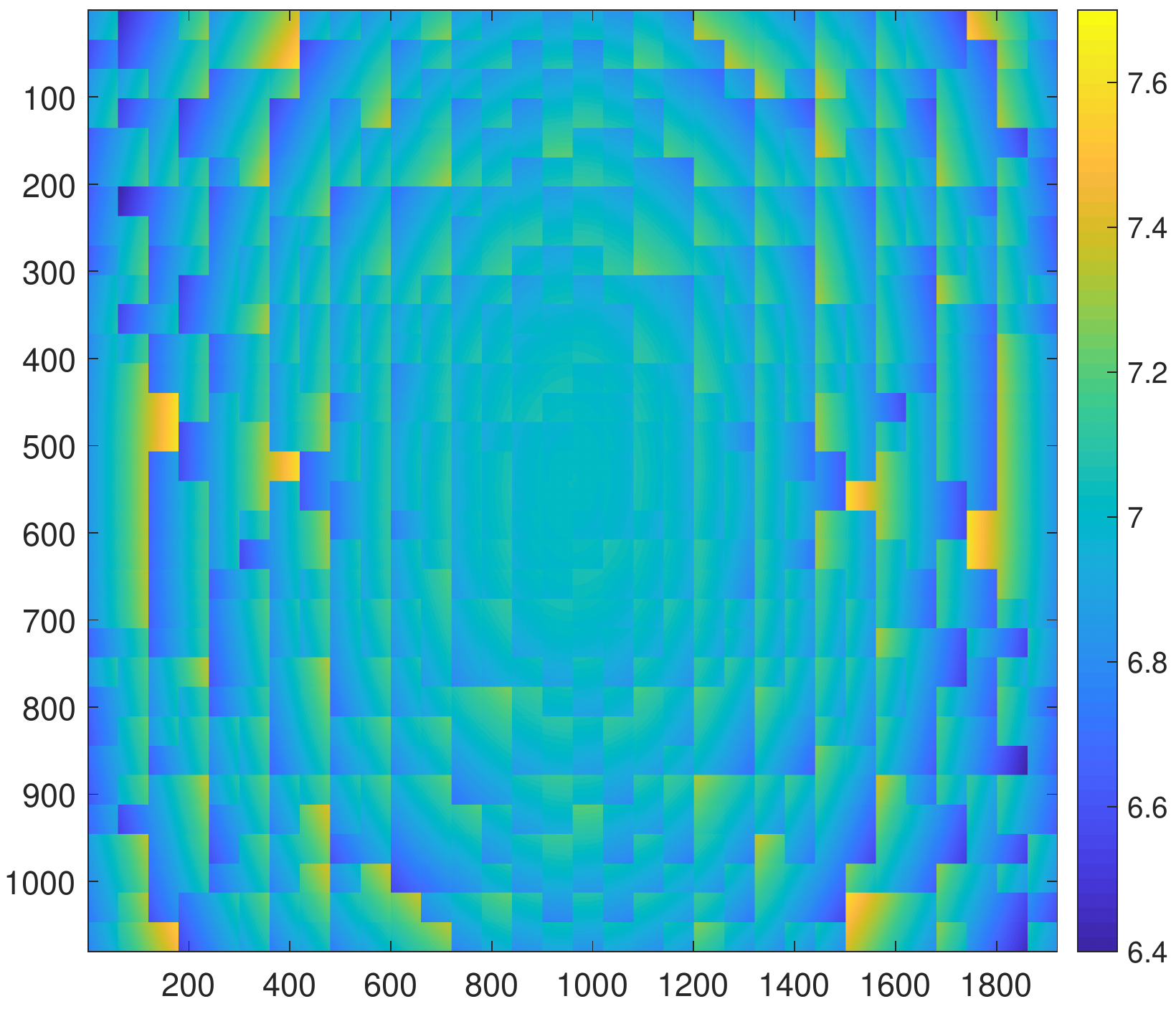}
		\caption{$2$-bit discrete phase shifts \linebreak Depth RMSE $=0.149$m \linebreak Depth MAE $=0.1111$m}
	\end{subfigure}
	\begin{subfigure} [t]{.29\textwidth}
		\captionsetup{justification=centering}
		\includegraphics [width=0.99\columnwidth]{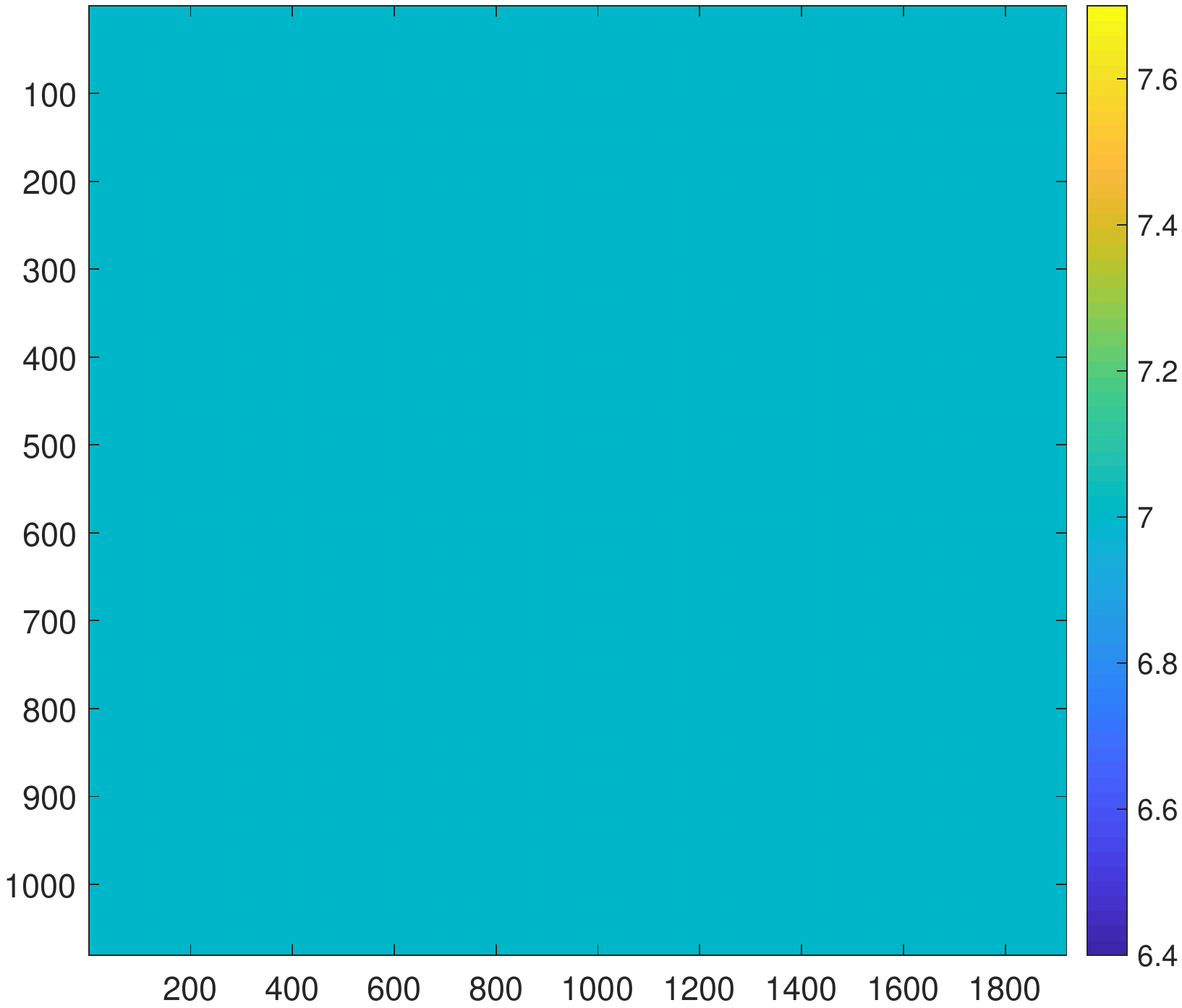}
		\caption{Ground truth}
	\end{subfigure}
	\caption{The $1080$p depth maps for the one wall scenario at $7$m separation distance are estimated for two cases of the RF phase shifters at the AR/VR device: (a) continuous phase shifts and (b) $2$-bit quantized phase shifts. $16\times16$ UPA is employed with codebook oversampling factors of two. Figures (c) illustrate the ground truth depth map.} 
	\label{fig:Onewall_maps_practicalPS}
\end{figure}

\begin{figure}[t] 
	\centering
	\begin{subfigure} [t]{.49\textwidth}
		\includegraphics [width=0.99\columnwidth]{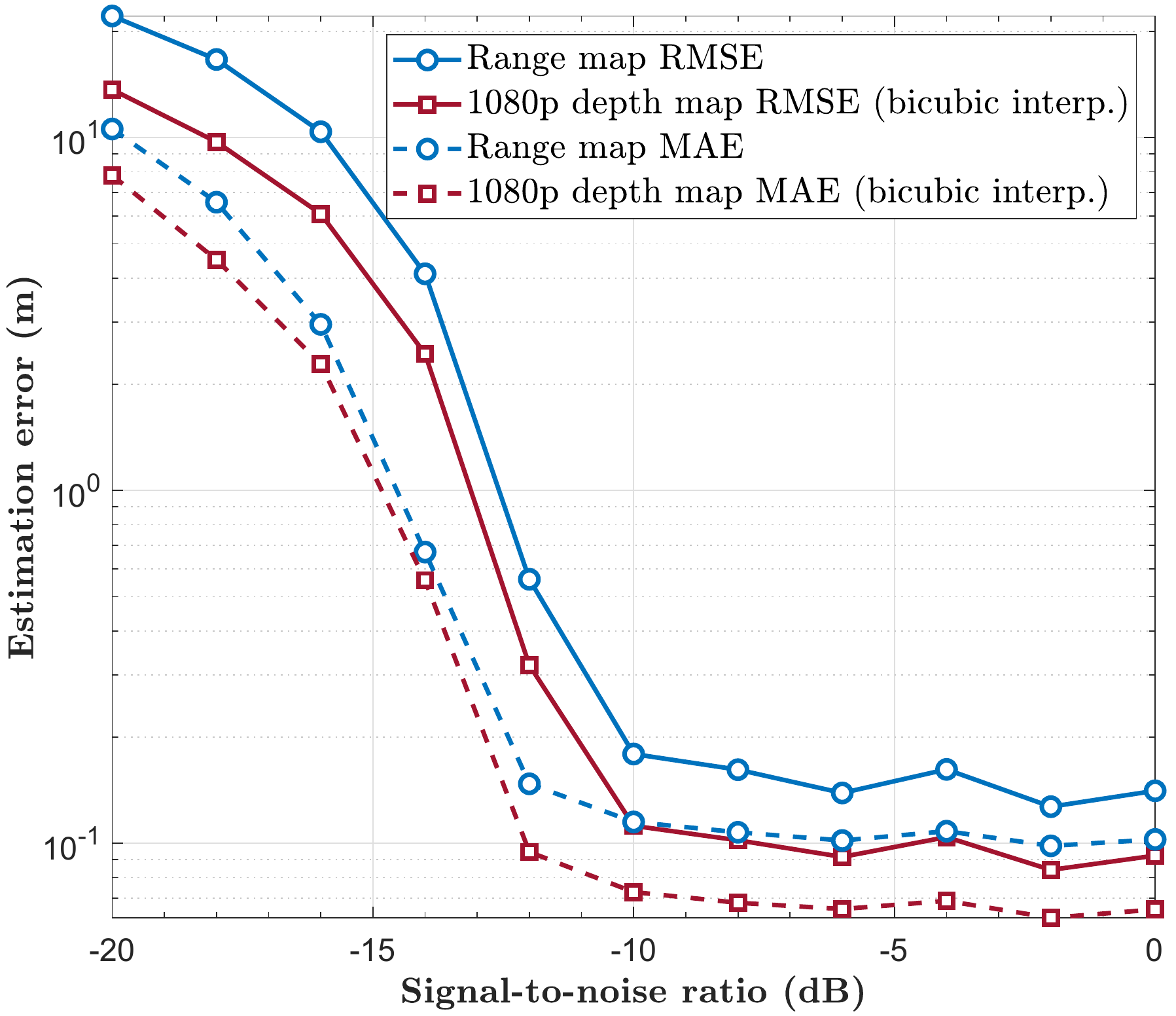}
		\caption{}
		\label{fig:Onewall_txpower_error}
	\end{subfigure}
	\begin{subfigure}[t]{.49\textwidth}
		\includegraphics[width=1\columnwidth]{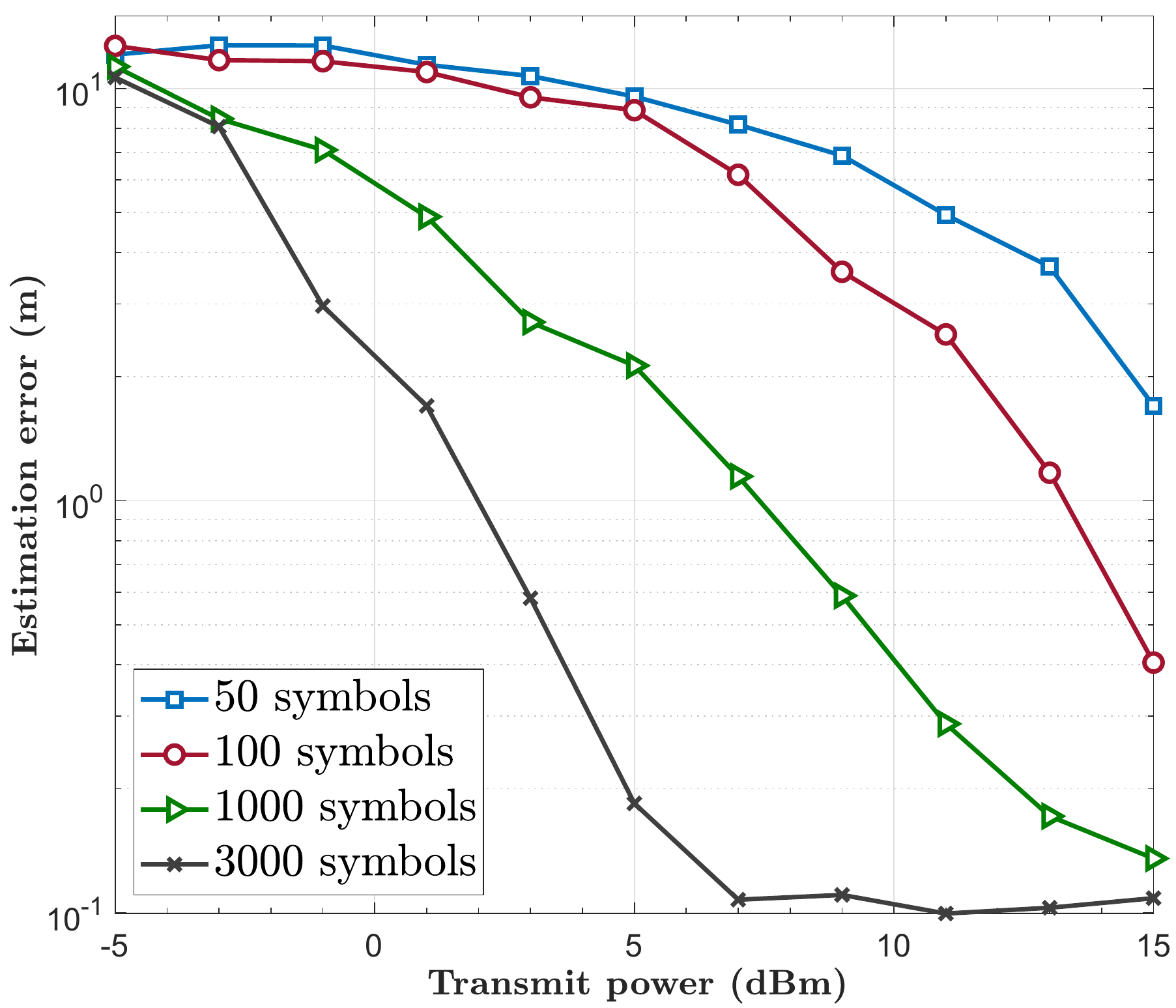}
		\caption{}
		\label{fig:Onewall_preamble_bc_mc}
	\end{subfigure}
	\caption{For the one wall scenario, the error performance of the proposed mmWave MIMO based depth estimation solution is evaluated under different error metrics in (a) and is evaluated for different preamble sequence lengths in (b). The wall is $7$ meters away from the AR/VR device with $16 \times 16$ UPAs. The figures show the robustness of the developed approach  under relatively low SNR regime. Note that the displayed transmit power range in (b) corresponds to average SNR range of  $-20.7$dB to $-0.7$dB.} 
	\label{fig:Dummy_name}
\end{figure}

\begin{figure}[t] \centerline{\includegraphics[scale=0.5]{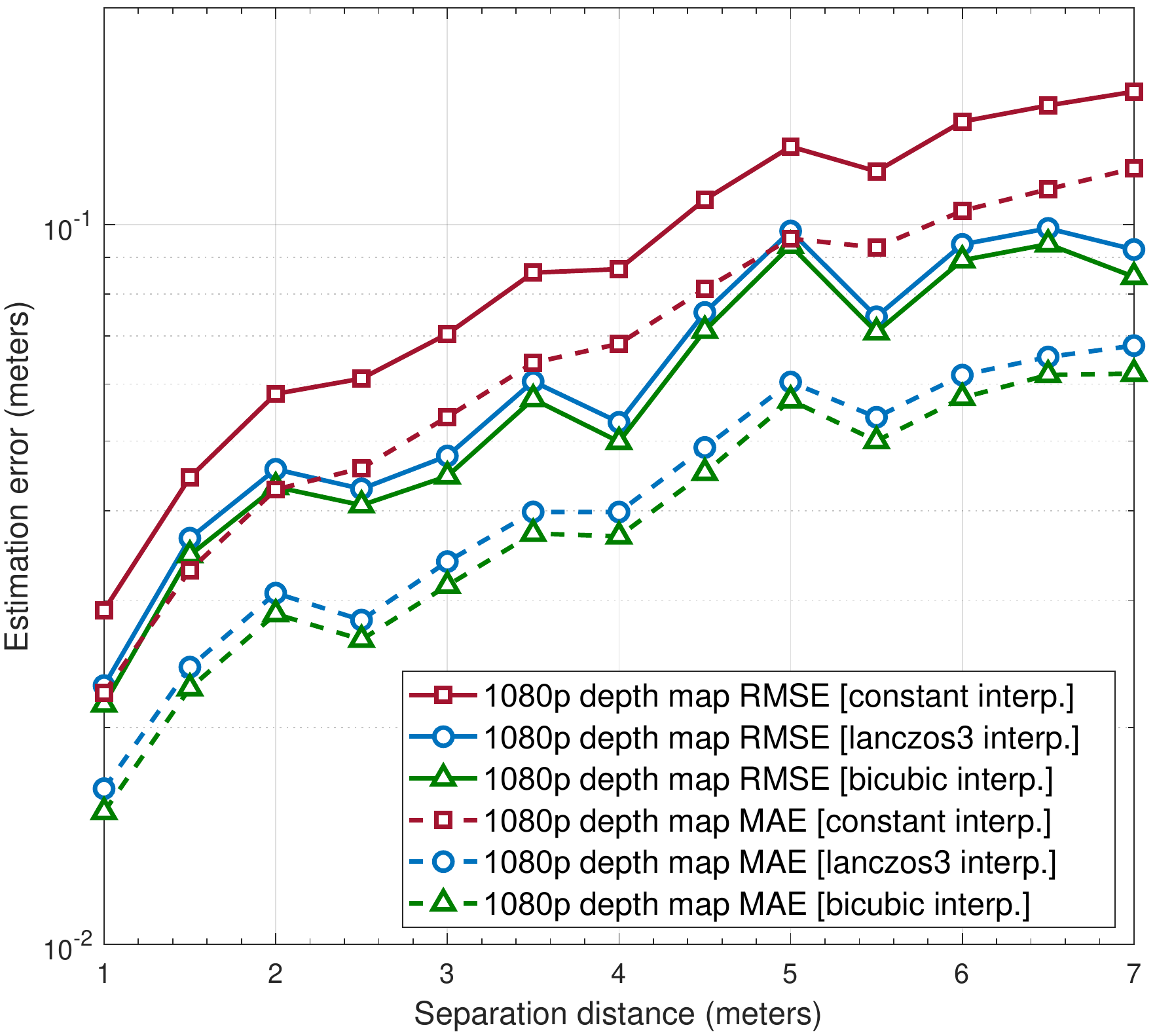}}
	\caption{The error performance of the proposed mmWave MIMO based depth estimation solution is evaluated across different separation distances for the one wall scenario. The estimation error starts from $\approx1.5$m at a $1$m distance and reaches around $10$cm at a $7$ meters distance.} 
	\label{fig:Onewall_separation_dist}
\end{figure}

\subsection{One wall scenario}  \label{subsec:one_wall}
The one wall scenario consists of an AR/VR transceiver facing a wall in free space propagation. Unless otherwise mentioned, the separation distance between the wall and the transceiver is $7$ meters and the wall building material is concrete. In \figref{fig:Onewall_maps}, we show the estimated range and depth maps for the one wall scenario compared to the ground truth maps. \figref{fig:Onewall_maps}(a) and \figref{fig:Onewall_maps}(b) show that the range map estimation error has an average MAE of  $0.098$m and RMSE of $0.127$m. Further,  the depth map estimation error \figref{fig:Onewall_maps}(c) and \figref{fig:Onewall_maps}(d) has an average  MAE of $0.12$m and RMSE of $0.153$m.  Overall, these figures show that the proposed approaches can accurately estimate the range/depth maps for a wall at 7m distance from the AR/VR device with around 10cm error, which highlights the effectiveness of this approach. 

\noindent \textbf{Impact of the important system parameters:} Next, we briefly evaluate the impact of the various system parameters on the performance of the proposed mmWave depth map estimation solution. 
\begin{itemize}
	\item \textbf{Number of antennas and sensing codebook beams:}  In \figref{fig:Onewall_maps_var_antennas}, we plot the estimated and ground-truth depth maps for different number of antennas and codebook oversampling factors. As illustrated, the depth estimation accuracy can generally improve by increasing the number of antennas and/or the codebook oversampling factors. This comes with the cost of deploying more antennas at the AR/VR device or employing more beams, which translates to a longer sensing time. In \figref{fig:Onewall_maps_practicalUPAsize}, we plot the estimated and ground-truth depth maps for different antenna configurations using the same number of antenna elements. As depicted, the depth estimation accuracy depends on the UPA configuration, with the best configuration being the $6\times4$ UPA because of its closeness to the $1080$p aspect ratio.
	
	\item \textbf{RF phase shift quantization:} As previously described in \sref{ssec:Sidelobe}, the phase quantization of the RF phase shifters in the AR/VR transceiver architecture produces a noticeable change in the radiation pattern shape of the sidelobes. To examine the effect of this phase quantization on the estimated depth maps, \figref{fig:Onewall_maps_practicalPS} shows the comparison of the estimated depth maps for two cases of the RF phase shifters at the AR/VR device: (a) continuous phase shift and (b) $2$-bit quantized phase shifts. As depicted, the phase quantization contributes with a small negative impact on the depth map estimation accuracy for the one wall scenario at a separation distance of $7$ meters.
	
	\item \textbf{Transmit sensing power:} In \figref{fig:Onewall_txpower_error}, we investigate the effect of changing the transmit power on the depth map estimation accuracy. The SNR value of $0$dB corresponds to a transmit power of $15$dBm. This figure shows that a transmit power  of $5$dBm (SNR of $-10$dB) could be sufficient to reach around 10cm error for the depth estimation accuracy. 
	
	\item \textbf{Preamble sequence length}: The estimation error versus  transmit power is depicted in \figref{fig:Onewall_preamble_bc_mc} for different values of preamble sequence lengths, namely preambles with 50, 100, 1000, and 3000 symbols. As shown in this figure, increasing the preamble sequence length improves the depth estimation accuracy at the expense of increased sensing time and post-processing complexity.

	\item \textbf{Separation distance between the AR/VR device and the facing wall:} \figref{fig:Onewall_separation_dist} investigates the impact of increasing the depth value on the depth estimation accuracy. 
	As shown in this figure, the larger the distance between the AR/VR device and the facing surface,  the larger the error in the depth estimate, which is expected. This figure also highlights some advantage for the  bicubic interpolation compared to the other interpolation methods. 
	
	\item  \textbf{The surface material:} Now, we evaluate the performance of the proposed approach for different surface materials. More specifically,  we summarize in Table \ref{table:Onewall_material_test} the range map MAE for different candidates of the wall material. Overall, we can notice some correlation between the estimation accuracy and the  \textit{scattered to incident power ratio} property of the materials, which are summarized in Table \ref{table:materials}.
\end{itemize}

\begin{table}[H]
	\caption{The estimation error results of the one wall scenario for different wall materials}
	\begin{center}
		\begin{tabular}{ | c | c | c | c | c | c | c | }
			\hline
			\textbf{Estimation Error (meters)} & \textbf{Concrete}  & \textbf{Ceilingboard} & \textbf{Wood} & \textbf{Floorboard}  & \textbf{Layered Drywall} & \textbf{Glass}  \\ \hline \hline
			Basic Correlator & $0.101$ & $0.099$ & $0.101$ & $0.0984$ & $0.103$ & $12.697$  \\ \hline
			Massive Correlator   & $0.0983$ & $0.097$ & $0.0983$ & $0.0983$ & $0.101$ & $15.498$ \\ \hline
		\end{tabular}
	\end{center}
	\label{table:Onewall_material_test}
\end{table}

\subsection{Two walls scenario}  \label{subsec:two_walls}

\begin{figure}[t] 
	\centering
	\begin{subfigure} [t]{.49\textwidth}
		\includegraphics [width=0.99\columnwidth]{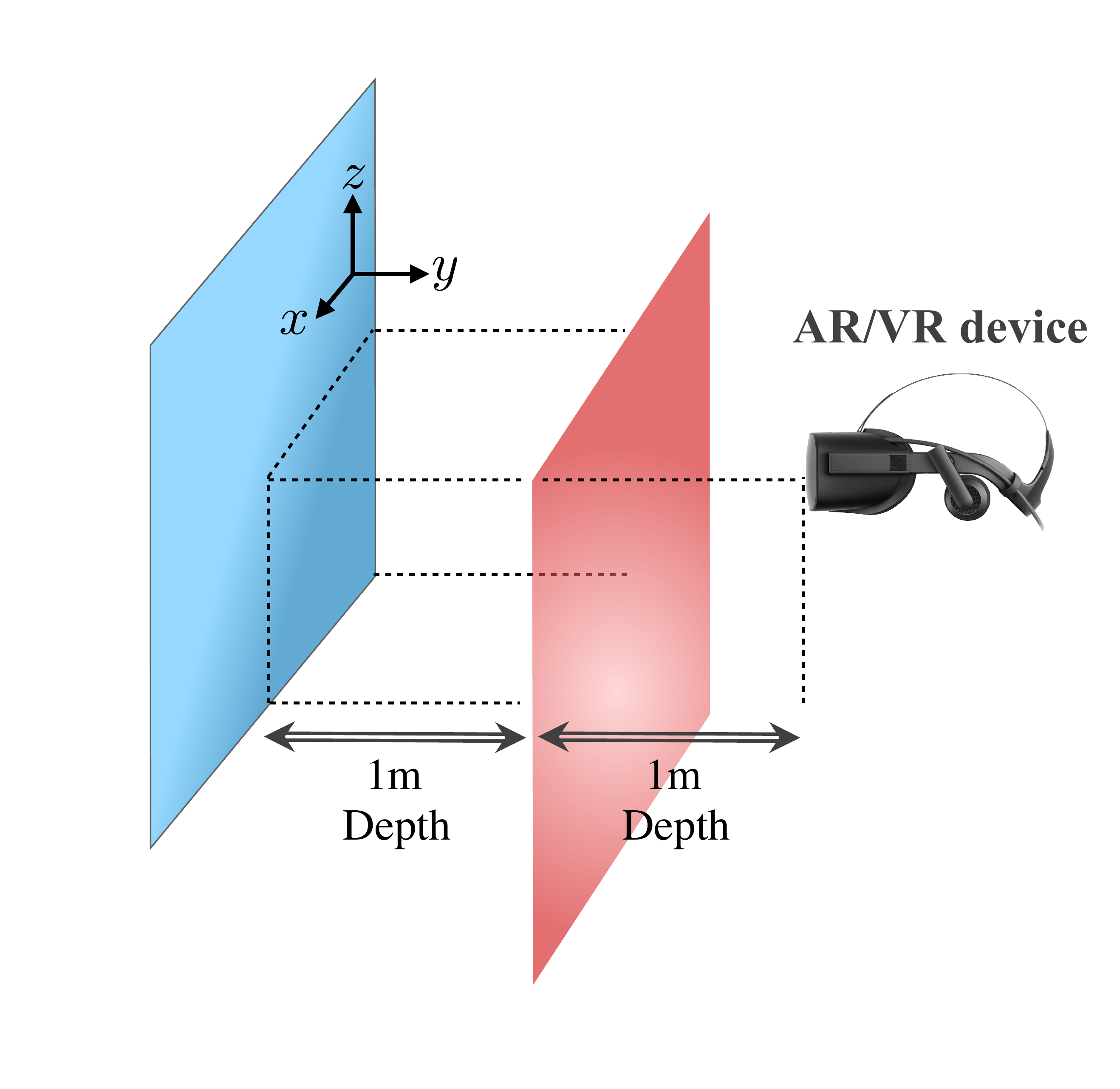}
		\caption{}
			\label{fig:Twowall_bird}
	\end{subfigure}
	\begin{subfigure}[t]{.49\textwidth}
		\includegraphics[width=1\columnwidth]{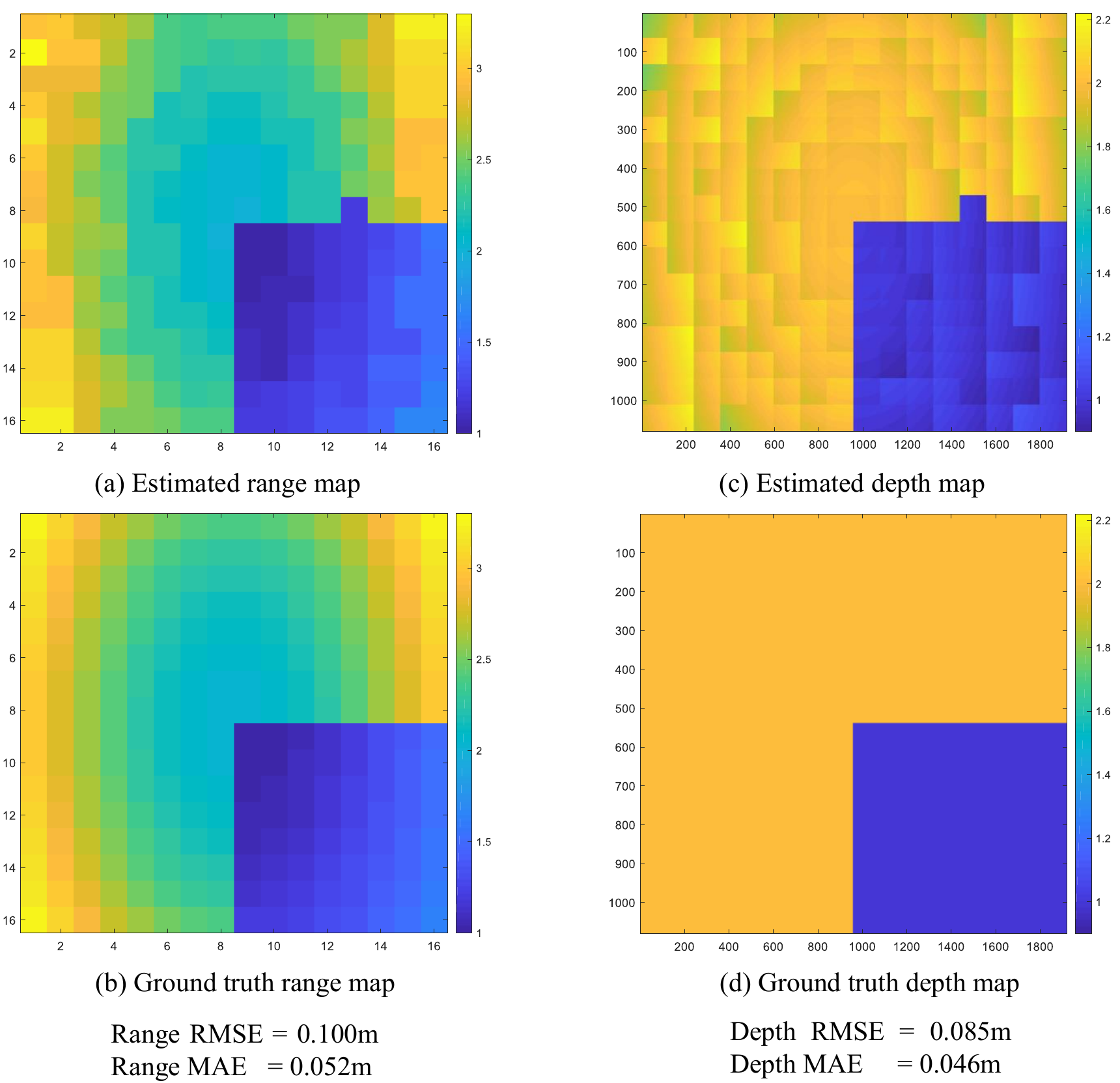}
		\caption{}
		\label{fig:Twowall_maps}
	\end{subfigure}
	\caption{(a) The adopted two walls scenario is illustrated. (b) The maps for the two walls scenario are depicted. The AR/VR device is employed with $16\times 16$ UPAs.  The depicted maps are the estimated maps (at the top), ground truth maps (at the bottom), range maps (on the left side), and $1080\rm{p}$ depth maps (on the right side). Comparing (a) with (b), the range map estimation error: MAE $= 0.052$m. Comparing (c) with (d), the depth map estimation error: MAE $= 0.046$m.} 
	\label{fig:Dummy_name_2}
\end{figure}

The two walls scenario consists of one AR/VR device facing two walls in free space propagation as depicted in \figref{fig:Twowall_bird}. The separation distance between the front wall and the AR/VR device is $1$m while the separation between the back wall  and the AR/VR device is $2$m. The walls' building material is concrete. Each wall consists of $2,048$ faceted faces, and each face contributing with at most one backscattered ray. The purpose behind studying this scenario is to test the alignment of the estimated map compared to the ground truth depth map. The results of this test are illustrated in \figref{fig:Twowall_maps}, where the estimated range and depth maps are compared to the ground truth maps. As shown in \figref{fig:Twowall_maps}, the two edges of the front wall in the estimated maps align reasonably well with the one displayed in the ground truth maps. This highlights the promising performance of proposed mmWave based depth estimation solution. 

\begin{figure}[t] 
	\centering
	\begin{subfigure} [t]{.49\textwidth}
		\includegraphics [width=0.85\columnwidth]{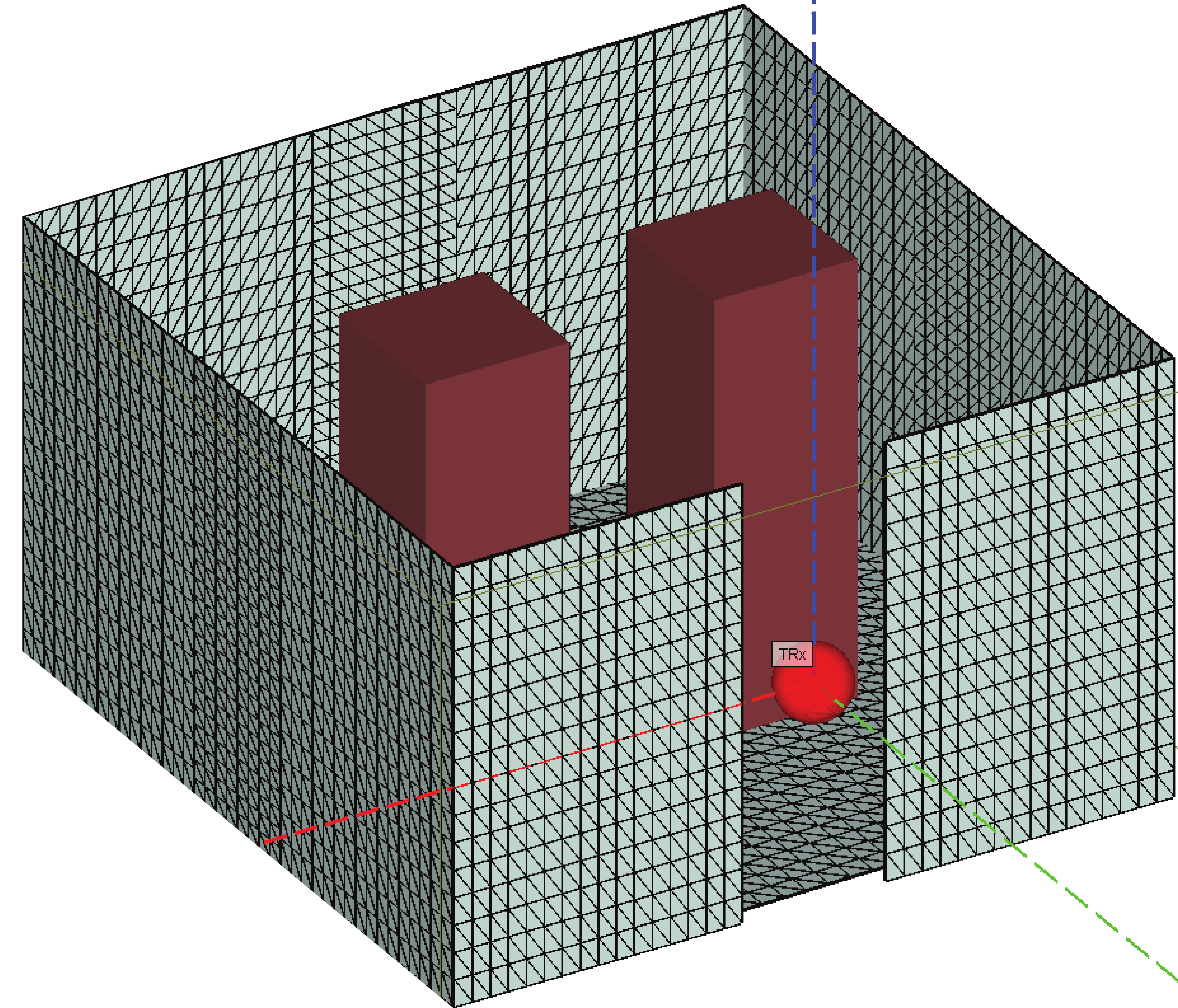}
		\caption{The room with two pillars}
	\end{subfigure}
	\begin{subfigure} [t]{.45\textwidth}
		\includegraphics [width=0.99\columnwidth,height=170pt]{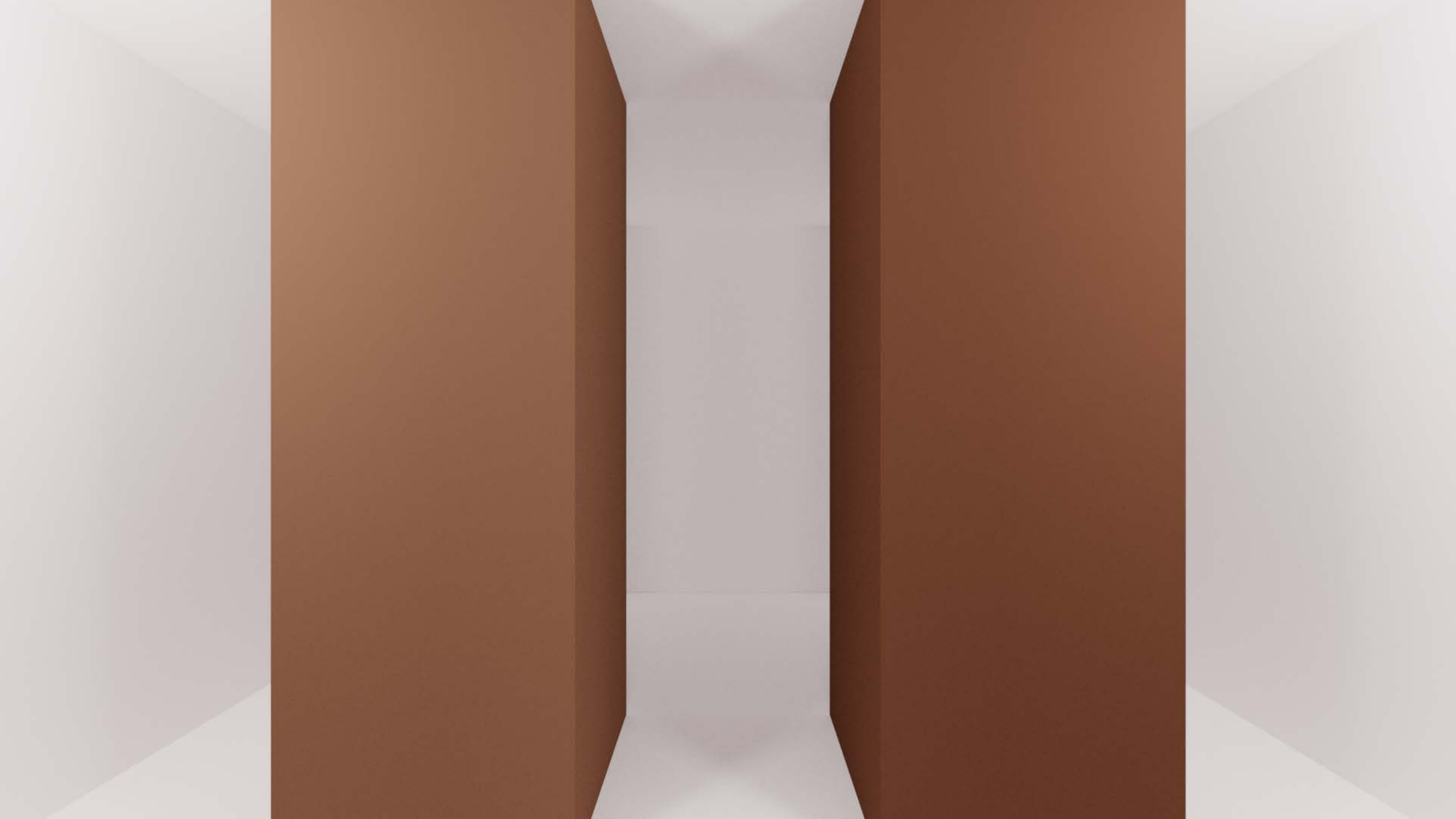}
		\caption{The room scene from the door position}
	\end{subfigure}
	\caption{Figure (a) illustrates the bird view of the room with two pillars. Figure (b) shows the scene from the AR/VR device position, centered at the front door. The $5$m$\times5$m room consists of a concrete floor plan with two wood pillars in the middle of the room. The wood pillars are at 2 meters distance from the AR/VR device.} 
	\label{fig:Indoor_room}
\end{figure}

\begin{figure}[t] \centerline{\includegraphics[scale=0.7]{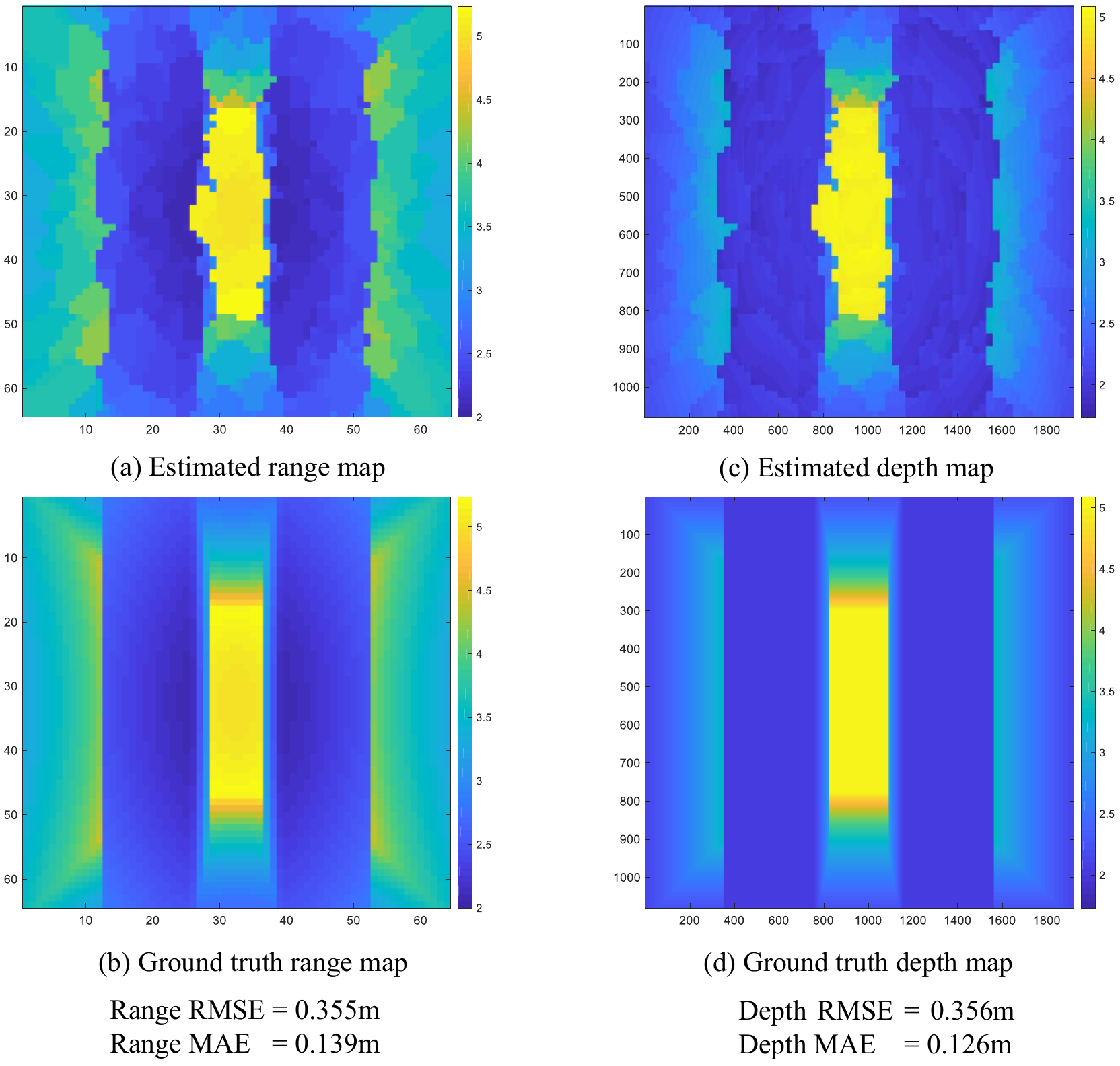}}
	\caption{The maps for the room with two pillars are depicted. $16\times 16$ UPAs are employed with codebook oversampling factors of four. The depicted maps are the estimated maps (at the top), ground truth maps (at the bottom), range maps (on the left side), and $1080\rm{p}$ depth maps (on the right side). Comparing (a) with (b), the range map estimation error: MAE $= 0.139$m. Comparing (c) with (d), the depth map estimation error: MAE $= 0.126$m.}
	\label{fig:Indoor_maps}
\end{figure}

\begin{figure}[t] \centerline{\includegraphics[scale=0.55]{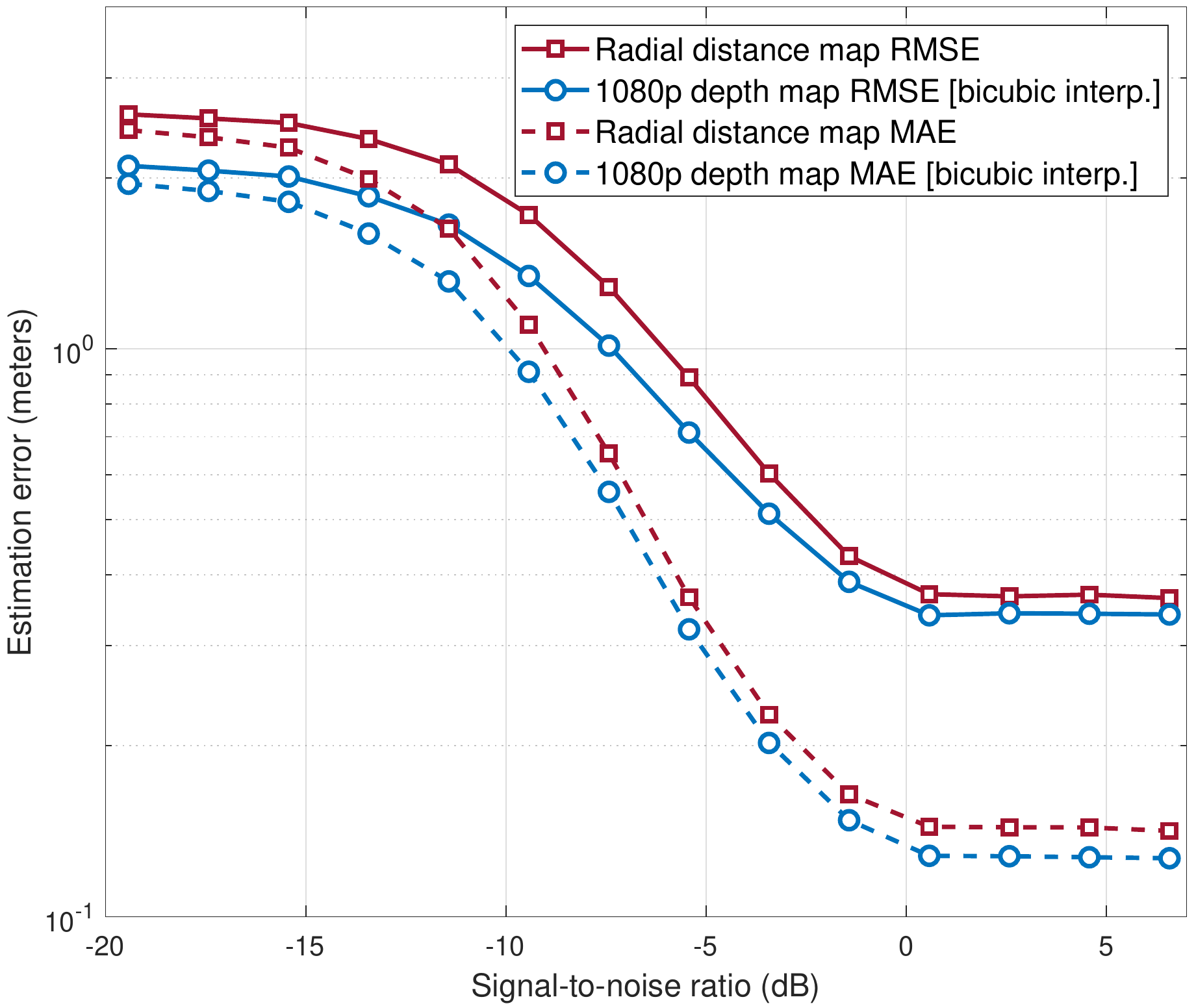}}
	\caption{For the room with two pillars, the error performance of the proposed mmWave MIMO based depth estimation is evaluated for different error metrics.  $16\times 16$ UPAs are employed with a codebook oversampling factors of four in both dimensions. This figure shows the robustness of the proposed mmWave MIMO based depth estimation under relatively low SNR regime.} 
	\label{fig:Indoor_txpower_error}
\end{figure}

\subsection{A room with two pillars}  \label{subsec:indoor_room}
In this scenario, we consider a $5$m$\times5$m room where one AR/VR device is centered at the front door of the room, as depicted in \figref{fig:Indoor_room}. The room consists of a concrete floor plan with two wood pillars in the middle of the room. The wood pillars are at 2 meters distance from the AR/VR transceiver. The floor plan consists of $15,488$ faceted faces whereas each of the wood pillars consists of $3,072$ faceted faces. Note that the ceiling of the floor plan is set to the invisible mode for visibility purposes only. 
For the estimation error assessment of the indoor space scenario, \figref{fig:Indoor_maps} shows the comparison between estimated and ground truth maps for $16 \times 16$ UPA antennas with a codebook oversampling factors of four in both azimuth and elevation dimensions. 
First, \figref{fig:Indoor_maps}(a) with \figref{fig:Indoor_maps}(b) show the estimate and ground truth range maps, which have a MAE of $0.139$m and RMSE of $0.355$m. For the depth maps, \figref{fig:Indoor_maps}(c) with \figref{fig:Indoor_maps}(d) represent  $1080\rm{p}$  maps with estimation error of(i) $0.126$m for the MAE and $0.356$m for the RMSE with nearest neighbor interpolation, and (ii) $0.123$m for the MAE and $0.328$m for the  RMSE with bicubic interpolation. From observing the difference in maps, the mmWave reasonably recover most of the depth information of the scene with low codebook resolution $(16 \times 16)$ compared to the ground truth $1080\rm{p}$ resolution. With narrower transmit and receive beams, i.e. more antenna elements, the estimation accuracy is expected to further improve.  

The depth map estimation accuracy for this scenario is also evaluated at different SNRs in \figref{fig:Indoor_txpower_error}. In this figure, we adopt the model and system parameters used in \figref{fig:Indoor_maps} with $16\times16$ UPAs and oversampling factors of four. It is also worth mentioning that $0$dB SNR corresponds to $-20$dBm transmit power in our setup. As shown in \figref{fig:Indoor_txpower_error}, the estimated depth maps have MAE of almost 10cm at $0$dB, which highlights the promising performance of our proposed depth map estimation approach at relatively low SNRs and in indoor room with several surfaces and different materials. This will be further emphasized in the following subsection. 

\begin{figure}[t] \centerline{\includegraphics[scale=0.75]{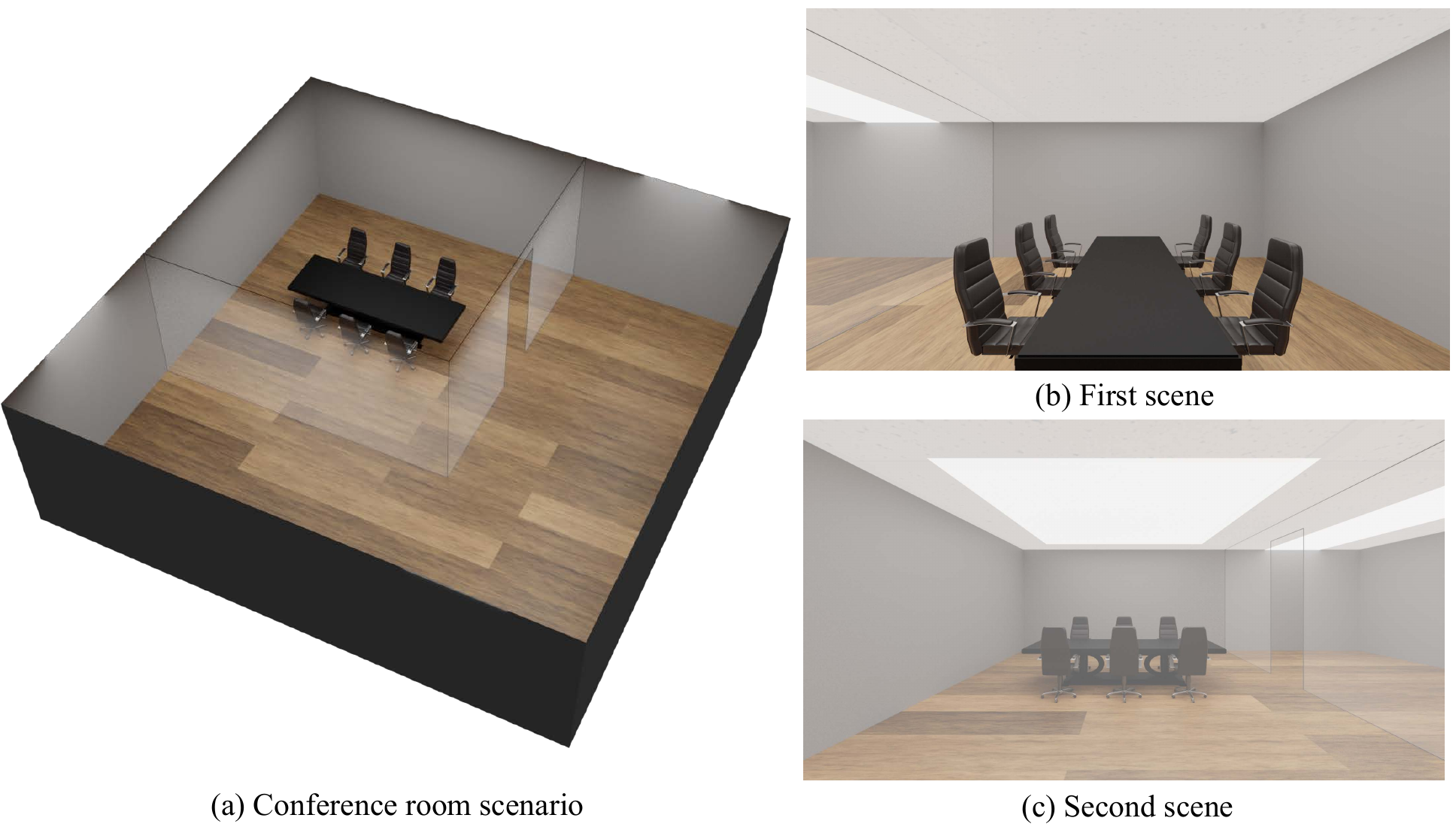}}
	\caption{(a) the bird view of the conference room scenario; (b) and (c) the scenes under study. The $10$m$\times10$m indoor space contain a $6$m$\times6$m conference room in glass. the indoor space walls are made from layered drywall, the ceiling is made from ceiling board and the floor is made from floorboard. The conference room chairs and tables are made from wood.} 
	\label{fig:conference_room}
\end{figure}

\begin{figure}[t] \centerline{\includegraphics[scale=0.7]{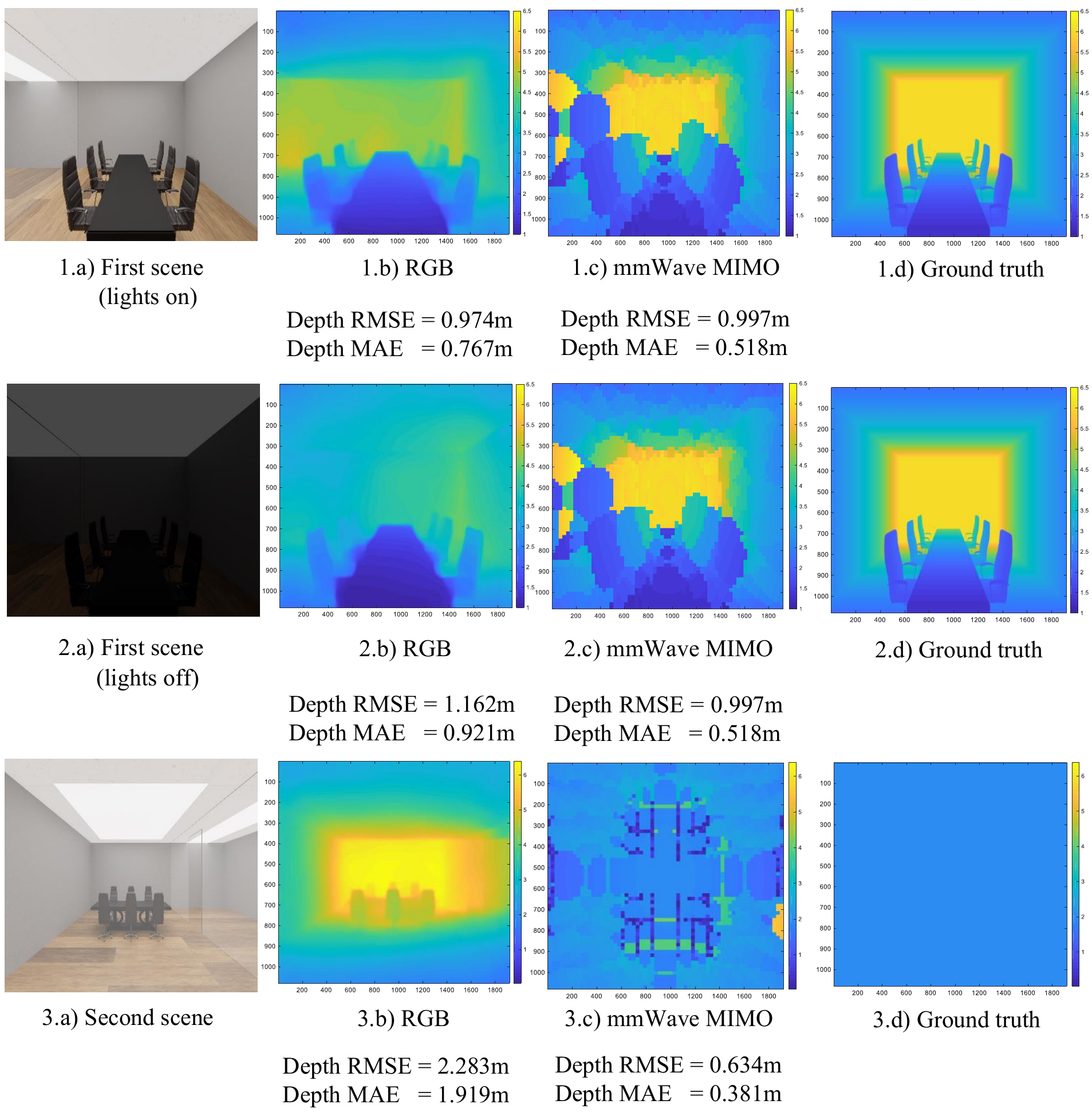}}
	\caption{For the conference room scenario, the proposed mmWave MIMO based depth estimation is compared with the RGB based depth estimation in \cite{Hu2019}. $16\times 16$ UPAs are employed with codebook oversampling factors of four. The depicted maps are the maps of the first scene with lights on/off (The top two rows) and the second scene (the bottom row). (a) The scenes under study; (b) the estimated maps from monocular RGB images; (c) the estimated  maps from our proposed solution; (d) the ground truth depth maps.}
	\label{fig:conference_maps}
\end{figure}

\subsection{Conference room scenario}  \label{subsec:conference_room}
In this scenario, we consider the conference room  shown in \figref{fig:conference_room}. The ceiling of the indoor space is set to the invisible mode for visibility purpose only. The $10$m$\times10$m indoor space has a $6$m$\times6$m conference room with glass walls. The indoor space walls are made of layered drywall, the ceiling is made of ceiling board, and the floor is made of floorboard. The conference room chairs and tables are made of wood. The conference room door opening is $1$m in width and $2.7$m in height. The number of facets for each item in the indoor space is as follows: $2,048$ facets for the layered drywall, $2,048$ facets for the floorboard, and $2,048$ facets for the ceiling board. In addition, the number of facets for each item in the conference room is as follows: $1,568$ facets for the glass wall, $4,446$ facets for the table, $21,192$ facets for the office chairs. The conference room scenario consists of two AR/VR devices  for two scenes under study --- the first device is centered at the front door of the conference room while the second transceiver is placed outside of the conference room facing the other glass facet. The scenes captured by the AR/VR camera for the two cases are shown in \figref{fig:conference_room}(b) and \figref{fig:conference_room}(c).

One main motivation for leveraging mmWave MIMO  to estimate the depth maps (compared to RGB based depth estimation approaches) is the expected higher efficiency in detecting transparent and dark objects. In \figref{fig:conference_maps}, we  compare our mmWave MIMO based depth estimation approach with the RGB based depth estimation approach, detailed in \cite{Hu2019}, for the two considered conference room scenarios. It's worth emphasizing here that the algorithms in \cite{Hu2019} achieve considerably good depth accuracy when tested on the NYU depth V2 dataset \cite{Silberman2012}. As shown in  \figref{fig:conference_maps}, the mmWave MIMO based estimator outperforms the RGB based estimator in recognizing transparent and dark objects. For the first scene, the glass wall was not detected by the RGB estimator. Also, in the presence of a scene with low illumination, the mmWave MIMO based estimator performance shows robustness in the estimation accuracy compared to the RGB based estimator. Figure 1.c) and 2.c) were generated with the aid of the SLR approach in \sref{ssec:Sidelobe}, with ${\delta_{\rm{H}}}=2,{\delta_{\rm{V}}}=3$. As for the second scene, the RGB based estimator is unable to detect the transparent glass compared to the mmWave MIMO based estimator. Interestingly, despite the fact that the glass scattering ratio is $0\%$ based on Table \ref{table:materials}, the conference room glass wall is partially recovered by the mmWave MIMO based estimator due to the boresight reflection path. This makes the wireless AR/VR experience safer by providing the ability to detect transparent surfaces.  All these promising results highlight the potential of leveraging the proposed mmWave MIMO based depth map estimation approaches for immersive AR/VR experience. 

\section{Conclusion}  \label{sec:Conclusion}
In this paper, we considered the problem of estimating accurate depth maps for AR/VR devices, which is an essential goal for immersive mixed-reality experience. For this problem, we proposed leveraging the mmWave communication systems that are deployed on the AR/VR devices to estimate and build high-resolution depth maps. We formulated the communication-constrained depth map sensing problem and proposed a comprehensive  framework for realizing this objective. The proposed framework includes (i) the construction of depth map specific sensing codebooks using practical mmWave antenna arrays and (ii) the development of efficient post-processing solutions for jointly processing the receive signals from the multiple sensing beams and estimating high-resolution depth maps. Simulations using accurate 3D ray-tracing models confirmed the promising accuracy of our proposed mmWave based depth map estimation approach in various environment scenarios.  In particular, the results show that the proposed approach can construct relatively high-resolution depth maps with less than 10cm error using practical mmWave systems. This highlights the potential of leveraging this solution to complement RGB-D based depth maps and realize immersive depth perception for wireless virtual/augmented reality systems. 

\section{Acknowledgment}  \label{sec:ack}
This work is sponsored by the Facebook Reality Lab. The authors would like to thank our WSLCL group members, Muhammad Alrabeiah and Andrew Hredzak, for their valuable help in building the Blender models and benchmarking the performance with the  RGB based depth estimation approaches.



\bibliographystyle{IEEEtran}

\begin{thebibliography}{10}
	\providecommand{\url}[1]{#1}
	\csname url@samestyle\endcsname
	\providecommand{\newblock}{\relax}
	\providecommand{\bibinfo}[2]{#2}
	\providecommand{\BIBentrySTDinterwordspacing}{\spaceskip=0pt\relax}
	\providecommand{\BIBentryALTinterwordstretchfactor}{4}
	\providecommand{\BIBentryALTinterwordspacing}{\spaceskip=\fontdimen2\font plus
		\BIBentryALTinterwordstretchfactor\fontdimen3\font minus
		\fontdimen4\font\relax}
	\providecommand{\BIBforeignlanguage}[2]{{%
			\expandafter\ifx\csname l@#1\endcsname\relax
			\typeout{** WARNING: IEEEtran.bst: No hyphenation pattern has been}%
			\typeout{** loaded for the language `#1'. Using the pattern for}%
			\typeout{** the default language instead.}%
			\else
			\language=\csname l@#1\endcsname
			\fi
			#2}}
	\providecommand{\BIBdecl}{\relax}
	\BIBdecl
	
	\bibitem{HTC}
	\BIBentryALTinterwordspacing
	VIVE, ``{VIVE Wireless Adapter}.'' [Online]. Available:
	\url{https://www.vive.com/us/wireless-adapter/}
	\BIBentrySTDinterwordspacing
	
	\bibitem{Hu2019}
	J.~Hu, M.~Ozay, Y.~Zhang, and T.~Okatani, ``{Revisiting Single Image Depth
		Estimation: Toward Higher Resolution Maps With Accurate Object Boundaries},''
	in \emph{Proc. of the IEEE Winter Conference on Applications of Computer
		Vision (WACV)}, 2019, pp. 1043--1051.
	
	\bibitem{Mal2018}
	F.~Mal and S.~Karaman, ``{Sparse-to-Dense: Depth Prediction from Sparse Depth
		Samples and a Single Image},'' in \emph{Proc. of International Conference on
		Robotics and Automation (ICRA)}.\hskip 1em plus 0.5em minus 0.4em\relax IEEE,
	2018, pp. 1--8.
	
	\bibitem{Riegler2019}
	G.~Riegler, Y.~Liao, S.~Donne, V.~Koltun, and A.~Geiger, ``{Connecting the
		Dots: Learning Representations for Active Monocular Depth Estimation},'' in
	\emph{Proc. of the IEEE Conference on Computer Vision and Pattern Recognition
		(CVPR)}, 2019, pp. 7624--7633.
	
	\bibitem{Zhang2018}
	Y.~Zhang, S.~Khamis, C.~Rhemann, J.~Valentin, A.~Kowdle, V.~Tankovich,
	M.~Schoenberg, S.~Izadi, T.~Funkhouser, and S.~Fanello, ``{Activestereonet:
		End-to-end Self-supervised Learning for Active Stereo Systems},'' in
	\emph{Proc. of the European Conference on Computer Vision (ECCV)}, 2018, pp.
	784--801.
	
	\bibitem{Cheng2019p}
	C.~Cheng, Y.~Wang, C.~Wei, C.~Chen, and L.~Lin, ``{IR Pattern Characteristics
		For Active Stereo Matching},'' US Patent 16/359,699, Oct.~3, 2019.
	
	\bibitem{Gruber2018}
	T.~Gruber, M.~Kokhova, W.~Ritter, N.~Haala, and K.~Dictmayer, ``{Learning
		Super-resolved Depth from Active gated Imaging},'' in \emph{Proc. of
		International Conference on Intelligent Transportation Systems (ITSC)}.\hskip
	1em plus 0.5em minus 0.4em\relax IEEE, 2018, pp. 3051--3058.
	
	\bibitem{Gruber2019}
	T.~Gruber, F.~Julca-Aguilar, M.~Bijelic, and F.~Heide, ``{Gated2depth:
		Real-time Dense Lidar From Gated Images},'' in \emph{Proc. of the IEEE
		Conference on Computer Vision and Pattern Recognition (CVPR)}, 2019, pp.
	1506--1516.
	
	\bibitem{IntelIRcamera}
	\BIBentryALTinterwordspacing
	Intel, ``{Intel RealSense Depth Camera D435}.'' [Online]. Available:
	\url{https://www.intelrealsense.com/depth-camera-d435/}
	\BIBentrySTDinterwordspacing
	
	\bibitem{Wang2019}
	Y.~Wang, W.-L. Chao, D.~Garg, B.~Hariharan, M.~Campbell, and K.~Q. Weinberger,
	``{Pseudo-lidar From Visual Depth Estimation: Bridging the Gap in 3D Object
		Detection for Autonomous Driving},'' in \emph{Proc. of the IEEE Conference on
		Computer Vision and Pattern Recognition (CVPR)}, 2019, pp. 8445--8453.
	
	\bibitem{Zhang_2018_CVPR}
	Y.~Zhang and T.~Funkhouser, ``{Deep Depth Completion of a Single RGB-D
		Image},'' in \emph{Proc. of the IEEE Conference on Computer Vision and
		Pattern Recognition (CVPR)}, June 2018.
	
	\bibitem{8108565}
	J.~A. {Zhang}, A.~{Cantoni}, X.~{Huang}, Y.~J. {Guo}, and R.~W. {Heath},
	``Joint communications and sensing using two steerable analog antenna
	arrays,'' in \emph{2017 IEEE 85th Vehicular Technology Conference (VTC
		Spring)}, June 2017, pp. 1--5.
	
	\bibitem{8114253}
	P.~{Kumari}, J.~{Choi}, N.~{González-Prelcic}, and R.~W. {Heath}, ``Ieee
	802.11ad-based radar: An approach to joint vehicular communication-radar
	system,'' \emph{IEEE Transactions on Vehicular Technology}, vol.~67, no.~4,
	pp. 3012--3027, April 2018.
	
	\bibitem{Soli}
	\BIBentryALTinterwordspacing
	J.~Lien, N.~Gillian, M.~E. Karagozler, P.~Amihood, C.~Schwesig, E.~Olson,
	H.~Raja, and I.~Poupyrev, ``Soli: Ubiquitous gesture sensing with millimeter
	wave radar,'' \emph{ACM Trans. Graph.}, vol.~35, no.~4, Jul. 2016. [Online].
	Available: \url{https://doi.org/10.1145/2897824.2925953}
	\BIBentrySTDinterwordspacing
	
	\bibitem{mmTrack}
	\BIBentryALTinterwordspacing
	T.~Wei and X.~Zhang, ``Mtrack: High-precision passive tracking using millimeter
	wave radios,'' in \emph{Proceedings of the 21st Annual International
		Conference on Mobile Computing and Networking}, ser. MobiCom ’15.\hskip 1em
	plus 0.5em minus 0.4em\relax New York, NY, USA: Association for Computing
	Machinery, 2015, p. 117–129. [Online]. Available:
	\url{https://doi.org/10.1145/2789168.2790113}
	\BIBentrySTDinterwordspacing
	
	\bibitem{Remcom}
	\BIBentryALTinterwordspacing
	Remcom, ``Wireless insite.'' [Online]. Available:
	\url{http://www.remcom.com/wireless-insite}
	\BIBentrySTDinterwordspacing
	
	\bibitem{Blender}
	\BIBentryALTinterwordspacing
	``Blender 2.80.'' [Online]. Available: \url{http://www.blender.org}
	\BIBentrySTDinterwordspacing
	
	\bibitem{BlenderKit}
	\BIBentryALTinterwordspacing
	``Blenderkit.'' [Online]. Available: \url{https://www.blenderkit.com/}
	\BIBentrySTDinterwordspacing
	
	\bibitem{BlenderDownloads}
	\BIBentryALTinterwordspacing
	``Blender demo files.'' [Online]. Available:
	\url{https://www.blender.org/download/demo-files/}
	\BIBentrySTDinterwordspacing
	
	\bibitem{TurboSquid}
	\BIBentryALTinterwordspacing
	``Turbosquid.'' [Online]. Available:
	\url{https://www.turbosquid.com/Search/3D-Models}
	\BIBentrySTDinterwordspacing
	
	\bibitem{Free3D}
	\BIBentryALTinterwordspacing
	``Free3d.'' [Online]. Available: \url{https://free3d.com/}
	\BIBentrySTDinterwordspacing
	
	\bibitem{Sabharwal2014}
	A.~Sabharwal, P.~Schniter, D.~Guo, D.~W. Bliss, S.~Rangarajan, and R.~Wichman,
	``{In-Band Full-Duplex Wireless: Challenges and Opportunities},'' \emph{IEEE
		Journal on Selected Areas in Communications}, vol.~32, no.~9, pp. 1637--1652,
	2014.
	
	\bibitem{Estep2014}
	N.~A. Estep, D.~L. Sounas, J.~Soric, and A.~Al{\`u}, ``{Magnetic-Free
		Non-Reciprocity and Isolation Based on Parametrically Modulated
		Coupled-Resonator Loops},'' \emph{Nature Physics}, vol.~10, no.~12, pp.
	923--927, 2014.
	
	\bibitem{Dinc2017}
	T.~Dinc and H.~Krishnaswamy, ``{A 28GHz Magnetic-Free Non-Reciprocal Passive
		CMOS Circulator Based on Spatio-Temporal Conductance Modulation},'' in
	\emph{Proc. of IEEE International Solid-State Circuits Conference
		(ISSCC)}.\hskip 1em plus 0.5em minus 0.4em\relax IEEE, 2017, pp. 294--295.
	
	\bibitem{Zhou2016}
	J.~Zhou, N.~Reiskarimian, and H.~Krishnaswamy, ``{Receiver with Integrated
		Magnetic-Free N-Path-Filter-Based Non-Reciprocal Circulator and Baseband
		Self-Interference Cancellation for Full-Duplex Wireless},'' in \emph{Proc. of
		IEEE International Solid-State Circuits Conference (ISSCC)}.\hskip 1em plus
	0.5em minus 0.4em\relax Institute of Electrical and Electronics Engineers
	Inc., 2016, pp. 178--180.
	
	\bibitem{Nagulu2019}
	A.~Nagulu and H.~Krishnaswamy, ``{Non-Magnetic 60GHz SOI CMOS Circulator Based
		on Loss/Dispersion-Engineered Switched Bandpass Filters},'' in \emph{Proc. of
		IEEE International Solid-State Circuits Conference (ISSCC)}.\hskip 1em plus
	0.5em minus 0.4em\relax IEEE, 2019, pp. 446--448.
	
	\bibitem{HeathJr2016}
	R.~W. Heath, N.~González-Prelcic, S.~Rangan, W.~Roh, and A.~M. Sayeed, ``An
	overview of signal processing techniques for millimeter wave {MIMO}
	systems,'' \emph{IEEE Journal of Selected Topics in Signal Processing},
	vol.~10, no.~3, pp. 436--453, April 2016.
	
	\bibitem{Alkhateeb2014d}
	A.~Alkhateeb, J.~Mo, N.~Gonzalez-Prelcic, and R.~Heath, ``{MIMO} precoding and
	combining solutions for millimeter-wave systems,'' \emph{IEEE Communications
		Magazine,}, vol.~52, no.~12, pp. 122--131, Dec. 2014.
	
	\bibitem{Richards2010}
	M.~A. Richards, J.~A. Scheer, and W.~A. Holm, \emph{{Principles of Modern
			Radar: Basic Principles}}.\hskip 1em plus 0.5em minus 0.4em\relax Raleigh,
	NC, USA: SciTech Publishing, 2010.
	
	\bibitem{Kumari2019}
	P.~Kumari, S.~A. Vorobyov, and R.~W. Heath~Jr, ``{Adaptive Virtual Waveform
		Design for Millimeter-Wave Joint Communication-Radar},'' \emph{arXiv preprint
		arXiv:1904.05516}, 2019.
	
	\bibitem{Spencer2000}
	Q.~Spencer, B.~Jeffs, M.~Jensen, and A.~Swindlehurst, ``{Modeling the
		Statistical Time and Angle of Arrival Characteristics of an Indoor Multipath
		Channel},'' \emph{IEEE Journal on Selected Areas in Communications}, vol.~18,
	no.~3, pp. 347--360, 2000.
	
	\bibitem{Smulders2009}
	P.~F. Smulders, ``{Statistical Characterization of 60-GHz Indoor Radio
		Channels},'' \emph{IEEE Transactions on Antennas and Propagation}, vol.~57,
	no.~10, pp. 2820--2829, 2009.
	
	\bibitem{ElAyach2014}
	O.~El~Ayach, S.~Rajagopal, S.~Abu-Surra, Z.~Pi, and R.~Heath, ``Spatially
	sparse precoding in millimeter wave {MIMO} systems,'' \emph{IEEE Transactions
		on Wireless Communications}, vol.~13, no.~3, pp. 1499--1513, Mar. 2014.
	
	\bibitem{Guerra2017}
	A.~Guerra, F.~Guidi, D.~Dardari, A.~Clemente, and R.~D'Errico, ``{A
		Millimeter-Wave Indoor Backscattering Channel Model for Environment
		Mapping},'' \emph{IEEE Transactions on Antennas and Propagation}, vol.~65,
	no.~9, pp. 4935--4940, 2017.
	
	\bibitem{Kay1993}
	S.~M. Kay, \emph{{Fundamentals of Statistical Signal Processing: Estimation
			Theory}}.\hskip 1em plus 0.5em minus 0.4em\relax Prentice Hall, 1993.
	
	\bibitem{Kumari2018}
	P.~Kumari, J.~Choi, N.~Gonz{\'a}lez-Prelcic, and R.~W. Heath, ``{IEEE
		802.11ad-Based Radar: An Approach to Joint Vehicular Communication-Radar
		System},'' \emph{IEEE Transactions on Vehicular Technology}, vol.~67, no.~4,
	pp. 3012--3027, 2018.
	
	\bibitem{Bidigare2012}
	P.~Bidigare, U.~Madhow, R.~Mudumbai, and D.~Scherber, ``{Attaining Fundamental
		Bounds on Timing Synchronization},'' in \emph{Proc. of International
		Conference on Acoustics, Speech and Signal Processing (ICASSP)}, 2012, pp.
	5229--5232.
	
	\bibitem{Herschfelt2018}
	A.~Herschfelt, H.~Yu, S.~Wu, H.~Lee, and D.~W. Bliss, ``{Joint
		Positioning-Communications System Design: Leveraging Phase-Accurate
		Time-of-Flight Estimation and Distributed Coherence},'' in \emph{Proc. of
		Asilomar Conference on Signals, Systems, and Computers (ACSSC)}, 2018, pp.
	433--437.
	
	\bibitem{Herschfelt2019D}
	A.~Herschfelt, ``{Simultaneous Positioning and Communications: Hybrid Radio
		Architecture, Estimation Techniques, and Experimental Validation},'' Ph.D.
	dissertation, Arizona State University, 2019.
	
	\bibitem{Alkhateeb2015a}
	A.~Alkhateeb and R.~W. Heath~Jr, ``Frequency selective hybrid precoding for
	limited feedback millimeter wave systems,'' \emph{submitted to IEEE
		Transactions on Communications, arXiv preprint arXiv:1510.00609}, 2015.
	
	\bibitem{Melvin2013}
	W.~L. Melvin and J.~A. Scheer, \emph{{Principles of Modern Radar Vol. II:
			Advanced Techniques}}.\hskip 1em plus 0.5em minus 0.4em\relax Edison, NJ,
	USA: SciTech Publishing, 2013.
	
	\bibitem{Dessouky2006}
	M.~I. Dessouky, H.~A. Sharshar, and Y.~A. Albagory, ``{Efficient Sidelobe
		Reduction Technique for Small-sized Concentric Circular Arrays},''
	\emph{Progress In Electromagnetics Research}, vol.~65, pp. 187--200, 2006.
	
	\bibitem{Albagory2007}
	Y.~A. Albagory, M.~Dessouky, and H.~Sharshar, ``{An Approach for Low Sidelobe
		Beamforming in Uniform Concentric Circular Arrays},'' \emph{Wireless Personal
		Communications}, vol.~43, no.~4, pp. 1363--1368, 2007.
	
	\bibitem{Grossi2019}
	E.~Grossi, M.~Lops, and L.~Venturino, ``{Detection and Localization of Multiple
		Targets in IEEE 802.11ad Networks},'' in \emph{Proc. of Asilomar Conference
		on Signals, Systems, and Computers (ACSSC)}, 2019.
	
	\bibitem{DeepMIMO2019}
	A.~Alkhateeb, ``Deepmimo: A generic deep learning dataset for millimeter wave
	and massive {MIMO} applications,'' in \emph{Proc. of Information Theory and
		Applications Workshop (ITA)}, San Diego, CA, Feb 2019, pp. 1--8.
	
	\bibitem{Silberman2012}
	N.~Silberman, D.~Hoiem, P.~Kohli, and R.~Fergus, ``{Indoor Segmentation and
		Support Inference from RGBD Images},'' in \emph{Proc. of European Conference
		on Computer Vision (ECCV)}, 2012.
	
\end{thebibliography}

\end{document}